\documentclass[%
aps,
prd,
amssymb,
amsmath,
amsfonts, 
eqsecnum,
showpacs,
nofootinbib,
floatfix,
twocolumn,
twoside,
a4paper,
superscriptaddress
]{revtex4-1}

\usepackage[T3,T1]{fontenc}
\usepackage{mathrsfs} 
\usepackage{bm} 
\makeatletter
\@ifclassloaded{beamer}
  { 
    \typeout{UsePackages: Detected beamer}
    \usepackage{tgheros}
    
  }
  { 
    \typeout{UsePackages: Did not detect beamer}
    \ifx\asybeamer\undefined 
    \typeout{UsePackages: Detected article}
    \usepackage[varg]{txfonts} 
    \usepackage{tgtermes} 
    \else 
    \typeout{Fonts: Detected asy for beamer}
    \usepackage{tgheros}
    
    \fi
  }
\usepackage{microtype} 
\def\MT@register@subst@font{
  \MT@exp@one@n\MT@in@clist\font@name\MT@font@list
  \ifMT@inlist@\else\xdef\MT@font@list{\MT@font@list\font@name,}\fi}
\makeatother

\usepackage{graphicx} %
\usepackage[x11names,svgnames,rgb]{xcolor} %
\usepackage{xspace}
\usepackage{braket}
\usepackage{accents}
\usepackage{siunitx}
\usepackage{mathtools}
\DeclareSymbolFontAlphabet{\mathrm}{operators}

\definecolor{CiteColor}{rgb}{0.18039, 0.18824, 0.57255}
\definecolor{UrlColor} {rgb}{0.741, 0.173, 0.000}
\definecolor{DarkUrlColor} {rgb}{0.500, 0.110, 0.000}
\definecolor{LinkColor}{rgb}{0.25098, 0.47843, 0.04706}

\makeatletter %
\newcommand{\ShowFont}{%
  \typeout{The main font is \f@encoding \space \f@family \space %
    \f@series \space \f@shape \space at \f@size pt.}%
  \typeout{The math font sizes are \tf@size pt (main), \sf@size pt %
    (script), and \ssf@size pt (scriptscript).}%
  \typeout{The linewidth is \the\linewidth}} %
\makeatother %

\usepackage{xargs}

\usepackage{graphicx}
\usepackage{dcolumn}
\usepackage{bm}
\usepackage{tabularx}
\usepackage{booktabs}
\usepackage{xspace}
\usepackage[utf8]{inputenc}

\usepackage[colorlinks]{hyperref}

\DeclareMathAlphabet{\mathbfsf}{\encodingdefault}{\sfdefault}{bx}{sl}

\newcommand{\n}{\newline}

\newcommand{\be}{\begin{equation}}
\newcommand{\ee}{\end{equation}}
\newcommand{\bea}{\begin{eqnarray}}
\newcommand{\eea}{\end{eqnarray}}

\newcommand{\phB}{\textsc{IMRPhenomB}\xspace}
\newcommand{\phC}{\textsc{IMRPhenomC}\xspace}
\newcommand{\phD}{\textsc{IMRPhenomD}\xspace}
\newcommand{\phX}{\textsc{IMRPhenomXAS}\xspace}
\newcommand{\phP}{\textsc{IMRPhenomP}\xspace}
\newcommand{\phHM}{\textsc{IMRPhenomHM}\xspace}
\newcommand{\phXHM}{\textsc{IMRPhenomXHM}\xspace}
\newcommand{\seobnr}{\textsc{SEOBNRv4}\xspace}
\newcommand{\seobnrrom}{\textsc{SEOBNRv4\_ROM}\xspace}

\newcommand{\fdamp}{f_{\rm{damp}}}
\newcommand{\frd}{f_{\rm{RD}}}

\newcommand{\vth} {{\boldsymbol{{\theta}}}}

\definecolor{ferngreen}{rgb}{0.31, 0.47, 0.26}
\definecolor{uniwienblue}{rgb}{0.23, 0.45, 0.67}
\definecolor{mediumpurple}{rgb}{0.58, 0.44, 0.86}
\definecolor{alizarincrimson}{rgb}{0.82, 0.1, 0.26}

\definecolor{dodgerblue}{HTML}{1E90FF}

\AtBeginDocument{%
  \hypersetup{
    citecolor=dodgerblue,
    linkcolor=dodgerblue,   
    urlcolor=dodgerblue}}

\newcommand{\UIB}{Departament de F\'isica, Universitat de les Illes Balears, IAC3 -- IEEC, Crta. Valldemossa km 7.5, E-07122 Palma, Spain}

\newcommand{\UoB}{School of Physics and Astronomy and Institute for Gravitational Wave Astronomy, University of Birmingham, Edgbaston, Birmingham, B15 9TT, United Kingdom}

\allowdisplaybreaks[1]

\begin{document}

\title[IMRPhenomX]
{Setting the cornerstone for the IMRPhenomX family of 
models for gravitational waves from compact binaries: The dominant harmonic for non-precessing quasi-circular black holes}


\author{Geraint Pratten}
\affiliation{\UoB}\affiliation{\UIB}

\author{Sascha Husa}
\affiliation{\UIB}

\author{Cecilio Garc{\'i}a-Quir{\'o}s}
\affiliation{\UIB}

\author{Marta Colleoni}
\affiliation{\UIB}

\author{Antoni Ramos-Buades}
\affiliation{\UIB}

\author{H{\'e}ctor Estell{\'e}s}
\affiliation{\UIB}


\author{Rafel Jaume}
\affiliation{\UIB}


\date{\today}

\begin{abstract}
In this paper we present \phX, a thorough overhaul of the \phD \cite{Husa:2015iqa,Khan:2015jqa} waveform model, which describes the dominant $l=2, \:\vert m \vert = 2$ spherical harmonic mode of non-precessing coalescing black holes in terms of piecewise closed form expressions in the frequency domain. Improvements include in particular the accurate treatment of unequal spin effects, and the inclusion of extreme mass ratio waveforms.
\phD has previously been extended to approximately include spin precession \cite{Hannam:2013oca} and subdominant spherical harmonics \cite{London:2017bcn}, and with its extensions it has become a standard tool in gravitational wave parameter estimation. Improved extensions of \phX 
are discussed in companion papers \cite{PhenXHM,PhenXP}.
\end{abstract}


\maketitle

\section{Introduction}
\label{sec:Introduction}

A key element of gravitational wave data analysis are waveform models, which serve as templates that detector data can be compared with, usually in the context of matched filter techniques, combined with template-bank based searches \cite{Usman:2015kfa,Sachdev:2019vvd}, or Bayesian inference \cite{Veitch:2014wba,Ashton:2018jfp}.
For general relativity, significant effort has been spent by the gravitational wave source modelling community to construct such models as approximate solutions of the Einstein equations, combining perturbative methods, numerical solutions, and qualitative insight.
The science case of gravitational wave astronomy is limited by the fidelity of the models to the complex physical processes they represent, and the computational efficiency of evaluating the models.
For GW150914 \cite{Abbott:2016blz,TheLIGOScientific:2016wfe}, the first detection and loudest binary black hole event of the first two observation runs \cite{LIGOScientific:2018mvr}, a detailed investigation of the effects of systematic errors in the waveform models was carried out in \cite{Abbott:2016wiq} for the models used for the analysis of the event: the time domain ``SEOBNR'' family of models based on the effective one-body approach \cite{Pan:2013rra,Taracchini:2013rva,Purrer:2015tud,Bohe:2016gbl,Babak:2016tgq}, and the phenomenological frequency domain model \phP \cite{Hannam:2013oca}. The latter extends the \phD model \cite{Husa:2015iqa,Khan:2015jqa} for
the dominant, i.e.~$l=2, \:\vert m \vert = 2$, spherical harmonic modes of quasicircular black hole binaries to include effects of spin-precession.
No evidence was found ``for a systematic bias relative to the statistical error of the original parameter recovery of GW150914 due to modeling approximations or modeling inaccuracies'', however more accurate models would be required for future observations.

In this paper we present \phX, a thorough update to the \phD model \cite{Husa:2015iqa,Khan:2015jqa}.
Following the phenomenological modelling framework, \phX is formulated in the frequency domain, and describes the waveform in terms of piecewise closed form expressions, with the aim to facilitate computationally efficient applications in gravitational wave data analysis. 
In a companion paper we present an extension to subdominant harmonics, which is aimed to supersede \phHM \cite{London:2017bcn}, which is only been calibrated to numerical relativity data for the dominant quadrupole spherical harmonic. In a second companion paper \cite{our_mb} we present a method to accelerate the evaluation of the waveform model, based on earlier work by Vinciguerra et al.~ \cite{Vinciguerra:2017ngf}.
In future work we will discuss extending our model to precession using the methods of \phP \cite{Hannam:2013oca} and \cite{Khan:2018fmp}.

The main elements of the model construction are chosen as follows:
The waveform is split into two real non-oscillatory functions, an amplitude and phase. Modelling then proceeds in two steps: first, closed form expressions are fitted to numerical waveform data for a set of calibration waveforms. These calibration waveforms are constructed as hybrid waveforms, appropriately gluing together approximate waveforms describing the inspiral, which in our case we take as the SEOBNRv4 version of an effective-one-body (EOB) model \cite{Bohe:2016gbl}, and numerical relativity waveforms, which describe the last orbits, merger, and ringdown of the system.
Finding appropriate analytical functions becomes easier as the frequency region for the fitting procedure is broken up into smaller regions. As with previous phenomenological waveform models, we choose three such regions, where the ansatz in each one and the choice of transition frequencies are guided by perturbative descriptions and physical intuition: (i) A low frequency inspiral regime, where the waveform can be described by adding additional terms to a post-Newtonian expansion. (ii) A high frequency regime where the waveform is dominated by quasi-normal ringdown.
(iii) An intermediate regime, which captures the complex physics of the merger and the transition between the physics of the inspiral and the ringdown,
where neither the post-Newtonian nor the quasi-normal-ringdown perturbative descriptions apply.

The result of the first step is a set of coefficients for each numerical waveform, which greatly compress the information used to accurately represent each waveform. In a second step, each coefficient is then modelled across the three-dimensional parameter space of non-precessing quasi-circular black hole binaries, described by mass ratio and the two spin components orthogonal to the orbital plane.
In contrast to previous frequency domain phenomenological models,
\phX captures the full spin dependence of the waveform: In the models preceding \phD \cite{Ajith:2009bn,Santamaria:2010yb}, a single effective spin was used to model the spin-dependence. In \phD, different effective spins were used to model the inspiral and high-frequency regime, which already significantly reduced the parameter bias for unequal spin cases \cite{Kumar:2016dhh}. In \phX, 
we use the hierarchical fitting method developed in \cite{Jimenez-Forteza:2016oae} to finally treat the full three-dimensional parameter space.

\phX also significantly increases the validity range in the mass ratio of any previous phenomenological model by including extreme mass ratio waveforms up to mass ratio 1000, which were computed by hybridizing numerical solutions of the perturbative Teukolsky equation \cite{Harms:2014dqa,Harms:2015ixa,Harms:2016ctx} as described in \cite{hybrids}. 
Due to improvements in the model construction and the larger number of input calibration waveforms, \phX offers a significant improvement in accuracy, showing $\sim 1$-$2$ orders of magnitude improvement in the mismatch compared to \phD. 
For a list of the key features implemented in \phX see Sec.~\ref{sec:summary}.

The paper is organized as follows: First we provide a detailed discussion of our conventions in Sec.~\ref{sec:conventions}.
Then we present our input waveforms in Sec.~\ref{sec:input}.
The mapping between phenomenological coefficients and physical parameters is discussed in Sec.~\ref{sec:parspacefits}, and the choice of transition frequencies between the model's three frequency regions is treated in Sec.~\ref{sec:regions}.
The model construction for the amplitude and phase is then described in Secs.~\ref{sec:amp} and \ref{sec:phase}, and an example of our use of the hierachical fitting procedure for parameter space fits is provided in Sec.~\ref{sec:worked}.
In  Sec.~\ref{sec:validation} we describe how we have validated our model, and we conclude with a summary and discussion of our work in Sec.~\ref{sec:summary}. Appendix \ref{appendix:TaylorF2} provides the details of the post-Newtonian TaylorF2 approximant as we use it.

\section{Conventions and Preliminaries}\label{sec:conventions}

\subsection{Intrinsic parameters conventions}\label{sec:intrinsic_parameters_conventions}

We consider binary systems of astrophysical black holes in general relativity, which do not exhibit spin precession and are quasi-circular (non-eccentric).
In the limit of large separation, each black hole is perfectly described by the Kerr solution, and the initial conditions for the dynamics are given by the position and velocity vectors (or equivalently momenta) of the two black holes. In this limit the momenta correspond to Newtonian particles in a circular orbit, and we will adopt the center-of-mass frame. The intrinsic parameters $\vth$ of such systems correspond to the 
dimensionless projections of the BH spins (intrinsic angular momenta) $\vec S_{1,2}$ in the preserved direction of the orbital angular momentum $\vec L$, and the masses $m_{1,2}$, where
\begin{equation}
\chi_i = \frac{\vec S_i \cdot \vec L}{m_i^2 \, \vert \vec L\vert}.
\end{equation}
We define the mass ratio $q = m_1/m_2 \geq 1$, total mass $M=m_1+m_2$, and symmetric mass ratio $\eta = m_1 m_2 /M^2$.

\subsection{Waveform conventions}\label{sec:wfconventions}

\phX models the $\ell=\vert m \vert = 2$ spherical harmonic modes of the coalescence of binary systems of non-precessing quasi-circular black holes. We assume a sense of rotation of the binary consistent with a right-handed coordinate system:
The orbital frequency vector $\vec \omega$ is chosen in the direction of the $z$-axis of a Cartesian coordinate system $(x,y,z)$. The black holes orbit in the plane $z=0$, and the spacetime, and thus the gravitational-wave signal, exhibits equatorial symmetry, i.e. the northern hemisphere $z \geq 0$ is isometric to the southern hemisphere $z \leq 0$.

We introduce a standard spherical coordinate system
\begin{equation}\label{eq:defcoords}
x = r \cos \varphi \cos\vartheta, \quad 
y = r \sin \varphi \cos\vartheta, \quad   
z = r \cos \vartheta,
\end{equation}
and  spherical harmonics $Y^{-2}_{\ell m}$ of spin-weight $-2$ (see e.g.~\cite{Wiaux:2005fm}), where here we will only require the modes:
\begin{equation}\label{eq:defYlm}
  Y^{-2}_{2\pm2} = \sqrt{\frac{5}{64\pi}} \left(1 \pm\cos \vartheta \right)^2
       e^{\pm2i\varphi}.
\end{equation}
The gravitational-wave strain $h$ depends on an inertial time coordinate $t$, the angles $\vartheta, \varphi$ in the sky of the source, and the source parameters $\vth$. It is written in terms of spherical harmonic modes $h_{\ell m}$, and alternatively in terms of gravitational wave polarizations $h_{+}$ and $h_{\times}$ as
\begin{align}\label{eq:h_harmonics}
    h(t,\vartheta,\varphi; \vth ) &=
    \displaystyle\sum_{m = -2,2} h_{2m} \, (t ; \vth) \; _{-2}Y_{2m} (\vartheta,\varphi), \\ \label{eq:h_polarizations}
    &= h_+(t , \vartheta , \varphi ; \vth )  - i \, h_{\times} \, (t , \vartheta , \varphi  ; \vth)
    .
\end{align}

We define the Fourier transform to be consistent with the conventions adopted in the LIGO Algorithms Library \cite{lalsuite}
\begin{equation}\label{eq:defFT}
\tilde h(f)=\int_{-\infty}^{\infty} h(t) \, e^{-i \, 2 \pi f t}\, dt .
\end{equation}
With this convention of the Fourier transform time derivatives are converted to multiplications by factors of $i 2 \pi f$ in the Fourier domain.
The frequency domain strain $\tilde{h}$ can then be written in the form of the time domain strain in (\ref{eq:h_harmonics},\ref{eq:h_polarizations}),
\begin{align}
\tilde{h} (f  , \vartheta , \varphi ; \vth) 
&= \displaystyle\sum_{m = -2,2} \tilde{h}_{2m} \, (f ; \vth) \; _{-2}Y_{2m} (\vartheta,\varphi) \\
&= \tilde{h}_+ \, (f , \vartheta , \varphi ; \vth )  - i \, \tilde{h}_{\times} \, (f  , \vartheta , \varphi; \vth ) .
\end{align}
%
The equatorial symmetry of non-precessing binaries implies
\begin{equation}
    h_{22}(t) = h^{*}_{2-2}(t), \quad 
\end{equation}
it is thus sufficient to model just one spherical harmonic.
For the Fourier transform this leads to    
\begin{equation}\label{eq:htilde_mode_relations}
    \tilde h_{22}(f) = \tilde h^{*}_{2-2}(-f).
\end{equation}
Our above choices imply that if the time domain modes are written in terms of a positive amplitude $a_{TD}(t)$ and a phase $\phi_{\rm{TD}}(t)$, then
\begin{equation}\label{def:time_amp_phi}
    h_{22}(t)  = a_{\rm{TD}}(t) e^{- i \phi_{\rm{TD}}(t)}, \quad  
    h_{2-2}(t) = a_{\rm{TD}}(t) e^{  i \phi_{\rm{TD}}(t)},
\end{equation}
Since we assume negligible eccentricity the frequency
time derivative of the phases of both modes are monotonic functions of $t$. We will assume right-handed circular motion for the binary, with the rotation axis being the z-axis defined by (\ref{eq:defcoords}). With the definitions (\ref{def:time_amp_phi},\ref{eq:defYlm}) this implies that $\phi_{\rm{TD}}(t)$ is then a monotonically {\em increasing} function of time and
the gravitational-wave polarizations in the time domain are given by
\begin{eqnarray}\label{def:polarizations_TD}
    h_{+}(t)      &=& +\sqrt{\frac{5}{4\pi}} \frac{1+\cos^2 \vartheta}{2} \cdot a_{\rm{TD}}  \cos(2\varphi - \phi_{\rm{TD}}), \\ 
    h_{\times}(t) &=& -\sqrt{\frac{5}{4\pi}} \cos\vartheta \cdot a_{\rm{TD}}  \sin(2\varphi - \phi_{\rm{TD}}).
\end{eqnarray}

With our convention for the Fourier transformation (\ref{eq:defFT}),
the definitions above imply that $\tilde h_{22}(f)$ is concentrated
in the negative frequency domain and $\tilde h_{2-2}(f)$ 
in the positive frequency domain. For the inspiral, this can be checked against the stationary phase approximation (SPA), see e.g. \cite{Finn:1992xs,Cutler:1994ys,Damour:2000gg} or the derivation in Appendix~\ref{appendix:SPA}.

We construct our model in the frequency domain, it is thus convenient to model the $\tilde h_{2-2}$, which is non-zero for positive frequencies. The mode $\tilde h_{22}$, defined for negative frequencies, can then be computed from (\ref{eq:htilde_mode_relations}).
We model the Fourier amplitude $A(f > 0,\vth)$, which is a positive function for positive frequencies, and zero otherwise, and the Fourier domain phase $\phi (f > 0,\vth)$, defined by 
\begin{align}
\tilde{h}_{2-2} (f,\vth) = A (f,\vth) \, e^{-i \phi (f,\vth)} .
\end{align}

The gravitational wave polarizations in the frequency domain are then given by
\begin{eqnarray}\label{def:polarizations_FD}
    \tilde h_{+}(f)      &=& +\sqrt{\frac{5}{16\pi}} \frac{1+\cos^2 \vartheta}{2} \cdot \tilde h_{2-2} e^{-i 2 \varphi}, \\ 
    \tilde h_{\times}(f) &=& -i \sqrt{\frac{5}{16\pi}} \cos\vartheta \cdot \tilde h_{2-2} e^{-i 2 \varphi}.
\end{eqnarray}
When one only carries out computations with the projections of the gravitational strain onto detectors, i.e.~specific polarizations, one only deals with Fourier transforms of real functions, and only positive frequencies are required.

Note that with the above definitions, for a face-on binary, i.e.~$\theta=0$, we get that $\tilde h = \tilde{h}_+ - i \:\tilde{h}_\times=0$. This does not mean that the signal vanishes for face-off binaries, but that when working with the full waveform without projection onto specific polarizations, one would also need to explicitly 
consider negative frequencies.

As a consequence of time derivatives being related to multiplication in
Fourier space, the conversion between the GW strain and the Newman-Penrose scalar $\psi_4$, where
\begin{equation}
\frac{d^2 h(t)}{dt^2} = \psi_4(t),
\end{equation}
is given by
\begin{equation}
\tilde{h}(f) = -\frac{\tilde{\psi}_4(f)}{4 \pi^2 f^2},
\end{equation}
and only affects the Fourier domain amplitude, but not the phase, up to a jump of $\pi$, and apart from possible effects specific to the numerical algorithm used to carry out this conversion. 

A time shift of the waveform is encoded only in the Fourier domain phase, but at the price of changing the shape of phase function. The Fourier transformation of a time and phase shifted function $h_{\varphi_0 , \tau} =  h(t - \tau) \, e^{i \varphi_0}$  is given by
\begin{align}
    \tilde{h}_{\varphi_0 , \tau} (\omega) &= \int^{\infty}_{-\infty} h(t - \tau) e^{i \varphi_0} \, e^{- i \omega t} dt \nonumber \\
    &= e^{i (\varphi_0 + \omega \tau)} \, \int^{\infty}_{-\infty} h(t^{\prime}) e^{- i \omega t^{\prime}} d t^{\prime} , \nonumber \\
    &= e^{i (\varphi_0 + \omega \tau)} \tilde{h} ,
\end{align}
and thus corresponds to an additional term in the phase, which is linear in frequency.

An additional ambiguity arises due to the choice of tetrad defining the polarizations. This is discussed in detail in \cite{Bustillo:2015ova,hybrids} and, for the dominant quadrupole mode, is equivalent to a fixed global rotation of the source.

\subsection{Waveform phenomenology preliminaries}

A detailed discussion of the phenomenology of the inspiral-merger-ringdown frequency domain waveforms modelled here has been given in \cite{Husa:2015iqa}. Here we summarize some of the key features which are most relevant for our modelling strategy.

Our goal is to describe the amplitude and phase of the waveform by piecewise closed form expressions, which are valid in some frequency interval. Using more such intervals makes the modelling of each interval simpler: e.g.
when using a sufficient number of intervals a cubic spline representation may be sufficient. Using a smaller number of intervals makes it harder to find an appropriate analytical form for each interval. 

In this work, we will use three regimes: For the inspiral, i.e.~for low frequencies, it is natural to describe the waveform in the framework of post-Newtonian theory (see e.g. \cite{Blanchet2014}) as a Taylor expansion in powers of $v/c$, where $v$ is an orbital velocity parameter and $c$ the speed of light, or equivalently a frequency $f$, where $\pi f= (v/c)^3$. One can then simply add higher order terms in $v/c$, often referred to as pseudo-post-Newtonian terms, where as-of-yet unknown coefficients are calibrated to the data set of numerical waveforms.
More concretely, we will base our inspiral description on the standard TaylorF2 approximant \cite{Damour:2001tu,Damour:2002kr,Arun:2004hn,Buonanno:2009zt}, which provides closed form expressions for the amplitude and phase of the Fourier transform of the gravitational wave strain for quasicircular inspirals, and is derived from time domain post-Newtonian expressions via the stationary phase approximation, see Appendices \ref{appendix:TaylorF2} and \ref{appendix:SPA}. We augment the known TaylorF2 series with higher order terms as described in Sec.~\ref{sec:amp_insp} for the amplitude, and in Sec.~\ref{sec:phase_insp} for the phase.

After the merger, the relaxation of the excited final black hole
to the Kerr solution can be described by black hole perturbation theory and quasinormal ringdown behaviour \cite{Kokkotas1999}. While the stationary phase approximation is not valid for the merger and ringdown, it has long been known that simple models of damped oscillations can be Fourier-transformed analytically, and thus can serve to analytically model key features of the ringdown in the frequency domain. We will briefly discuss such models below, and how they can be used to form the basis of the closed-form frequency domain model we want to construct.

We use a third, ``intermediate", frequency regime to capture the transition between the inspiral and ringdown regimes.
This transition regime roughly corresponds to the merger, and models the complex physics that occurs when the spacetime is highly dynamical and so far eludes a perturbative treatment.
A crucial element of modelling this intermediate regime is to find an appropriate start frequency, when the inspiral breaks down, 
in the sense that an inconveniently large number of post-Newtonian orders would be required for an accurate description.
For extreme mass ratios, the innermost circular orbit gives a good estimate of this frequency, but it is not appropriate for 
comparable masses, where we have found the minimal energy circular orbit (MECO) as defined by \cite{Cabero:2016ayq} provides a good estimate.

We now return to the description of the ringdown in the frequency domain, and will discuss simple analytical models to motivate how we proceed.
For the simple damped oscillation
\begin{equation}\label{eq:RDFTexample}
h(t) = \Theta(t) e^{2 \pi t (i \frd - t \fdamp)},
\end{equation}
where $\Theta(t)$ denotes the Heaviside theta function, the Fourier
transform is
\begin{equation}
\tilde h(f) =  \frac{1}{2\pi \left(\fdamp + i (f  - \frd)\right)} \,,  
\end{equation}
with absolute value
\begin{equation}
\vert \tilde h(f) \vert =  \frac{1}{2 \pi \sqrt{\fdamp^2 + (f - \frd)^2}}.
\end{equation}
and phase derivative
\begin{equation}
\frac{d\arg \tilde h(f)}{df} = -\frac{1}{2\pi} \frac{\fdamp}{\fdamp^2 + (f - \frd)^2}.
\end{equation}
The quasi-normal mode frequencies are thus imprinted on the Fourier domain amplitude and phase derivative through a Lorentzian function
for the phase derivative, and its square root for the amplitude, with a falloff of $1/f$ for high frequencies.
The physical waveform should however fall off faster than any polynomial due to its smoothness, since for smooth functions $h(t)$ one has that (see e.g.~\cite{TrefethenWeb1996})
\begin{equation}
\tilde h(f) = O(\vert  f \vert^{-\vert M \vert}) \quad as \quad \vert f\vert \rightarrow \infty \quad\mbox{for all}\, M .
\end{equation}
If a function $h(t)$ only has $p$ continuous derivatives in $L^2$ for some $p\geq 0$  and a $p$th derivative in $L^2$ of bounded variation, then \cite{TrefethenWeb1996} (see also the discussion in \cite{Husa:2015iqa}):
\begin{equation}\label{eq:FourierFalloff}
\tilde h(f) = O(\vert  f \vert^{-p-1}) \quad as \quad \vert f\vert \rightarrow \infty.
\end{equation}
Other variants of the example of Eq.~(\ref{eq:RDFTexample}) are e.g. replacing $t$ by $\vert t \vert$ and dropping the theta function, or replacing the complex exponential by a sine or cosine (see e.g.~
 \cite{Luna:2006gw,Mehta:2019wxm}), which leads to minor modifications in the results, such as a faster polynomial falloff of the amplitude.

This has motivated to model the frequency domain ringdown amplitude as a Lorentzian for the early phenomenological waveform models \cite{Ajith:2007qp,Ajith:2007kx,Ajith:2009bn,Santamaria:2010yb}.
For the \phD model the Lorentzian amplitude ansatz has been modified with a decaying exponential to be consistent with the falloff expected from smooth functions, and the falloff rate has been calibrated to numerical relativity waveforms. A rough estimate of the falloff rate can be obtained with a smooth ansatz for the time domain waveform. Inspecting the Newman-Penrose quantity $\psi_4$ around the merger for numerical relativity waveforms, one finds that it is roughly symmetric around the peak. This symmetry has also been found in a recent approximate analytical calculation \cite{McWilliams:2018ztb}. Following \cite{McWilliams:2018ztb} for the amplitude ansatz for $\psi_4$, but making the unrealistic assumption that the gravitational wave frequency is constant around the amplitude peak, we get that for
\begin{equation}\label{eq:RDFTexamplepsi4}
\psi_4(t) = \frac{e^{2 \pi t (i \frd -  \fdamp)}}{e^{2 \pi t \fdamp} +  e^{-2 \pi t \fdamp} }
\end{equation}
the Fourier transform is
\begin{equation}
\tilde \psi_4(f) = \frac{1}{2\pi \left(\fdamp + i (f  - \frd)\right)} \,,  
\end{equation}
For large frequencies the amplitude falls off as
\begin{equation}
\tilde \psi_4(f) \sim e^{-\pi(f-\frd)/(2 \fdamp)}. 
\end{equation}
We have compared the asymptotic falloff rate of $-\pi/(2\fdamp)$
with our hybrid data set, and find that it typically overestimates the numerical data, but only by a factor within $1.32 - 1.38$ for 90\% of the cases, which is surprisingly good giving the crudeness of the model (\ref{eq:RDFTexamplepsi4}).
For the ringdown amplitude, we will thus essentially follow the \phD ansatz of a Lorentzian, multiplied with an exponential damping factor.

\begin{figure}[h!]
\begin{center}
\includegraphics[width=\columnwidth]{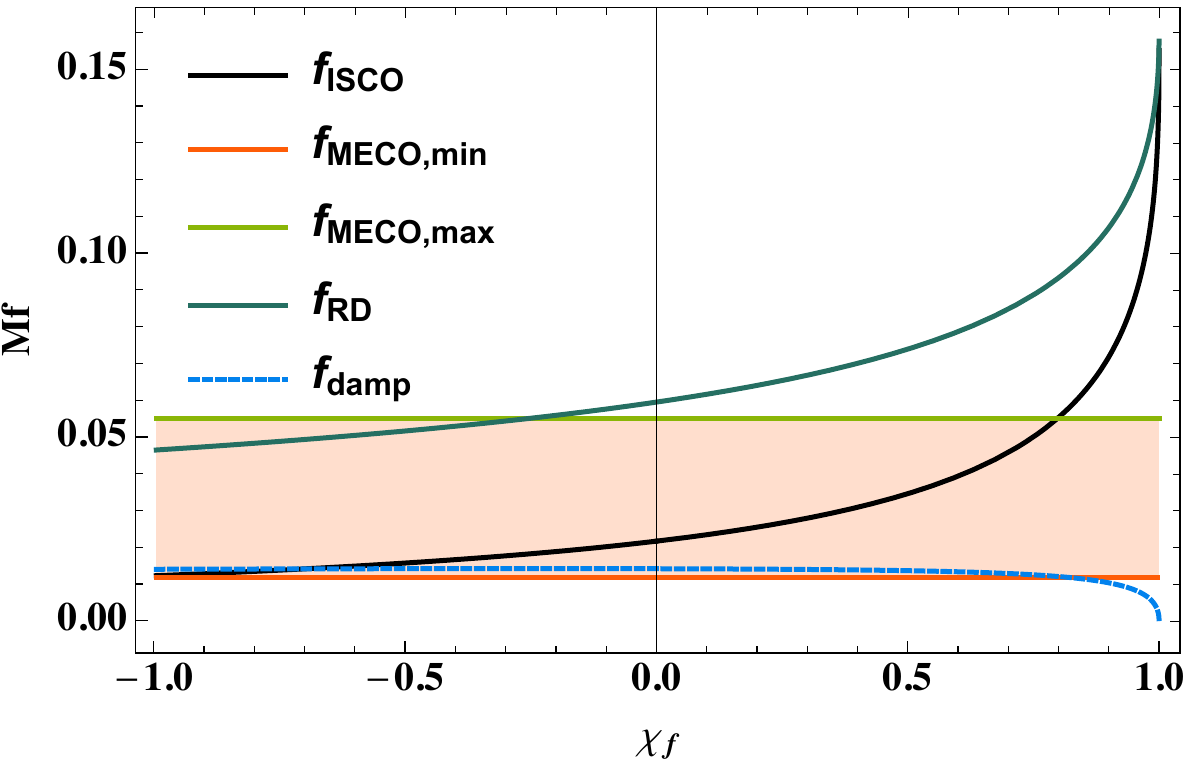}
\caption{
The ringdown and damping frequencies $\frd$ and $\fdamp$, and the ISCO frequency are plotted as functions of the dimensionless Kerr parameter $\chi_f$ of a black hole, a negative sign of $\chi_f$ indicates that the orbital angular momentum points in the opposite direction as the spin of the (final) black hole.
Also plotted are the maximal and minimal MECO frequencies across the non-precessing parameter space (the maximum occurs for equal black holes with maximal component spins aligned with the orbital angular momentum, and the minimum for the extreme mass ratio limit with maximal anti-aligned component spins.
}
\label{fig:frequencies}
\end{center}
\end{figure}

For the phase derivative in the high frequency regime, modifications of the simple model leads to Lorentzians with added background terms, in the form
\begin{equation}\label{eq:basicLorentzian}
    \frac{d\phi}{df} = \frac{a}{(f-\frd)^2 + (2 \fdamp)^2} + \mbox{background},
\end{equation}
which are consistent with our numerical data. For \phD the ringdown regime was thus modelled as a Lorentzian, plus a polynomial in $f^{-1}$. We will follow the same strategy with two main modifications: First we will modify how to represent the polynomial that models the ``background" term. Second, we note that $\fdamp$ has a very large dynamic range. For negative spins, $\fdamp$ is quite large and leads to very broad Lorentzians, which are not confined to the ringdown region.
An overview of frequencies that play a crucial role in designing our modelling approach is shown in Fig.~\ref{fig:frequencies}.
As the figure shows, all the frequencies exhibit significant variation, which requires a corresponding dynamic range in the transition frequencies between the three frequency regions of the model.

The loss of accuracy and gradual breakdown of the post-Newtonian series expansion for high frequencies as the merger is approached also determines the frequency regime where numerical solutions are required to provide unambiguous approximate solutions and error estimates. The computational cost of such simulations increases drastically as the initial frequency is lowered, the leading order post-Newtonian estimate for the time to merger $T$ for start frequency $f_0$ is
\begin{equation}
    T \propto \frac{f_0^{-8/3}}{\eta}.
\end{equation}
Covering densely the parameter space that we want to model with numerical relativity waveforms with start frequencies lower than the sensitive band of our detectors is still prohibitively expensive, and we will thus use ``hybrid" waveforms as our input data set, where numerical relativity waveforms are appropriately glued to an inspiral description derived from the post-Newtonian expansion, as discussed below in Sec.~\ref{sec:input}.

\section{Input waveforms}\label{sec:input}

The primary input data we use for developing, calibrating, and evaluating the \phX waveform model is the set of hybrid waveforms described in \cite{hybrids}. These waveforms are constructed by appropriately gluing together numerical waveforms, which cover the last orbits, merger and ringdown, with an inspiral model. For comparable masses up to mass ratio 18, the numerical waveforms have been computed by solving the full Einstein field equations using the methods of numerical relativity. The inspiral model is taken to be an EOB resummation of post-Newtonian waveforms. For extreme mass ratios 
we use numerical solutions for linearized gravitational waves in a Kerr background, sourced by EOB dynamics, as described below.

The EOB approach provides a framework to extend the validity of post-Newtonian results \cite{Buonanno:1998gg,Buonanno1999} with resummation techniques, and to incorporate additional information, such as a calibration to numerical relativity results, which has lead to families of time-domain models for the complete waveform, from inspiral to ringdown 
\cite{Taracchini:2013rva,Bohe:2016gbl,Nagar:2018zoe}. Here we use the recent SEOBNRv4 \cite{Bohe:2016gbl} EOB model to hybridize with numerical relativity waveforms.
For \phD, the SEOBNRv2 approximant \cite{Taracchini:2013rva} was used, removing however the calibrations to numerical relativity in order to decrease the dependence between the 
two models. Here we use the original calibrated SEOBNRv4 model, as our goal is to maximize the accuracy of the resulting waveform model.

SEOBNRv4 has in fact been calibrated to numerical waveforms, and describes the complete waveform from inspiral to ringdown. We could
thus also augment our calibration data set with SEOBNRv4 waveforms in regions of the parameter space where numerical relativity waveforms are sparse. In the model presented here, we only use such SEOBNRv4 waveforms at low frequencies in the inspiral, i.e. well below the MECO frequency, where very little to no NR information is present. For the intermediate and merger-ringdown regions, only EOB-NR hybrids and test-particle hybrids are used to calibrate the waveform model. 

The comparable mass numerical relativity waveforms used in the hybrid data set have been produced with SpEC~\cite{Ossokine:2013zga,Hemberger:2012jz,Szilagyi:2009qz,Scheel:2008rj,Boyle:2007ft,Mroue:2013xna, Buchman:2012dw}),
which uses pseudo-spectral numerical methods and black-hole excision, as well as with the BAM \cite{Bruegmann:2006at,Husa:2007hp} and Einstein Toolkit \cite{EinsteinToolkit:2019_10} codes. 

The BAM code solves the $3+1$ decomposed Einstein field equations using the $\chi$-variant \cite{Campanelli:2005dd} of the moving-punctures implementation of the BSSN formulation \cite{Nakamura:1987zz,Shibata:1995we,Baumgarte:1998te}. Spatial derivatives are computed using sixth-order accurate finite differencing stencils \cite{Husa:2007hp}. Kreiss-Oliger dissipation terms converge at fifth order \cite{}, and a fourth-order Runge-Kutta algorithm is used for the time evolution. BBH  puncture  initial  data  \cite{Brandt:1997tf,Bowen:1980yu}  are  calculated  with  a pseudo-spectral  elliptic  solver  described  in  Ref \cite{Ansorg:2004ds}. The GWs are calculated using the Newman-Penrose scalar $\psi_4$ and extracted at a finite distance from the source.

The Einstein-Toolkit simulations use Bowen-York initial data \cite{Bowen:1980yu,Brandt:1997tf} computed using the \texttt{TwoPunctures} thorn \cite{Ansorg:2004ds}. Time evolution is performed using the $W$-variant \cite{Marronetti:2007wz} of the BSSN formulation of the Einstein field equations as implemented by \texttt{McLachlan} \cite{Brown:2008sb}. The BHs are evolved using standard moving punctures gauge-conditions \cite{Baker:2005vv,Campanelli:2005dd}. The lapse is evolved according to the $1+\log$ condition \cite{Bona:1994dr} and the shift according to the hyperbolic $\tilde{\Gamma}$-driver \cite{Alcubierre:2002kk}. Simulations are performed using 8th order accurate finite difference stencils with Kreiss-Oliger dissipation\cite{KreissOliger}. Adaptive mesh refinement is provided by \texttt{Carpet} \cite{Schnetter:2003rb,Schnetter:2006pg}, with the wave-extraction zone being computed on spherical grids using the \texttt{Llama} multipatch infrastructure \cite{Pollney:2009yz}. Low eccentricity initial data is produced following the procedure outlined in \cite{Ramos-Buades:2018azo}. Further details will be given in \cite{hybrids}. 

SpEC is a multi-domain pseudo-spectral code \cite{Lindblom:2005qh,Szilagyi:2009qz,Hemberger:2012jz} that uses excision to remove the BH interiors, thereby removing the BH singularity from the computational domain. The code evolves the Generalised Harmonic coordinate formulation of the Einstein field equations \cite{Friedrich:1985aa,Pretorius:2004jg,Garfinkle:2001ni,Lindblom:2005qh} with constraint damping. Initial data is constructed using the Extended Conformal Thin Sandwich (XCTS) equations \cite{York:1998hy,Pfeiffer:2002iy,Pfeiffer:2002wt}, with newer simulations typically choosing the conformal metric and trace of the extrinsic curvature to be a weighted superposition of two single BHs in Kerr-Schild coordinates \cite{Lovelace:2008tw}. Boundary conditions imposed on the excision boundaries ensure that these boundaries are apparent horizons \cite{Szilagyi:2009qz,Scheel:2008rj,Hemberger:2012jz,Lindblom:2005qh}. Further details can be found in \cite{Boyle:2019kee}. 

We used 186 waveforms from the public SXS catalog as of 2018 \cite{Mroue:2013xna}. After the release of the latest SXS collaboration catalog, \cite{Boyle:2019kee}, we extended the dataset to incorporate 355 SpEC simulations and updated the parameter space fits for the phase accordingly. We opted not to update the amplitude fits to incorporate the latest SpEC simulations as this is anticipated to have a smaller impact on the overall accuracy of the waveform model. 

The 95 BAM waveforms consists of previously published and new waveforms. The Einstein Toolkit simulations have been recently produced by the authors. For further details on the BAM and 
Einstein Toolkit waveforms see \cite{hybrids}.

The key data sets that determine the calibration range of our waveform model are
the BAM waveforms for a range of spins at mass ratio 1:18, high-spin BAM and SXS waveforms at mass ratios 4 and 8, and equal mass SXS data sets at very high spins of $-0.95$ and $+0.994$. 

\begin{figure}[h!]
\begin{center}
\includegraphics[width=\columnwidth]{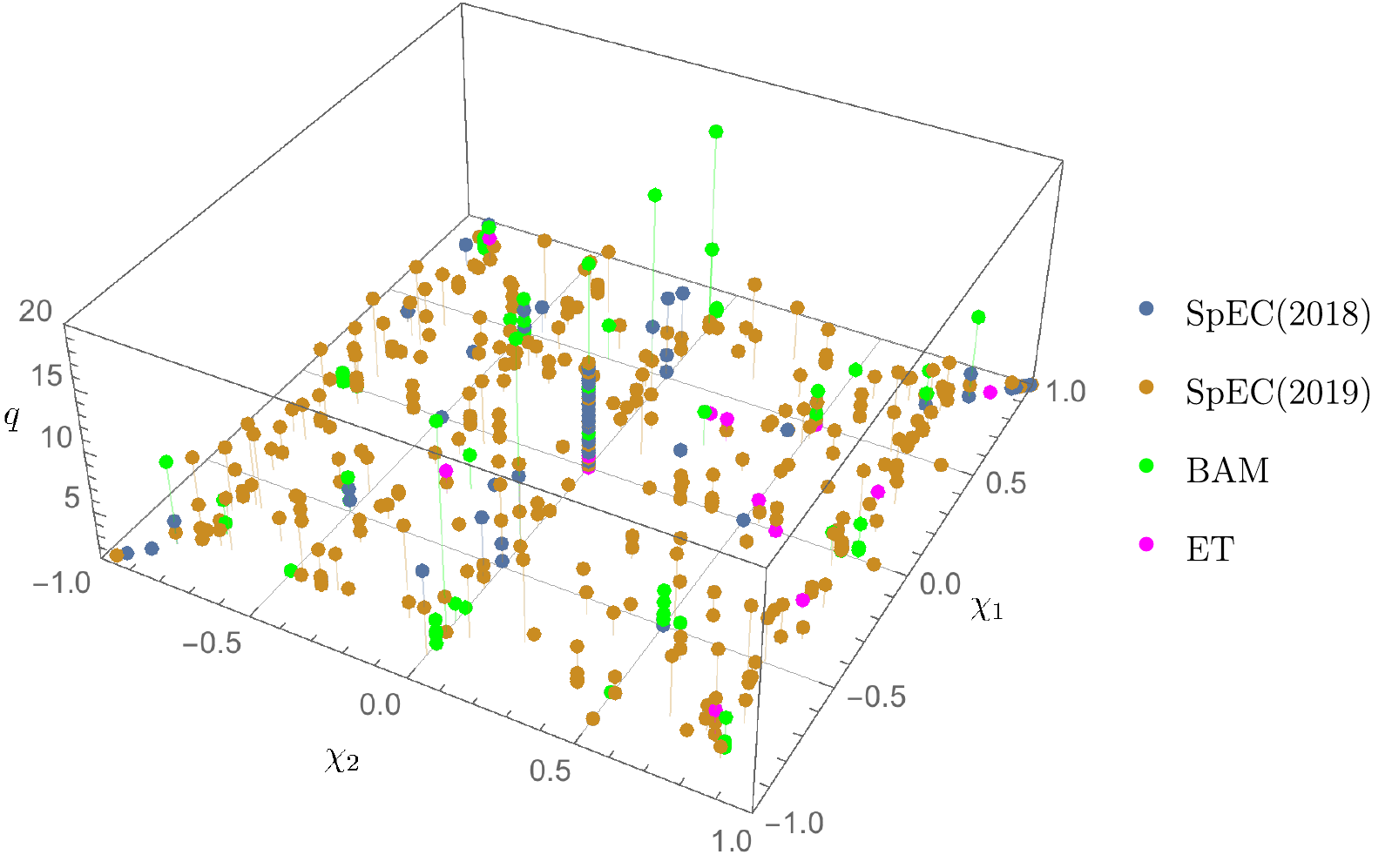}
\caption{
The mass ratio and spins for the NR waveforms used in the calibration of \phX. SXS simulations are shown in blue (\cite{Mroue:2013xna}) and orange (\cite{Boyle:2019kee}), BAM simulations in green and Einstein Toolkit simulations in pink. 
}
\label{fig:hyb_parspace}
\end{center}
\end{figure}

The coverage of the comparable mass parameter space is shown in Fig.~\ref{fig:hyb_parspace}. 

As the computational cost of NR simulations diverges rapidly as $\eta \rightarrow 0$, no systematic NR simulations are available for mass ratios $q \geq 18$. This severely limits the parameter space against which we can calibrate a waveform model to NR. Constraining the asymptotic behaviour of the parameter space fits in the extreme-mass-ratio limit is essential for well-behaved extrapolation and to reduce uncertainty in the waveform model for intermediate-mass-ratio binaries, where NR coverage is extremely sparse. For many of the coefficients appearing in our waveform model, no fully analytical knowledge, with complete spin dependence, is available and we instead opt to constrain the fits by calibrating against semi-analytical waveforms in the test-particle limit. 

As in \cite{Keitel:2016krm}, the simulations for BBH mergers in the test-particle limit are produced using \texttt{Teukode} \cite{Harms:2014dqa,Harms:2015ixa,Harms:2016ctx}, which combines a semi-analytical description of the dynamics with a time-domain numerical approach for computing the full multipolar waveform. The dynamics of the binary are prescribed using EOB dynamics, where conservative geodesic motion has been augmented with a linear-in-$\eta$ radiation reaction \cite{Nagar:2006xv,Damour:2007xr}. This makes use of the factorized and resummed circularized waveform introduced in \cite{Damour:2008gu,Pan:2013rra} and uses PN information up to 5.5PN. The fluxes are computed by solving the Regge-Wheeler-Zerilli (RWZ) $1+1$ equations (non-spinning) or the Teukolsky $2+1$ equations (spinning). These equations are solved in the time domain using a hyperboloidal foliation and horizon-penetrating coordinates that allow for the unambiguous extraction of radiation at $\mathscr{I}^+$ (future null infinity) \cite{Bernuzzi:2010xj,Bernuzzi:2011aj,Harms:2014dqa}. 

As with the NR simulations detailed above, the test-particle waveforms are hybridized against a longer EOB inspiral. For the calibration of \phX, we use two sets of waveforms: one set at $q = 200$ and the other set at $q = 1000$. The spin of the primary BH spans an interval $\left[-0.9 , 0.9 \right]$ and the secondary BH is taken to be non-spinning. 

The waveforms in the test-particle limit should only be treated as approximate as $\mathcal{O}(\eta)$ effects are neglected in the conservative dynamics and the 5PN-accurate EOB-resummed analytical multipolar waveforms, used to build the radiation reaction force, show relatively poor performance. A more detailed discussion on some of the observed discrepancies between the comparable-mass limit and the extreme-mass-ratio limit will be given below. 

A recently proposed framework for the factorization and resummation of the residual waveform amplitudes \cite{Nagar:2016ayt,Messina:2018ghh,Nagar:2019wrt} is expected to improve the self-consistency of the test-particle waveforms and hence the self-consistency of the calibration. A detailed discussion of different approaches to resummation and the radiation reaction was presented in \cite{Nagar:2019wds}. A detailed study of the consistency of \phX in the test-particle limit will be presented elsewhere. 

\section{Mapping Phenomenological Coefficients to Physical Parameters}\label{sec:parspacefits}
\begin{centering}
\begin{figure}
\includegraphics[width=\columnwidth]{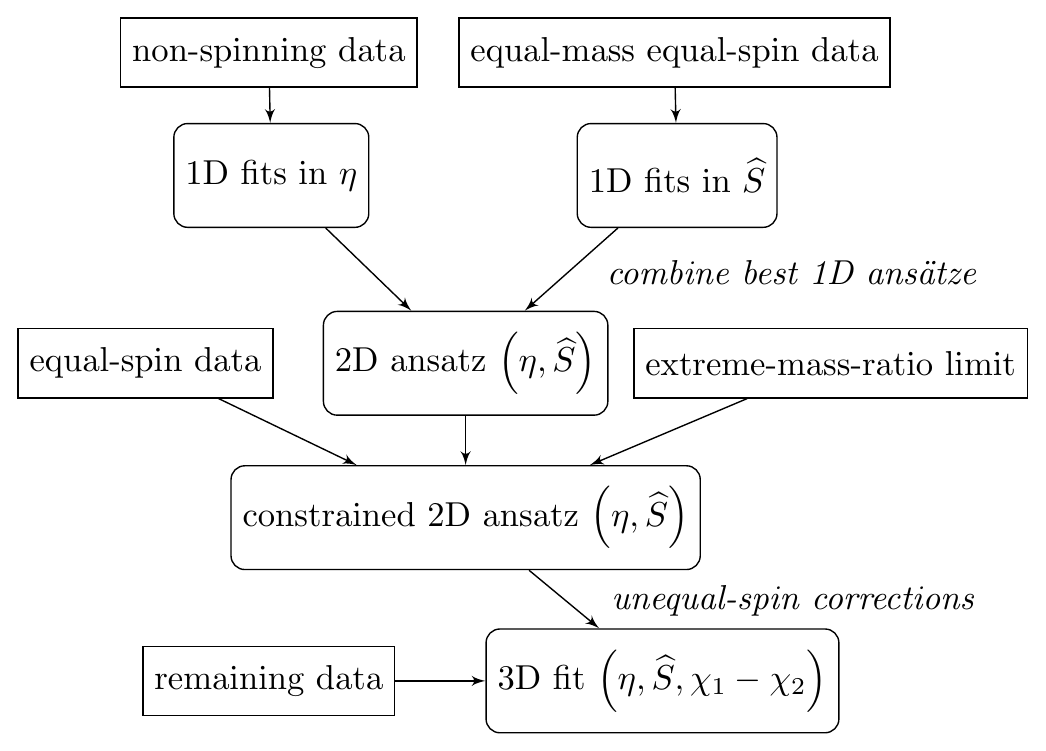}
  \caption{ \label{fig:fits_flowchart} %
Flowchart for the hierarchical parameter space fits,
taken from Fig. 1 of \cite{Jimenez-Forteza:2016oae}.
The starting point are one-dimensional sub-manifolds, taken to be the equal mass and spin limit and the non-spinning limit. The next step is to use a given spin parameterisation to perform an expansion about the 1D fits. Finally, the residuals between the data and the 2D fits are used to fit for unequal spin contributions.}
\end{figure}
\end{centering}

The model here has 8 amplitude coefficients and 13 phase coefficients, meaning that  there are 21 phenomenological coefficients that must be mapped to the physical 3D parameter space $(\eta,\hat{S},\delta \chi)$, where $\hat{S}$ is an effective spin parameterisation of our choice and $\delta \chi = \chi_1 - \chi_2$ is the linear-in-spin difference. The mapping procedure detailed here is a generalisation of the approach taken in previous phenomenological models and pioneered in fits to the radiated energy, final mass and final spin \cite{Jimenez-Forteza:2016oae,Keitel:2016krm}. Here we use a hierarchical, bottom-up approach to calibrate fits to numerical relativity waveforms. As in previous post-Newtonian studies, the dominant parameter dependencies are on the mass ratio and effective spin parameterisations. The remaining unequal spin contribution is sub-dominant and can be effectively modelled by working to linear order in the spin difference. We provide a representative example of this workflow in Sec.~\ref{sec:worked} and a flowchart of the logic behind the hierarchical fitting procedure is shown in Fig.~\ref{fig:fits_flowchart}.

\subsection{Collocation Points}\label{sec:collocationPoints}
Direct calibration of the phenomenological coefficients to the hybrid data can often be problematic due to poor numerical conditioning. This is notable during inspiral, where both poor convergence and eventual breakdown of the PN series can lead to numerical instabilities for the pseudo-PN coefficients used to capture higher frequency behaviour. Such coefficients typically alternate in sign leading to significant numerical cancellations that must be captured accurately across the parameter space in order for the model to remain accurate. 


In constructing a phenomenological waveform model, the aim is to calibrate a model for the amplitude or phase within a given domain. Often this reduces to constructing a polynomial fit, $P_n (x)$, to the hybrid data $f (x)$. As stated by the Weierstrass theorem \cite{Weierstrass:1885}, for any continuous real valued function on an interval $\left[ a,b \right]$, there exists a polynomial $P_n (x)$ with $\epsilon > 0$ such that for all $x \in \left[ a,b\right]$ we have $| f(x) - p(x) | < \epsilon$. However, a well known caveat to this theorem is that the result is highly dependent on the set of polynomials used and on their convergence. In particular, use of equidistant nodes when constructing $P_n (x)$ can lead to oscillatory divergences from $f(x)$ as we increase the degree of the polynomial. This is known as Runge's phenomena and results in unphysical oscillations that can impact the accuracy of the model. In order to help alleviate such issues, there are a number of possible options. For instance, we could help tame oscillatory behaviour by fitting to a lower degree polynomial or we could construct multiple overlapping subintervals constructed with low-degree polynomials, i.e. piecewise polynomial interpolation. Instead, the strategy adopted here, and in \cite{PhenXHM}, is to choose the interpolation nodes $\lbrace x^i \rbrace^{n}_{i=0}$ such that the maximum error $||e_n (x) ||_{\rm{\infty}}$ is minimized. This can be achieved by selecting $(n+1)$ sample points for the polynomial $P_n (x)$ at the roots of the Chebyshev polynomial
\begin{align}
    x_k &= \cos \left[ \frac{(2k + 1) \pi}{2 n} \right] .
\end{align}
\newline
There are a few key advantages to using Chebyshev nodes when constructing such phenomenological fits. First, the error will be the smallest for all polynomials of degree $n$. Secondly, the error can often be more uniformly distributed over the interval in which we perform the fit. Finally, the error decreases exponentially with $n$, leading to spectral convergence of the fit
\begin{align}
    || e_n (x) || &\leq \frac{|| f^{(n+1)} ||_{\rm{\infty}}}{2^n (n+1)!} .
\end{align}
\newline
 In contrast, the error from using equidistant nodes scales approximately as $|| f(x) - P_n (x) || \propto \mathcal{O}(2^n)$. For these reasons, we find it optimal to use collocation points evaluated at the Chebyshev nodes in the domain of interest. 

The value of these collocation points is then fit across the entire parameter space using the hierarchical procedure discussed in the next section. Using the values of the collocation points, or their differences, we can reconstruct the underlying phenomenological ansatz by solving a system of linear equations using standard methods, such as an LU factorization. 

\subsection{Hierarchical Fitting Procedure: One-Dimensional Subspace Fits}
\label{sec:1d_fits}
The hierarchical fits are performed by first sampling the non-spinning parameter space, as this is the simplest and best-sampled subspace in the NR data set. All fits are constructed using Mathematica's \texttt{NonlinearModelFit} function. We use both polynomial and rational ans{\"a}tze. As in \cite{Jimenez-Forteza:2016oae,Keitel:2016krm}, high-dimensional polynomials are used to construct a Pad\'e approximant to the desired order, with the coefficients of the approximant being used as the starting values for the rational-function fits. The use of Pad\'e approximants to pre-condition the \texttt{NonLinearModelFit} helps to alleviate issues related to non-convergence, which can arise due to singularities in the rational function. Following \cite{Jimenez-Forteza:2016oae,Keitel:2016krm}, rational functions with a numerator of polynomial order $m$ and a denominator of polynomial order $k$ will be denoted as an ansatz of order $(m,k)$. The use of rational functions offers numerous advantages to high-dimensional polynomials. In particular we find that rational functions are smoother, less prone to unphysical oscillations and extrapolate in a more controlled manner. 

\subsection{Hierarchical Fitting Procedure: Two-Dimensional Subspace Fits}
\label{sec:2d_fits}
The next step is to construct the two-dimensional fits spanning the $(\eta , \hat{S})$ subspace. The ansatz from the 1D fits will be supplemented with a polynomial of order $J$ in order to capture the 2D curvature associated to $\hat{S}$-dependent terms via
\begin{align}
b_i \rightarrow b_i \displaystyle\sum^{j = J}_{j = 0} \, f_{ij} \, \eta^j .
\end{align}
\n
The general 2D ansatz for a phenomenological coefficient is therefore
\begin{align}
\lambda (\eta,\hat{S}) = \lambda (\eta,0)  \; - \; \lambda(0.25,0)  \; + \;  \lambda (0.25,\hat{S},f_{ij} ) .
\end{align}
\n
The order to which we expand in $\eta$ is dependent upon the behaviour of the phenomenological coefficient that is being fitted. Typically we find that expanding to third order in $\eta$ ($J = 3$) is the lowest order that leaves sufficient freedom to incorporate the constraints from the 1D fits and the extreme mass ratio limit as well as to adequately capture all the features of the data set. At higher order in $\eta$, numerous pathologies outside the calibration regime can start to develop, leading to a significant degradation in the performance of the calibrated model. In order to avoid potential singularities, appropriate care must be taken to remove pathological coefficients from the denominator of the rational ansatz. 

\subsection{Hierarchical Fitting Procedure: Unequal Spin Contributions and 3D Fits}
\label{sec:3d_fits}
The final stage in the hierarchical approach is to incorporate the subdominant effect of unequal spins. Here we parameterise this effect by $\Delta \chi = \chi_1 - \chi_2$. The residuals are defined by subtracting the 2D equal spin fit from the fit against the unequal-spin NR cases:
\begin{align}
\Delta \lambda \, (\eta, \hat{S}, \Delta \chi ) &= \lambda \, (\eta, \hat{S},\Delta \chi) - \lambda \, (\eta, \hat{S}) .
\end{align}
This procedure can be done at discrete points in the symmetric mass ratio provided that sufficient unequal spin NR simulations are available. 

At a given mass ratio, the residuals form a 2D surface $(\hat{S},\Delta \chi, \Delta \lambda)$ which can be used to informatively construct an ansatz for the unequal spin effects. As with many aspects of phenomenological waveform modelling, insight can be taken from studying the structure of the post-Newtonian equations. For example, if we consider the next-to-leading order (NLO) spin-orbit (SO) contribution to the flux
\begin{align}
\mathcal{F}_{\rm SO}^{\rm NLO} &\propto \left( - \frac{9}{2} + \frac{272}{9} \eta \right) \\
\nonumber &\qquad \sqrt{1 - 4 \eta} \; \left( - \frac{13}{16} + \frac{43}{4} \eta \right) \; \Sigma_{\ell} ,
\end{align}
\n
where 
\begin{align}
S_{\ell} &= m^2_1 \; {\chi}_{1 \ell} \; + \; m^2_2 \; {\chi}_{2 \ell} ,\\
\Sigma_{\ell} &= \left( \; m_2 \; {\chi}_{2 \ell} \; - \; m_1 \; {\chi}_{1 \ell} \; \right).
\end{align}
\n
By inspection, the linear-in-spin difference contribution is killed by a factor of $\delta = \sqrt{1-4\eta}$ in the equal mass limit. Away from equal masses, the unequal spin contribution is a simple polynomial function in $\eta$. By comparison, the leading-order (LO) spin-spin (SS) term is given by
\begin{align}
\mathcal{F}^{\rm LO}_{\rm SS} &\propto x^2 \left(  {8 S_{\ell}^2}+ 8 \; \delta \; \Sigma_{\ell} \; S_{\ell} + \left(\frac{33}{16} - 8 \eta \right) \; \Sigma_{\ell}^2 \right),
\end{align}
\n
where in the equal mass limit the mixed term $\Sigma_{\ell} \, S_{\ell}$ is killed by a factor of $\delta$ but we still have a non-vanishing quadratic-in-spin-difference term $\Sigma^2_{\ell}$. 

In practice, we find that the 2D surfaces are typically close to flat, suggesting that the unequal spin effects are dominanted by a linear dependence on $\Delta \chi$ and a possible mixture term $\hat{S} \Delta \chi$. This linear dependence will break down in the equal mass limit as, under an exchange of $\chi_1$ and $\chi_2$, terms linear in $\Delta \chi$ will vanish. In this limit, the surface is approximately parabolic and well-modelled by a quadratic term. 

Based on the above considerations, we use a general ansatz with three spin-difference terms 
\begin{align}
\label{eq:full_spin_diff_ansatz}
\Delta \lambda (\eta,\hat{S},\Delta \chi) = A_1 (\eta) \Delta \chi + A_2 (\eta) \Delta \chi^2 + A_3 (\eta) \Delta \chi \, \hat{S} .
\end{align}
\n 
The resulting full 3D ansatz is therefore given by
\begin{align}
\lambda (\eta, \hat{S}, \Delta \chi) &= \lambda (\eta,\hat{S}) + \Delta \lambda (\eta,\hat{S},\Delta \chi) .
\end{align}
\n
Additional higher order terms in the effective spin or spin difference are not used as there is no motivation from either PN or visual inspection of the residuals. In addition, the intrinsic error of the NR simulations begins to dominate and caution is required to ensure that we do not overfit noisy data.  As a check, we follow the approach in \cite{Jimenez-Forteza:2016oae} and perform four fits in $\Delta \chi$ for the values of $A_i$: linear, linear+quadratic, linear+mixed and the sum of all three contributions.

\subsection{Choice of Spin Parameterisation}
A choice that must be made when constructing the fits across the parameter space is the spin parameterisation, $\hat{S}$, employed. The choice of parameterization can help minimize errors when building fits on a subspace of the data. One of the most widely used spin parameterizations is the effective aligned spin \cite{Damour:2001tu,Racine:2008qv,Ajith:2009bn}
\begin{align}
    \chi_{\rm{eff}} &= \frac{m_1 \chi_1 + m_2 \chi_2}{M} .
\end{align}
\n 
This choice was made in early Phenomenological waveform models \phB \cite{Ajith:2009bn} and \phC \cite{Santamaria:2010yb}. In \phD, an alternative spin parameterization was used based on the reduced spin parameter, that describes the leading order spin-orbit term at 1.5PN \cite{Poisson:1995ef,Ajith:2011ec}
\begin{align}
    \chi_{\rm{PN}} &= \chi_{\rm{eff}} - \frac{38 \eta}{113} \left( \chi_1 + \chi_2 \right) ,
\end{align}
\n 
normalized to $\left[ -1 , 1 \right]$ for any mass ratio
\begin{align}
    \hat{\chi}_{\rm{PN}} &= \frac{\chi_{\rm{PN}}}{1 - 76 \eta / 113} .
\end{align}
\n 
This PN motivated parameterization is particularly suited to use in IMR waveform models \cite{Purrer:2013xma} and was also found to best capture spin-orbit contributions to the binding energy \cite{Ossokine:2017dge}. We will adopt $\hat{\chi}_{\rm{PN}}$ as our spin parameterization of choice for the inspiral regime. 

For the final state, however, the underlying physics is best captured by the linear spin combination $S_1 + S_2$. We therefore find it useful to employ an effective total spin parameter
\begin{align}
\label{eq:stot}
    \hat{S}_{\rm{tot}} &= \frac{S}{m^2_1 + m^2_2} , \quad \textrm{with} \quad S = m^2 \chi_1 + m^2 \chi_2 ,
\end{align}
\n
which was found to work well for final-state quantities \cite{Husa:2015iqa,Jimenez-Forteza:2016oae}. In \phX, we will use $\hat{S}$ to parameterize the fits to the intermediate and merger-ringdown coefficients. A detailed study of the impact of different spin parameterizations is beyond the scope of this paper. 

\section{Matching Regions}
\label{sec:regions}
\begin{figure}[h!]
\begin{center}
\includegraphics[width=\columnwidth]{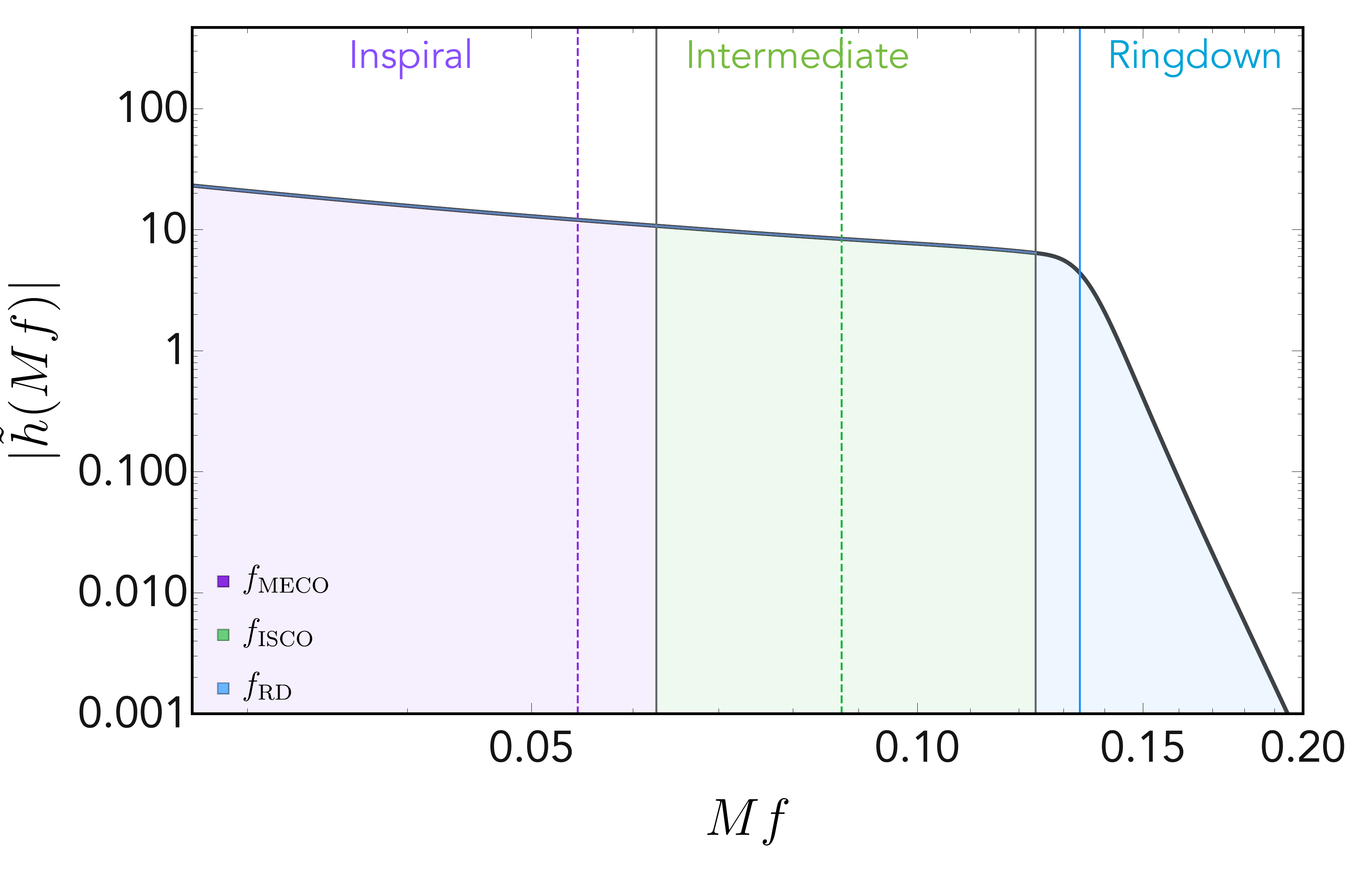}
\includegraphics[width=\columnwidth]{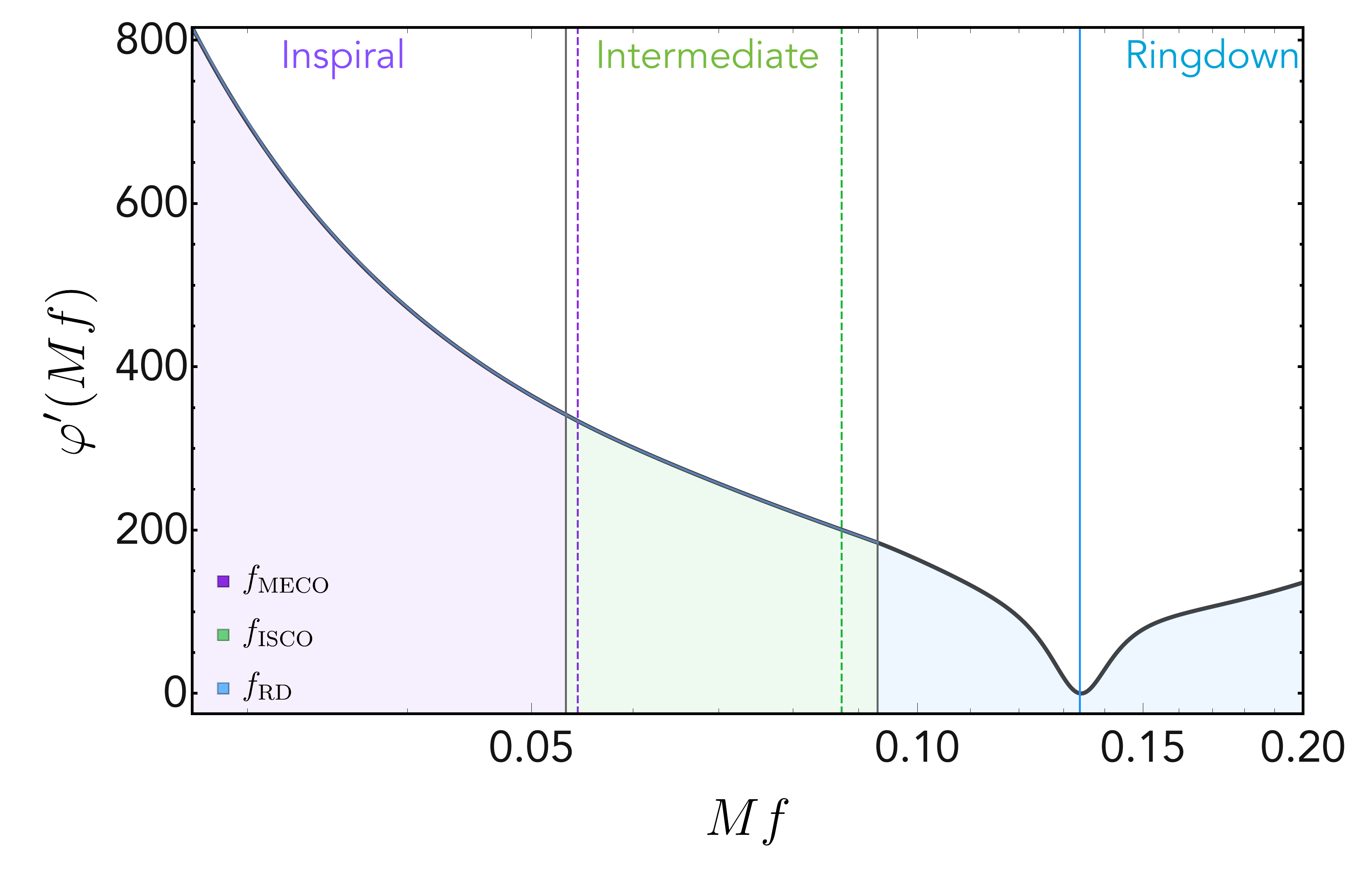}
\caption{
Transition regions for the amplitude and phase derivative $\varphi^{\prime} = \partial_f \varphi (f)$. The purple shaded area shows the \textit{inspiral} region, green shaded the \textit{intermediate} and the blue shaded the \textit{merger-ringdown}. The three colored lines show the MECO (purple), ISCO (green) and ringdown (blue) frequencies respectively.
}
\label{fig:WaveformRegions}
\end{center}
\end{figure}

Following the strategy adopted in previous phenomenological waveform models \cite{Ajith:2007kx,Ajith:2009bn,Santamaria:2010yb,Husa:2015iqa,Khan:2015jqa}, we split the waveform into three frequency regions and model each of these regimes separately. This is done for both the amplitude and the phase derivative.  In this section we define these regions and explicitly highlight the calibration range used in determining the fits as well as the transition windows used when reconstructing the phenomenological model. These regions are highlighted in Fig.~\ref{fig:WaveformRegions}. 

\subsection{Inspiral}
Two of the key improvements of \phX over \phD concern the frequency region over which the model is calibrated to the hybrid data. For \phD this region was 
\begin{equation}
0.0035 \leq   Mf \leq 0.018\,.
\end{equation}
In \phX the lower frequency has been reduced to 
\begin{equation}
    f_{0} = 0.0026\,
\end{equation}
which corresponds to lowering the starting frequency from $71.1 Hz$ to $52.8 Hz$ for a binary of total mass of 10 $M_{\odot}$, or to lowering the binary mass for which the calibration completely covers frequencies above 10 Hz from $71.1 M_{\odot}$ to $52.8 M_{\odot}$. We find that this change significantly improves matches between hybrids and the model for lower masses. The reason for not lowering the starting frequency further has been twofold: First, we build our hybrid waveforms in the time domain, and need to Fourier transform the hybrids on an equispaced frequency grid. In order to simplify our setup, we choose this frequency grid to be the same for all our waveforms. In order to achieve sufficient resolution at high frequencies, we choose a time step of $t/M = 0.5$. The start frequency is chosen to conveniently fit the complete hybrid generation in the time and frequency domain into the RAM available on a laptop for mass ratios up to about 500, as described in \cite{hybrids}. Allowing for windowing and robustly cutting away different types of artifacts, restricts the low frequency limit that can be achieved. In future editions of the model, a more flexible approach will be used, to further reduce the start frequency. The second reason is that a further reduction of start frequency would only be useful with further studies of how to best model the inspiral (e.g. regarding the order of pseudo-PN terms used, and the number of collocation points used). These studies are outside the scope of the present paper, but will be required for a more accurate representation of the extreme mass ratio limit, and possibly for further increases in accuracy.

In \phD, the maximal frequency for the inspiral description in terms of a modified post-Newtonian ansatz, was also fixed, irrespective of the binaries mass ratio or spin. In the extreme mass ratio case, an appropriate choice of transition frequency is given by the ISCO (innermost stable circular orbit) frequency, which can be evaluated in closed form, and ranges from $Mf\approx 0.006$ for inspiral into an extreme Kerr black hole with orbital angular momentum anti-aligned with the spin of the large black hole, to $Mf\approx 0.08$ for the aligned case. A fixed transition frequency from the inspiral to the intermediate regime is thus clearly not appropriate for extreme mass ratios, but also not for comparable masses where the dynamical range is smaller.

A natural termination frequency for the inspiral, which also applies to comparable masses, can be based on the minimum energy circular orbit (MECO) frequency. In a standard binary black hole inspiral, the orbital energy will gradually decrease until it reaches some minimum. The MECO is defined to be the orbit at which the orbital energy reaches its minimum value. Naturally, the MECO is implicitly tied to the PN order under consideration, which can be problematic in the extreme mass ratio limit where the PN approximation is poorly convergent. In order to alleviate such problems, \cite{Cabero:2016ayq} implemented a hybrid-MECO in which test-particle dynamics are folded into the PN approximation in order to provide a well-defined MECO condition valid for all spins. Schematically, the hybrid energy is constructed by replacing the test-particle limit of the PN energy with the exact orbital energy per unit mass for a test-particle around a Kerr black hole \cite{Cabero:2016ayq}
\begin{align}
E^{\rm{Hybrid}} = \frac{E^{\rm{n-PN}}}{\eta} - \left( \displaystyle\sum^{x=2n}_{x=0} E^{\rm{Kerr}} (v^x) \right) + E^{\rm{Kerr}} 
\end{align}
\n
where
\begin{align}
E^{\rm{Kerr}}  = \left( \frac{1 - 2 w + \chi w^{3/2}}{\sqrt{1 - 3w + 2 \chi w^{3/2}}} - 1 \right) ,
\end{align}
\n
and $w = v^2 / (1 - \chi v^3)^{2/3}$. This expression was shown to have a minimum for currently known PN orders. In practice, we use a phenomenological fit to the hybrid-MECO as a natural PN approximation to the end of the inspiral. This alleviates the necessity of performing a root-finding operation when evaluating the waveform model. 

The inspiral calibration range for the amplitude $(A)$ and the phase $(\varphi)$ is taken to be
\begin{align}
f^{\varphi}_{C,\rm{Ins}} &\in \left[ 0.0026 , 1.02 f_{\rm{MECO}} \right], \\
f^{A}_{C,\rm{Ins}} &\in \left[ 0.0026, 1.025 f^{A}_{T} \right] ,
\end{align}
\n
where the $C$ explicitly denotes the calibration domain and
\begin{align}
    f^{A}_{T} &= f_{\rm{MECO}} + \frac{1}{4} \left( f_{\rm{ISCO}} - f_{\rm{MECO}} \right).
\end{align}
However, when building the waveform, the inspiral region interval is defined by
\begin{align}
f^{\varphi}_{\rm{Ins}} &\in \left( 0 ,  f_{\rm{MECO}} - \delta_{R} \right)  \\
f^{A}_{\rm{Ins}} &\in \left( 0,  f^{A}_T \right] ,
\end{align}
\n
where 
\begin{align}
\label{eq:im_transit}
    \delta_{R} &= 0.03 \left( f^{\varphi}_{R,3} - f_{\rm{MECO}} \right), \\
    f^{\varphi}_{T} &= 0.6 \left( \frac{1}{2} f_{\rm{RD}} + f_{\rm{ISCO}} \right) .
\end{align}
\n 
The inspiral region corresponds to the purple-shaded region in Fig.~\ref{fig:WaveformRegions}.

\subsection{Intermediate Regime}
The intermediate regime is introduced in order to phenomenologically bridge the gap between the post-Newtonian regime and the perturbative black hole ringdown regime. The start of this region is determined by the breakdown of post-Newtonian theory and the end of the region is set relative to the ISCO and ringdown frequencies. This enables us to implicity incorporate a natural hierarchy of frequencies in a standard binary black hole inspiral: $f_{\rm{MECO}} < f_{\rm{ISCO}} < f_{\rm{ring}}$. 

For the intermediate region, the calibration domain is taken to be
\begin{align}
  f^{\varphi}_{C,\rm{Int}} &\in \left[ f_{\rm{MECO}} - \delta_R , f^{\varphi}_{T} + 0.5 \delta_R \right], \\
f^{A}_{C,\rm{Int}} &\in \left[ 0.98 f^{A}_{T} , 1.02 f_{\rm{peak}} \right] ,  
\end{align}
\n
where $f_{\rm{peak}}$ is the analytical location of the peak of the ringdown \cite{Husa:2015iqa,Khan:2015jqa}
\begin{align}
    f_{\rm{peak}} &= \left| \frd + \fdamp \sigma \frac{\sqrt{1 - \lambda} - 1}{\lambda} \right| .
\end{align}
The intermediate interval when building the waveform is defined by 
\begin{align}
    f^{\varphi}_{\rm{Int}} &\in \left[ f_{\rm{MECO}} - \delta_{R} , f^{\varphi}_{T} + \delta_{R} \right] \\
f^{A}_{\rm{Int}} &\in \left( f^A_T,  f_{\rm{peak}} \right] .
\end{align}
\n 
The intermediate region corresponds to the green-shaded region in Fig.~\ref{fig:WaveformRegions}.

\subsection{Merger-Ringdown Regime}
Finally, the merger-ringdown regime is particularly well described in terms of the ringdown and damping frequency of the remnant BH. The calibration interval for the merger-ringdown is taken to be
\begin{align}
  f^{\varphi}_{C,\rm{MR}} &\in \left[ 0.985 f^{\varphi}_T , f_{\rm{RD}} + 1.25 f_{\rm{damp}} \right], \\
f^{A}_{C,\rm{MR}} &\in \left[ f_{\rm{RD}} - \frac{(1 + 4 \eta)}{5} f_{\rm{damp}} , f_{\rm{RD}} + 3.25 f_{\rm{damp}} \right] ,  
\end{align}
\n
note that the factor of $(1+4\eta)$ has been added to help control the fits in the extreme-mass-ratio limit, where the amplitudes at the peak of the rescaled data can become particularly flat and no clear merger-ringdown can be defined in a morphological sense. 

The merger-ringdown frequency interval when reconstructing the waveform is defined by
\begin{align}
  f^{\varphi}_{\rm{MR}} &\in \left( f^{\varphi}_{T} , 0.3 M f \right), \\
f^{A}_{\rm{MR}} &\in \left( f_{\rm{peak}} , 0.3 M f \right) ,  
\end{align}
\n
where $0.3 M f$ is an arbitrary high-frequency cutoff frequency implemented for \phX in \texttt{LAL}. The merger-ringdown region corresponds to the blue-shaded region in Fig.~\ref{fig:WaveformRegions}.

\section{Amplitude Model}\label{sec:amp}
When calibrating the amplitude model of \phX to the hybrid data, we factor out the leading order PN behaviour $f^{-7/6}$. We opt to normalize the data such that as $f \rightarrow 0$ the data tends to unity. This normalization is motivated by the Newtonian limit
\begin{align}
\lim_{f \rightarrow 0} \, \left[ f^{7/6} \, A_{\rm{PN}} (f) \right] \rightarrow \sqrt{\frac{2 \eta}{3 \pi^{1/3}}} ,
\end{align}
\n
with the resulting normalization factor being
\begin{align}
A_0 \equiv  \sqrt{\frac{2 \eta}{3 \pi^{1/3}}} \; f^{-7/6} .
\end{align}

%

\subsection{Inspiral}\label{sec:amp_insp}
\begin{figure}[h!]
\begin{center}
\includegraphics[width=\columnwidth]{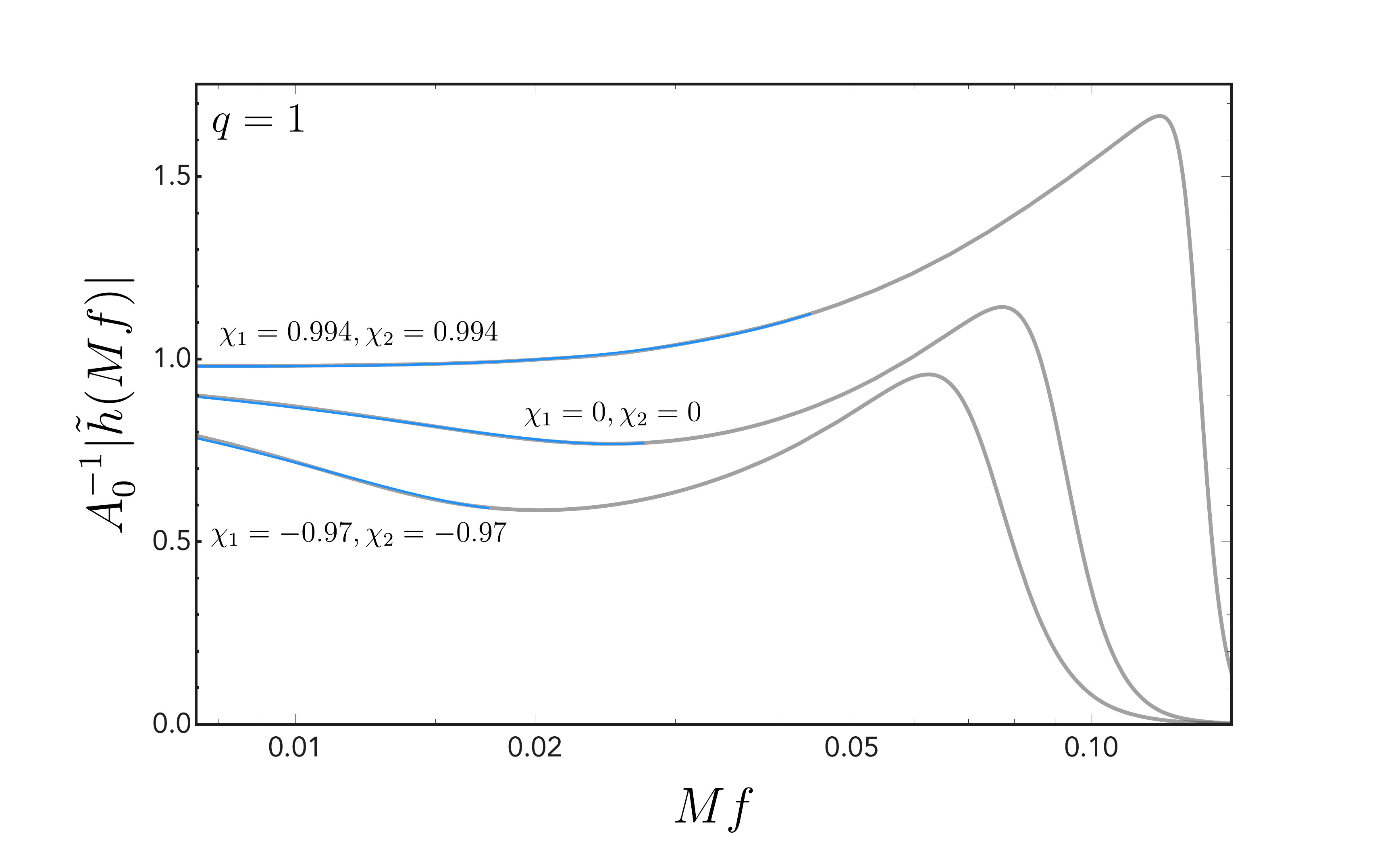}
\caption{
Amplitude inspiral fit for a series of $q = 1$ SXS simulations.
}
\label{fig:Amp_Ins_q1}
\end{center}
\end{figure}

The inspiral model is based on a PN re-expanded TaylorF2 amplitude augmented with pseudo-PN terms that are calibrated to the hybrid data
\begin{align}
A_{\rm{Ins}} (f) = A_{\rm{PN}} + A_0 \displaystyle\sum_{i=1}^3 \, \rho_i \, (\pi \, f )^{(6+i)/3} ,
\end{align} 
\n
where $A_{\rm{PN}}$ constitutes the known PN terms 
\begin{align}
\label{eq:insp_amp_ansatz}
A_{\rm{PN}} (f) = A_0 \displaystyle\sum_{i=0}^6 \, \mathcal{A}_i \, (\pi \, f )^{i/3} ,
\end{align} 
\n
and $\rho_i$ are the \textit{pseudo}-PN coefficients, where $A_0$ is the normalization factor corresponding to the leading order PN term $f^{-7/6}$. An example of the calibrated inspiral amplitude compared to the hybrid data is shown in Fig.~\ref{fig:Amp_Ins_q1}.

The pseudo-PN coefficients are constructed by calibrating collocation points at the nodes
\begin{align}
    \lbrace 0.5, 0.75, 1.0 \rbrace f_{\rm{MECO}} ,
\end{align}
\n 
and analytically solving the system of equations generated by evaluating the pseudo-PN terms in  Eq.~\ref{eq:insp_amp_ansatz} at the above nodes. 

\subsection{Intermediate}
\begin{figure}[h!]
\begin{center}
\includegraphics[width=\columnwidth]{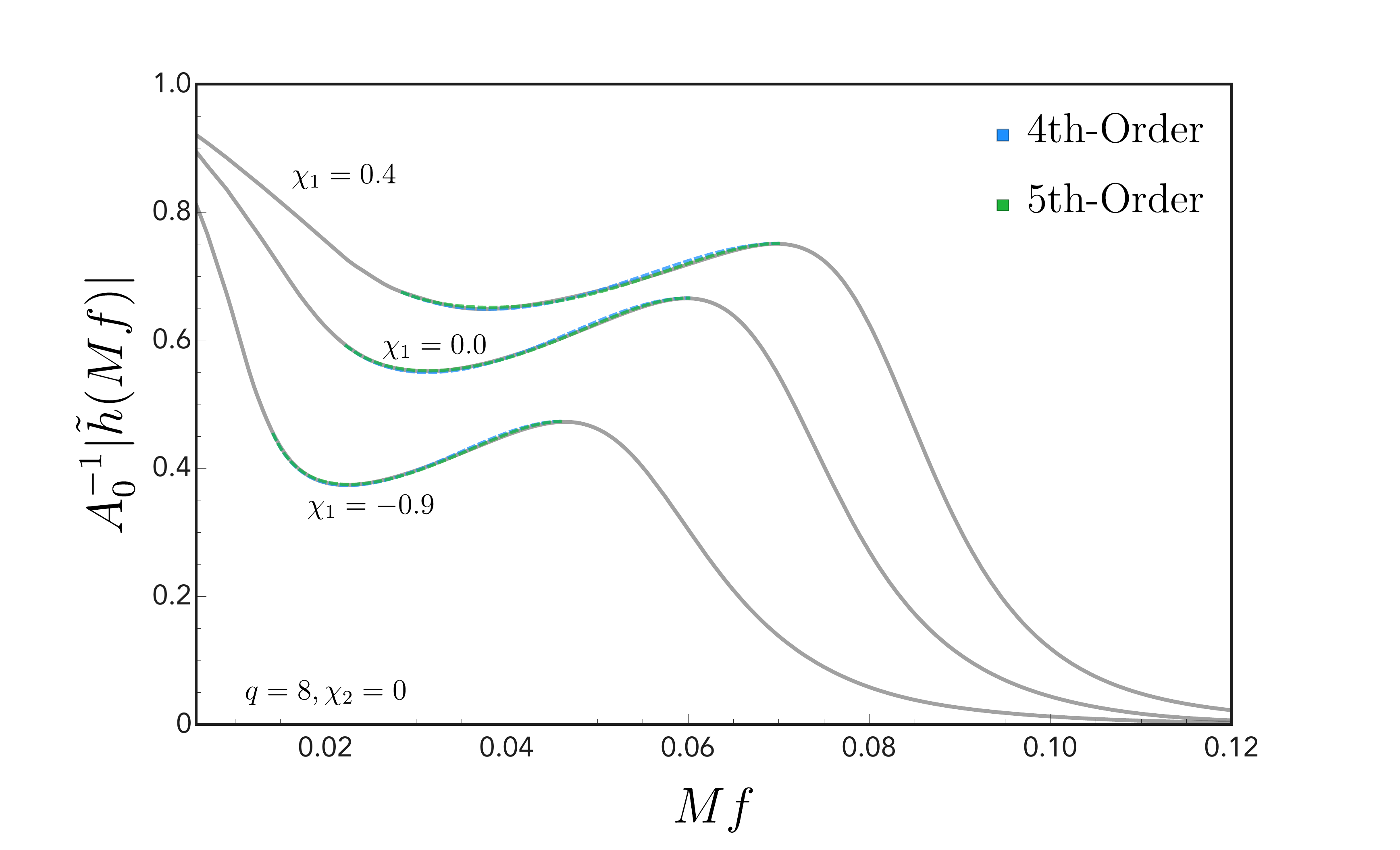}
\caption{
The fit to the intermediate amplitude using a 4th-order and 5th-order ansatz using 3 SXS simulations at $q = 8$. The 5th-order ansatz is able to more accurately fit the features in the waveform but has poor extrapolation compared to the 4th order ansatz.
}
\label{fig:Amp_Int_q8}
\end{center}
\end{figure}

As an example of the modularity of \phX, we implement two different models for the intermediate amplitude. The first model is based on the inverse of a fourth-order or fifth-order polynomial 
\begin{align}
A_{\rm{Int}} = \frac{A_0}{ \alpha_0 + \alpha_1 f + \alpha_2 f^2 + \alpha_3 f^3 + \alpha_4 f^4} ,
\end{align} 
\n 
and the second model on the inverse of a fifth-order polynomial
\begin{align}
    A_{\rm{Int}} = \frac{A_0}{ \alpha_0 + \alpha_1 f + \alpha_2 f^2 + \alpha_3 f^3 + \alpha_4 f^4 + \alpha_5 f^5} .
\end{align}
\n
For an ansatz with $n$ coefficients, we require $n$ pieces of information in order to reconstruct the underlying function. For the fifth-order polynomial, the function requires six input parameters, given by the value of the amplitude at two collocation points together with four boundary conditions: two for the amplitude and two for the first derivative of the amplitude. The amplitude is therefore $C^1$ continuous by construction. A similar argument holds for the fourth-order function, though using 5 coefficients. The collocation points used for both models are detailed in Tables~\ref{tab:amp_int_coll_5} and \ref{tab:amp_int_coll_4}.
For the fifth-order polynomial, the coefficients $\alpha_i$ are the solution to the system of equations
\begin{align}
A_{\rm{Ins}} (f_1) &= v_1 , \\
A_{\rm{Hyb}} (f_2) &= v_2 , \\
A_{\rm{Hyb}} (f_3) &= v_3 , \\
A_{\rm{MR}} (f_4) &= v_4 , \\
A^{\prime}_{\rm{Ins}} (f_1) &= d_1 , \\
A^{\prime}_{\rm{MR}} (f_4) &= d_4 .
\end{align}
\n
For the fourth-order polynomial, the system of equations is analagous to \cite{Khan:2015jqa}
\begin{align}
A_{\rm{Ins}} (f_1) &= v_1 , \\
A_{\rm{Hyb}} (f_2) &= v_2 , \\
A_{\rm{MR}} (f_4) &= v_3 , \\
A^{\prime}_{\rm{Ins}} (f_1) &= d_1 , \\
A^{\prime}_{\rm{MR}} (f_3) &= d_3 .
\end{align}
The fifth-order ansatz allows us to capture more dramatic features in the amplitude morphology, which is particular important as we extend to higher mass ratios. For aligned-spin binaries, the system is highly adiabatic and there are many quasi-circular orbits before the smaller black hole plunges into the larger black hole. For anti-aligned spins, the system is not adiabatic and the system evolves through to the plunge phase much quicker, especially at high mass ratios. For the un-adiabatic case, the binary shows a distinct morphology in which the amplitude drops as we rapidly transition from the inspiral to the merger-ringdown. A comparison between the fourth-order and fifth-order intermediate ansatz against $q = 8$ hybrid data is shown in Fig.~\ref{fig:Amp_Int_q8}. 

For the remainder of this paper, we will work with the fifth-order intermediate ansatz unless otherwise stated.

\begin{table}
\centering
\caption{Location of collocation points $f_i$ for the fifth-order intermediate ansatz. The coefficients $v_1, v_4, d_1, d_4$ are constrained by the inspiral and merger-ringdown model. The free coefficients $v_2$ and $v_3$ must be fit to the data.}
\label{tab:amp_int_coll_5}
\begin{tabularx}{.5\textwidth}{@{}llll@{}}
\toprule
Collocation Points                              & Value                      & Derivative                          &  \\ \midrule
$f_1 = f^{W}_1$                                 & $v_1 = A_{\rm{Ins}} (f_1)$ & $d_1 = A^{\prime}_{\rm{Ins}} (f_1)$ &  \\
$f_2 = (f^A_T + f_{\rm{peak}}) / 3$ & $v_2 = A_{\rm{Hyb}} (f_2)$ &                                     &  \\
$f_3 = 2(f^A_T + f_{\rm{peak}}) / 3$ & $v_3 = A_{\rm{Hyb}} (f_3)$ &                                     &  \\
$f_4 = f_{\rm{peak}}$                            & $v_4 = A_{\rm{MR}} (f_4)$ & $d_4 = A^{\prime}_{\rm{MR}} (f_4)$ &  \\ \bottomrule
\end{tabularx}
\end{table}

\begin{table}
\centering
\caption{Location of collocation points $f_i$ for the fourth-order intermediate ansatz. The coefficients $v_1, v_3, d_1, d_3$ are constrained by the inspiral and merger-ringdown model, whereas $v_2$ must be fit to the data. }
\label{tab:amp_int_coll_4}
\begin{tabularx}{.5\textwidth}{@{}llll@{}}
\toprule
Collocation Points                              & Value                      & Derivative                          &  \\ \midrule
$f_1 = f^{A}_T$                                 & $v_1 = A_{\rm{Ins}} (f_1)$ & $d_1 = A^{\prime}_{\rm{Ins}} (f_1)$ &  \\
$f_2 = (f^A_T + f_{\rm{peak}}) / 2$ & $v_2 = A_{\rm{Hyb}} (f_2)$ &                                     &  \\
$f_3 = f_{\rm{peak}}$                            & $v_3 = A_{\rm{MR}} (f_3)$ & $d_3 = A^{\prime}_{\rm{MR}} (f_3)$ &  \\ \bottomrule
\end{tabularx}
\end{table}

\subsection{Merger-Ringdown}
\begin{figure}[h!]
\begin{center}
\includegraphics[width=\columnwidth]{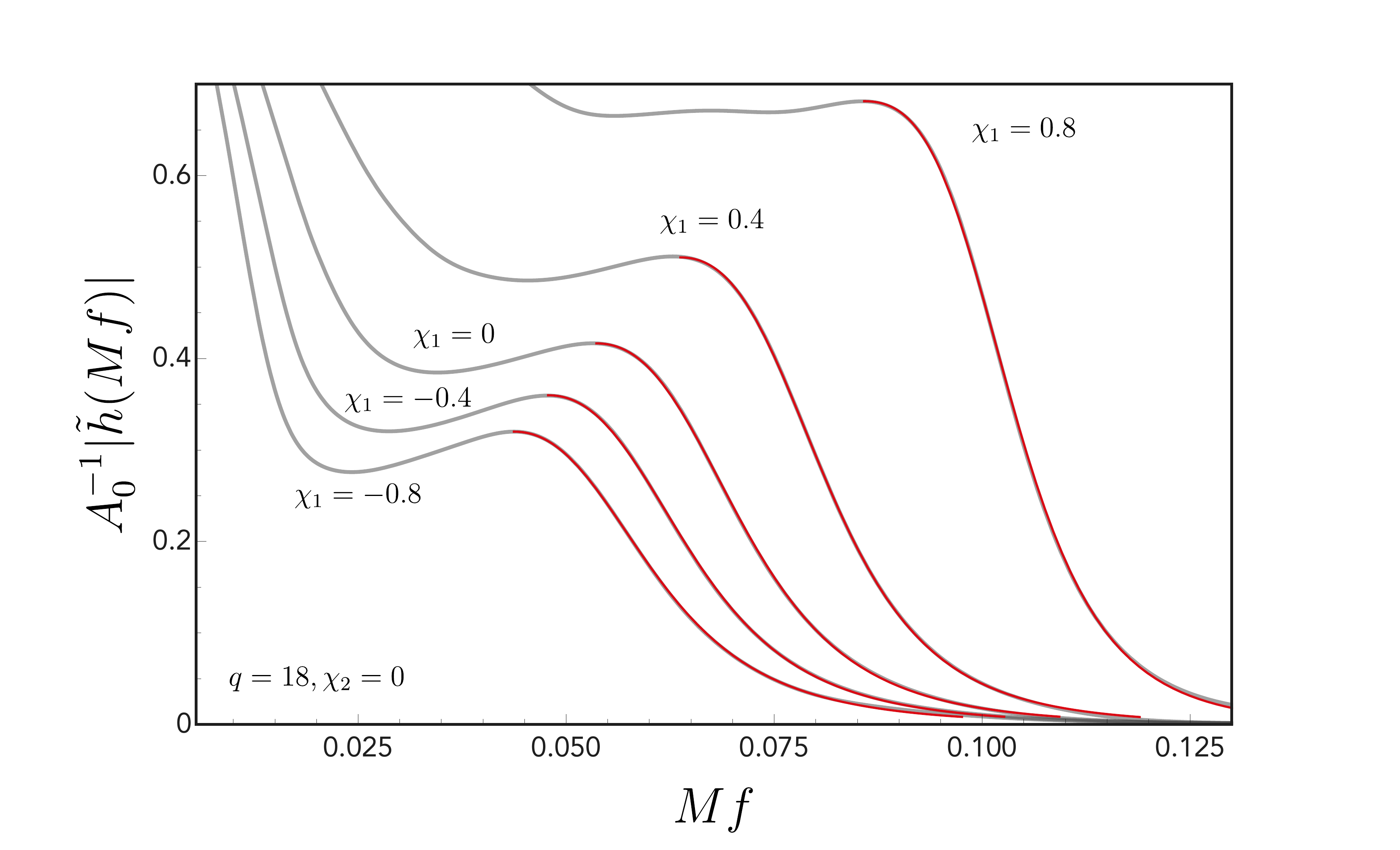}
\caption{
The fit to the merger-ringdown amplitude for 5 BAM simulations at $q = 18$ corresponding to spins on the primary BH of $\chi_1 \in \lbrace -0.8, -0.4, 0, 0.4, 0.8 \rbrace$. The secondary BH is non-spinning. 
}
\label{fig:Amp_MR_q18}
\end{center}
\end{figure}

The merger-ringdown ansatz is modelled using a deformed Lorentzian, corresponding to the Fourier transform of a two-sided exponential decay function. The ansatz adopted is given by \cite{Husa:2015iqa,Khan:2015jqa}
\begin{align}
A_{\rm{MR}} &= \left[ \frac{a_R \; (\fdamp \; \sigma) }{ (f - \frd)^2 + (\fdamp \; \sigma)^2} \right] e^{- \lambda (f - \frd) / (\fdamp \; \sigma)} .
\end{align}
\n 
Unlike the other two regions, here we choose to calibrate $\lambda$ and $\sigma$ directly. In order to solve the amplitude coefficient $a_R$, we further calibrate a collocation point at the defined at $f_{\rm{peak}}$. Together with $\lambda$ and $\sigma$, we can use the collocation point to solve a trivial system of equations for $a_R$. An example of the ringdown fit applied to the BAM $q = 18$ data is shown in Fig.~\ref{fig:Amp_MR_q18}. 

\section{Phase Model}\label{sec:phase}
\subsection{Inspiral}\label{sec:phase_insp}
The inspiral phase model is based on TaylorF2, derived under the stationary phase approximation, augmented with pseudo-PN coefficients that are calibrated to the hybrid data. The full TaylorF2 phase can be written as
\begin{align}
\varphi_{\rm{Ins}} &= \varphi_{\rm{TF2}}  (M f ; \vth) \\ 
\nonumber &\quad + \frac{1}{\eta} \Bigg( \sigma_0 + \sigma_1 f + \frac{3}{4} \sigma_2 f^{4/3} + \frac{3}{5} \sigma_3 f^{5/3} \\ \nonumber &\qquad \qquad + \frac{1}{2} \sigma_4 f^{6/3} + \frac{3}{7} \sigma_5 f^{7/3} \Bigg) .
\end{align}
\n
where $\varphi_{\rm{TF2}}$ is the analytically known TaylorF2 phase 
\begin{align}
    \varphi_{TF2} &= 2 \pi f t_c - \varphi_c - \frac{\pi}{4} + \\ \nonumber
    &\qquad \frac{3}{128 \eta} \left( \pi M f \right)^{-5/3} \displaystyle\sum^{7}_{i = 0} \varphi_i \left( \vth \right) \, \left( \pi M f \right)^{i/3} ,
\end{align}
\n
and $\varphi_i (\vth)$ are known PN coefficients that are functions of the intrinsic parameters of the binary. The coefficients $\sigma_i$ are the pseudo-PN coefficients that we calibrate against the hybrid dataset and supplementary SEOBNRv4 waveforms. A detailed discussion of the PN information used in \phX is given in Appendix~\ref{appendix:TaylorF2}. 

The calibration of the pseudo-PN coefficients is performed by subtracting a given TaylorF2 approximant  from the full hybrid phase and factoring out the leading order frequency power
\begin{align}
\label{eq:residual}
\mathcal{R}_{\rm insp} (f) = f^{-8/3} \, \left( \varphi^{\prime}_{\rm Hybrid} (f) \, - \, \varphi^{\prime}_{\rm TF2} (f) \right) .
\end{align} 
\n
By performing such a rescaling, we numerically condition the data such that we more accurately capture the un-modelled higher-PN contributions to the phase. 

Note that, by construction, the calibrated pseudo-PN coefficients are implicitly tied to the specific TaylorF2 approximant used. If we incorporate additional analytical PN information, we would need to recalibrate the pseudo-PN coefficients on a case-by-case basis. 

In this paper, we demonstrate the flexibility of the \phX framework by producing four different calibrated inspiral models. The first two models adopt a \textit{canonical} 3.5PN TaylorF2 phase using recent cubic-in-spin and quadratic-in-spin corrections \cite{Marsat:2014xea,Bohe:2015ana} but use 4 or 5 pseudo-PN coefficients, with $\sigma_0$ being fixed by imposing $C^1$ continuity in the phase. The model with 4 additional coefficients requires 4 collocation points whereas the model with 5 additional coefficients requires 5 collocations points. The final set of models are all based on an \textit{extended} TaylorF2 phase that incorporates some recent results at 4PN \cite{Damour:2014jta,Bernard:2015njp,Bernard:2016wrg,Damour:2017ced,Bernard:2017bvn,Marchand:2017pir}, 4.5PN \cite{Marchand:2016vox} and a recently identified tail-induced, spin-spin term in the flux \cite{Messina:2018ghh}. As before, we produce two variants with 4 and 5 pseudo-PN coefficients respectively. 

\begin{figure}[h!]
\begin{center}
\includegraphics[width=\columnwidth]{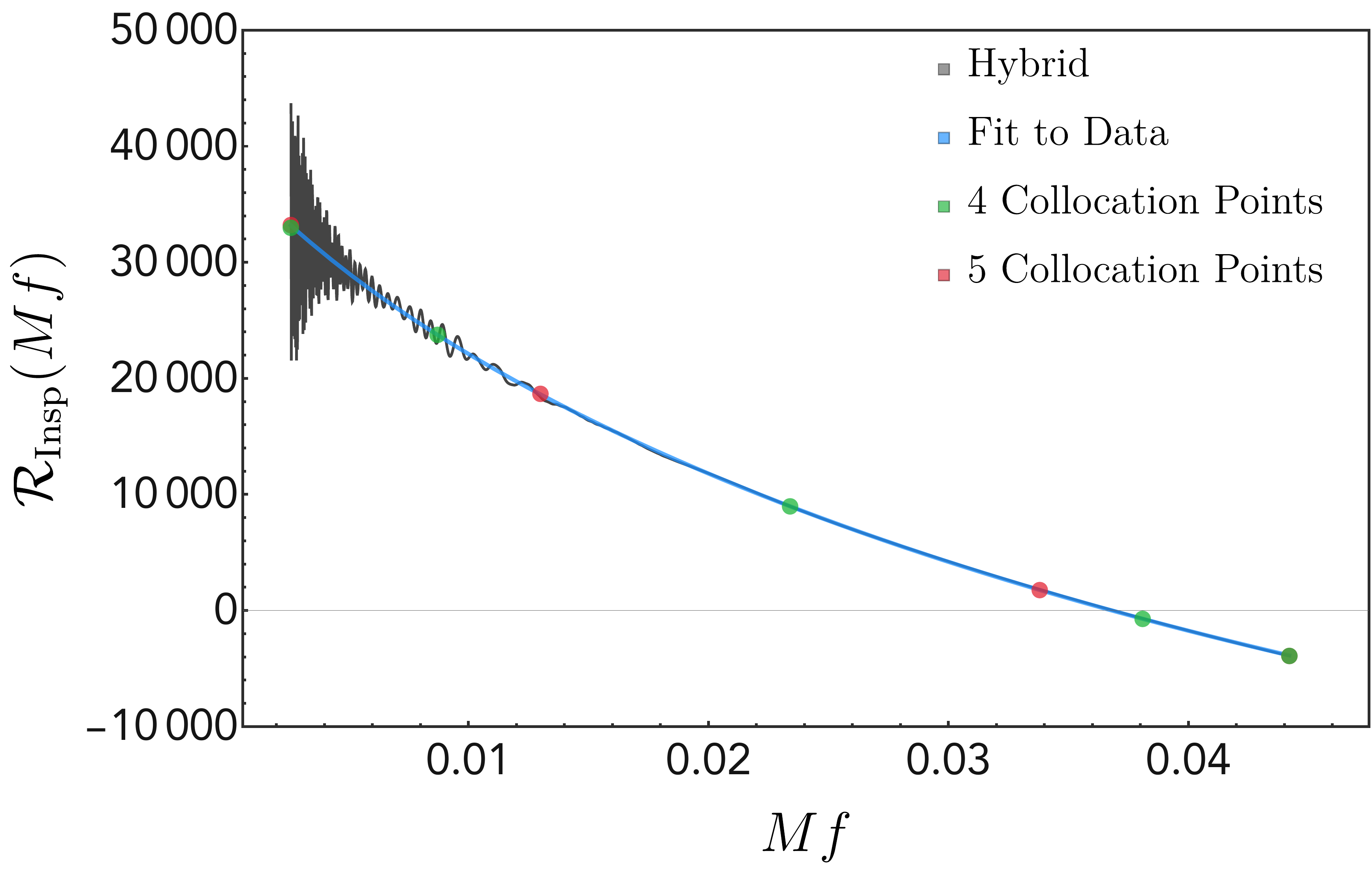}
\caption{
The pseudo-PN coefficients are fit to the the hybrid data by subtracting a given TaylorF2 approximant and factoring out the leading order frequency, $f^{-8/3} \left[ \varphi^{\rm{Hyb}} (f) - \varphi^{\rm{TF2}} (f) \right]$. Such data conditioning helps the fit to capture the physical behaviour of the inspiral waveform up to the MECO frequency. The green circles denote the location of the sampling points for 3 psuedo-PN coefficients and the blue squares for 4 pseudo-PN coefficients. The sampling points are chosen based on a Gauss-Chebyshev aimed at reducing Runga's phenomena near the boundaries. Here we use \textrm{SXS-BBH-0153} hybridized against SEOBNRv4 as a representative example.
}
\label{fig:InspCollocation}
\end{center}
\end{figure}

The collocation points for this system are set by the Gauss-Chebyshev nodes. When using 4 collocation points, these nodes are given by
\begin{align}
    v_i^{\rm{Int}} &= \left\lbrace f_L , \frac{1}{4} \delta^{\rm{Im}}_{\varphi} , \frac{3}{4} \delta^{\rm{Im}}_{\varphi} , f_H \right\rbrace
\end{align}
\n 
where $f_H = 1.02 f_{\rm{MECO}}$, $f_L = 0.0026$ and $\delta^{\rm{Im}}_{\varphi} = f_H - f_L$. For 5 collocation points, the nodes occur at
\begin{align}
    v_i^{\rm{Int}} &= \Bigg\lbrace f_L , \frac{1}{2} \left( 1 - \frac{1}{\sqrt{2}} \right) \delta^{\rm{Im}}_{\varphi} + f_L , f_L + \frac{1}{2} \delta^{\rm{Im}}_{\varphi}  , \\ \nonumber &\qquad \qquad \frac{1}{2} \left( 1 + \frac{1}{\sqrt{2}} \right) \delta^{\rm{Im}}_{\varphi} + f_L 
    , f_H \Bigg\rbrace .
\end{align}
\newline
The location of the nodes when using 4 and 5 collocation points is demonstrated in Fig.~\ref{fig:InspCollocation}, where we fit the residual ansatz in Eq.~\ref{eq:residual} to the hybrid data.

Unless otherwise stated, we adopt the \textit{canonical} TaylorF2 ansatz with \textit{4 pseudo-PN coefficients} as the default inspiral model. The performance of the different inspiral models is discussed in Sec.~\ref{sec:validation} and the mismatches, as defined in Eq.~\ref{eq:mismatch}, shown in Fig.~\ref{fig:mismatcheshist}. 

\subsection{Intermediate}\label{sec:phase_int}
We now consider the phenomenological intermediate region. As in \cite{Husa:2015iqa,Khan:2015jqa}, we adopt a polynomial ansatz but add a Lorentzian term to smoothly match the phase to the merger-ringdown ansatz. As with the inspiral, we provide two models as an example of the modularity of \phX. The functional form for the general ansatz used for the intermediate phase is
\begin{align}
\eta \; \varphi^{\prime}_{\rm{Int}} &= b_{0} + b_4 f^{-4} + b_3 f^{-3} + b_2 f^{-2} + b_1 f^{-1} \\ 
&\qquad \qquad \nonumber - \frac{4 c_0 \, a_{\varphi} }{(f - \frd)^2 + (2 \fdamp)^2} ,
\end{align}
\n
where the terms in the Lorentzian are implicitly set by the merger-ringdown model. The first model adopts 4 collocation points and sets $b_3 = 0$. The second model uses 5 collocation points and retains all 5 coefficients $\lbrace b_0, b_1, b_2, b_3, b_4, b_5 \rbrace$. Unlike \phD, we impose additional constraints on the intermediate ansatz and use the value of the inspiral and merger-ringdown fits respectively to determine the boundary collocation points. The 4-coefficient model therefore requires two calibrated collocation points and the 5-coefficient model 3 calibrated terms. The Gauss-Chebyshev nodes for 4 collocation points occur at
\begin{align}
    v_i^{\rm{Int}} &= \left\lbrace f_L , \frac{1}{4} \delta^{\rm{Im}}_{\varphi} , \frac{3}{4} \delta^{\rm{Im}}_{\varphi} , f_H \right\rbrace ,
\end{align}
\n 
where $f_H = f^{\varphi}_{T} + 0.5 \delta_R$, $f_L = f_{\rm{MECO}} - \delta_R$ and $\delta^{\rm{Im}}_{\varphi} = f_H - f_L$. Similarly, for the 5 collocation points, the nodes occur at
\begin{align}
    v_i^{\rm{Int}} &= \Bigg\lbrace f_L , \frac{1}{2} \left( 1 - \frac{1}{\sqrt{2}} \right) \delta^{\rm{Im}}_{\varphi} + f_L , f_L + \frac{1}{2} \delta^{\rm{Im}}_{\varphi}  , \\ \nonumber &\qquad \qquad \frac{1}{2} \left( 1 + \frac{1}{\sqrt{2}} \right) \delta^{\rm{Im}}_{\varphi} + f_L 
    , f_H \Bigg\rbrace .
\end{align}
\n 
In order to help numerically condition the collocation points, we opted to fit the difference with respect to $v_4^{\rm{MR}}$, the value of the merger-ringdown phase at the ringdown frequency. Such a strategy is particularly beneficial when extrapolating to higher mass ratios and high-spin configurations, where the sparsity of available NR simulations can lead to poor constraints on the parameter space fits. For the 4-coefficient model, the two free coefficients are $v_2$ and $v_3$. We therefore require parameter space fits for 
\begin{align}
    \delta_{2,\rm{RD} 4} &= v_2^{\rm{Im}} - v_4^{\rm{RD}} , \\
    \delta_{3,\rm{RD} 4} &= v_3^{\rm{Im}} - v_4^{\rm{RD}} , 
\end{align}
\n 
which could be used in conjunction with a fit for $v_4^{\rm{RD}}$ to reconstruct $v_2^{\rm{Im}}$ and $v_3^{\rm{Im}}$ respectively. In order to help tame unphysical behaviour at extremely large mass ratios and large spins, we use a weighted average between the above fit and a direct fit to $v^{\rm{Im}}_2$ in the final model
\begin{align}
    v^{\rm{Im}}_2 &= 0.75 (\delta_{2,\rm{RD4}} + v^{\rm{RD}}_4) + 0.25 v^{\rm{Im}}_2 . 
\end{align}

\begin{figure}[h!]
\begin{center}
\includegraphics[width=\columnwidth]{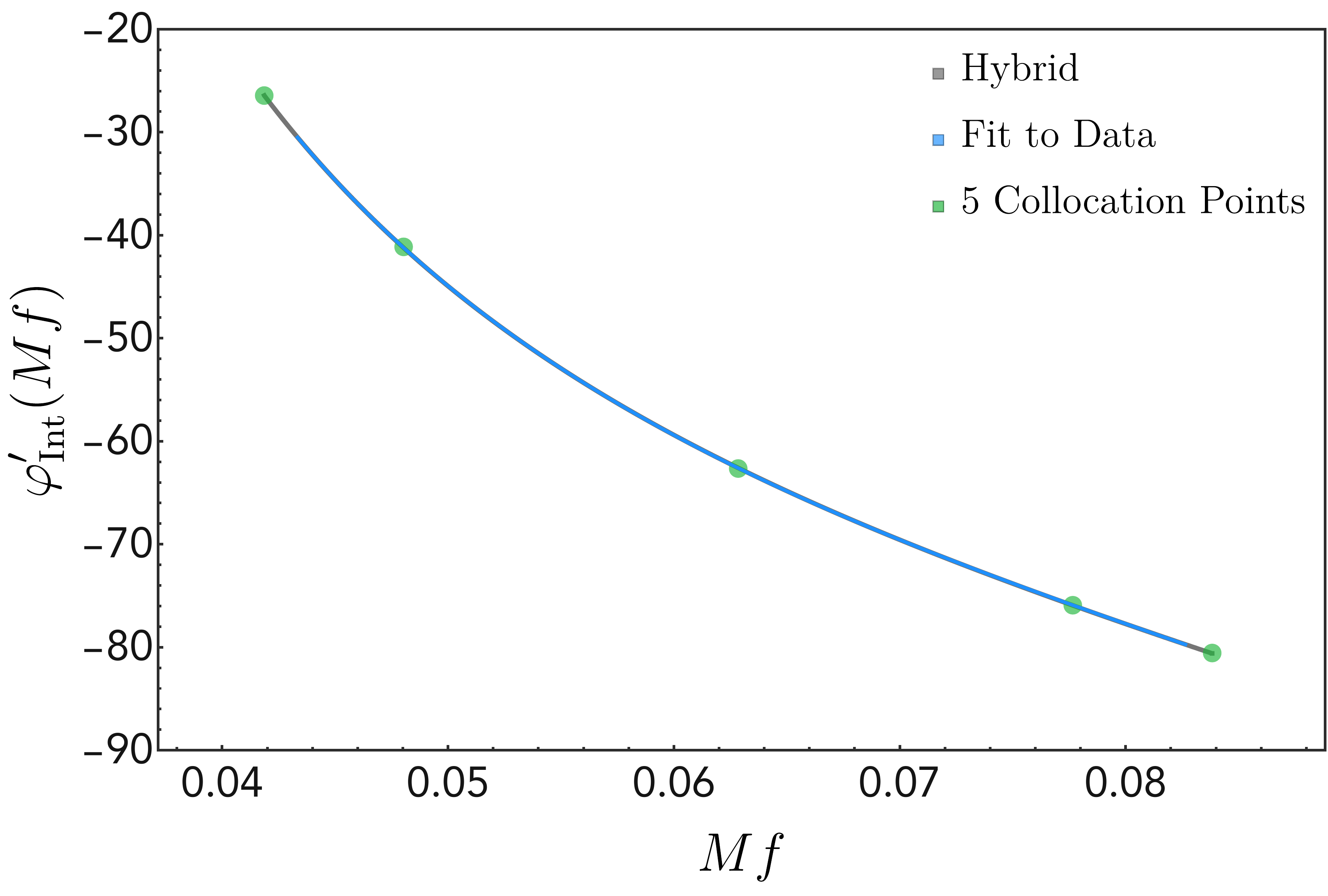}
\caption{
Phase derivative for \textrm{SXS-BBH-0153} hybridized against SEOBNRv4 in the intermediate region and the fit to the data (blue) reconstructed from a system of collocation points (green). 
}
\label{fig:IntCollocation}
\end{center}
\end{figure}

An example of the reconstructed intermediate phase derivative against hybrid data, along with the collocation points used, is shown in Fig.\ref{fig:IntCollocation}. Unless otherwise stated, we take the 5th order polynomial ansatz as the default model for the intermediate phase derivative. 

\subsection{Merger-Ringdown}\label{sec:phase_mr}
As in previous phenomenological waveform models, the ansatz for the merger-ringdown is based on a Lorentzian embedded in a background arising form the late inspiral and merger. In order to capture the steep inspiral gradient, negative powers of the frequency were added to the Lorentzian. 
\begin{align}
\eta \; \varphi^{'}_{\rm{RD}} &= c_{\rm{RD}} + \displaystyle\sum^n_{i} c_i \, f^{-{p_i}} + \frac{c_0 \, a_{\varphi}}{\fdamp^2 + (f - \frd)^2} 
\end{align}
\n
In \phD, the leading contribution was taken to be $p_2 = 2$ and an additional term $p_3 = - 1/4$ was added in order to reduce residuals across the parameter space. However, for \phX, we find that we no longer require the coefficient $\alpha_5$, defined in Eq.~13 of \cite{Khan:2015jqa}, to correct for the ringdown frequency. Using the recent recalibration of the final mass and spin fits in \cite{Jimenez-Forteza:2016oae}, the values of the ringdown and damping frequency are sufficiently accurate that we are able to drop this coefficient. This allows us to calibrate an additional coefficient without increasing the dimensionality of the fit. The inclusion of an additional polynomial coefficients is of particular importance in correctly modelling the gradient of the merger-ringdown in the extremal spin limit. For \phX, we use three polynomial coefficients with powers of $-4, -2$ and $-1/3$ respectively. 

\begin{figure}[h!]
\begin{center}
\includegraphics[width=\columnwidth]{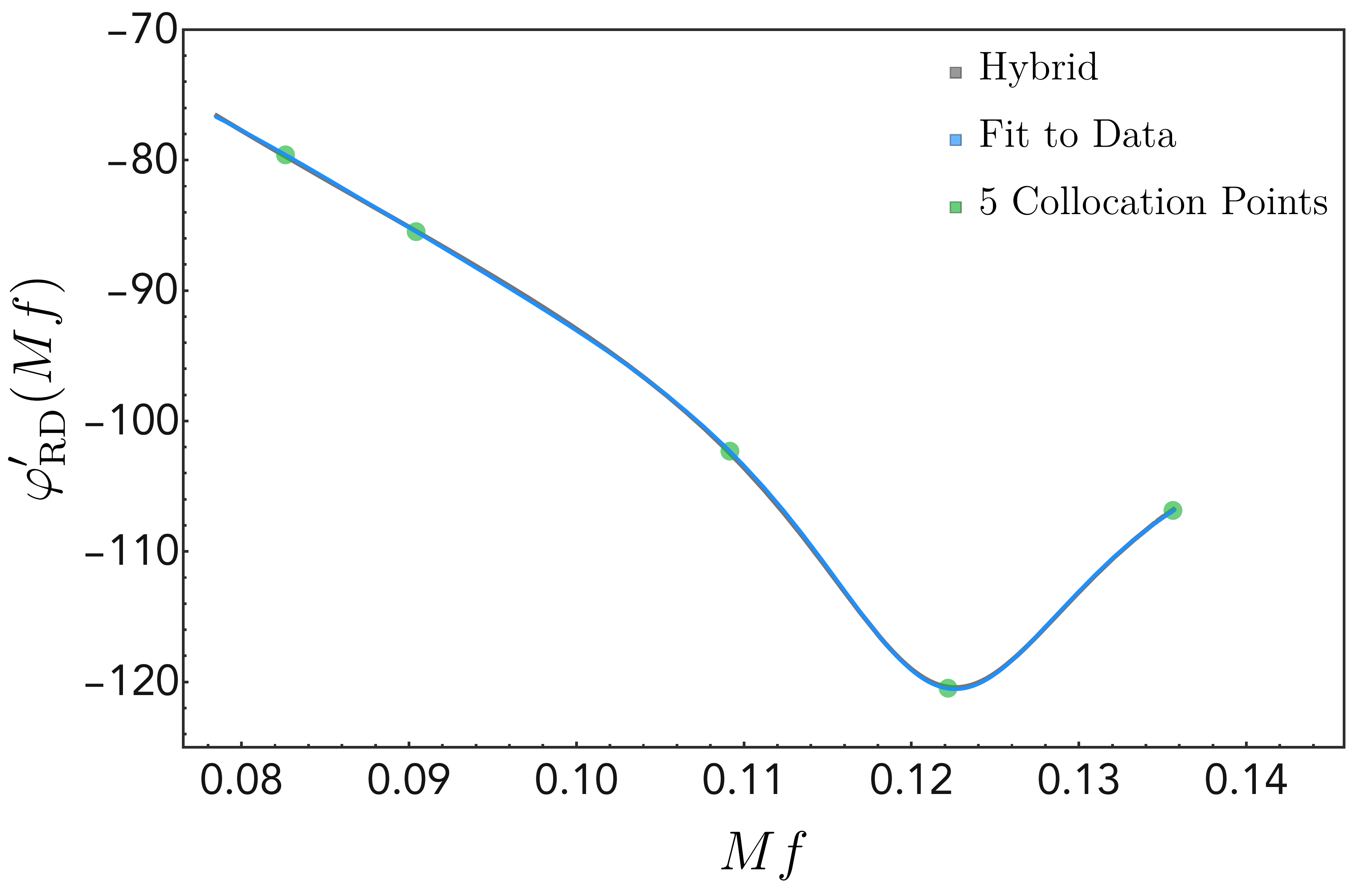}
\caption{
Phase derivative for \textrm{SXS-BBH-0153} hybridized against SEOBNRv4 in the merger-ringdown region along with the fit to the data (blue) reconstructed from a system of collocation points (green). 
}
\label{fig:RDCollocation}
\end{center}
\end{figure}

As with the other regions, we use Gauss-Chebyshev nodes to fix the collocation points but set the 4th node to occur at the ringdown frequency. The ringdown frequency approximately correponds to the peak of the Lorentzian, as can be seen in Fig.~\ref{fig:RDCollocation}. Whilst this may impact the optimality of reconstructing the underlying function, we find that $v_4^{\rm{MR}}$ is very-well conditioned and can be fit to high accuracy. The collocation points nodes for the merger-ringdown phase are therefore taken to be
\begin{align}
    v_i &= \left\lbrace f_L , \frac{1}{2} \left( 1 - \frac{1}{\sqrt{2}} \right) \delta^{\rm{MR}}_{\varphi} + f_L , f_L + \frac{1}{2} \delta_{\varphi}^{\rm{MR}} , f_{\rm{RD}}, f_H \right\rbrace ,
\end{align}
where $f_H = f_{\rm{RD}} + \frac{5}{4} f_{\rm{damp}}$, $f_L = f^{\varphi}_T$ and $\delta^{\rm{MR}}_{\varphi} = f_H - f_L$. 

An example of the reconstructed merger-ringdown phase derivative against hybrid data, along with the collocation points used, is shown in Fig.\ref{fig:RDCollocation}. The fit detailed here is as implemented in the final model. 

\subsection{Final State}\label{sec:final_state_fits}
As was highlighted earlier, \phX implicitly benefits from a recalibration of the fits to the final state and how this is mapped from the progenitor system \cite{Jimenez-Forteza:2016oae}. The total angular momentum pre-merger can be written in terms of the physical spins $\vec{S}_1, \vec{S}_2$ and the orbital angular momentum $\vec{L}$ as 
\begin{align}
    \vec{J} &= \vec{L} + \vec{S}_1 + \vec{S}_2 ,
\end{align}
\n 
where, due to symmetry arguments, we can approximate the Kerr parameter of the remnant BH as $a_f = J_f / M^2_f$. The final spin is approximated using the fits given in \cite{Jimenez-Forteza:2016oae}. To determine the dimensionless angular ringdown, $M \omega_{\rm{RD}}$, and damping, $M \omega_{\rm{damp}}$, frequencies as functions of the final spin $a_f$, we use rational functions fit to the dataset from \cite{Berti:2009kk} extended to better capture extremal spin behaviour. The dimensionful ringdown frequency can then be written in terms of the final mass
\begin{align}
    \omega_{\rm{RD}} &= \frac{M \omega_{\rm{RD}}}{M_f} = \frac{M \omega_{\rm{RD}}}{M - E_{\rm{rad}}} .
\end{align}
\n 
In order to accurately determine the final mass, we use the recently recalibrated fit to the radiated energy of \cite{Jimenez-Forteza:2016oae}.

\section{Model Calibration, A Worked Example}
\label{sec:worked}
In this section we provide a worked example of the hierarchical fitting procedure used to calibrate \phX. Here we detail the calibration of $v^{\rm{MR}}_4$, the phase derivative evaluated at the ringdown frequency, which effectively captures the value of the phase derivative at the peak of the Lorentzian. We use available SXS, BAM and ET waveforms supplemented by the test-particle waveforms. As was also observed in \cite{Husa:2015iqa} and \cite{Jimenez-Forteza:2016oae}, an effective spin parameterization defined in terms of the dimensionful spin components ${S}_i$ most accurately reflects the physics driving the merger-ringdown. The spin parameterization, $\hat{S}$, that we will use to calibrate $v_4^{\rm{MR}}$ is the total effective spin, $\hat{S}_{\rm{tot}}$, defined in Eq.~\ref{eq:stot}.

\subsubsection{1D Fits}
As described above, the starting point is a 1D fit to the non-spinning subspace. We follow the general procedure of first producing a high-degree polynomial and using this to construct a Pad\'e approximant that can be used to pre-condition a rational fit using the \texttt{NonLinearModelFit} package in \texttt{Mathematica}. The resulting rational function fit is of the following form
\begin{align}
f \left( \eta \right) &= \frac{a_0 + a_1 \eta + a_2 \eta^2 + a_3 \eta^3 + a_4 \eta_4 + a_5 \eta_6}{1 + a_6 \eta},
\end{align}
\newline
with numerical coefficients
\begin{align}
\label{eq:a_coeff}
    a_0 &= 85.8606,\\ \nonumber
    a_1 &= -4616.74, \\ \nonumber
    a_2 &= -4925.76, \\ \nonumber
    a_3 &= 7732.06, \\ \nonumber
    a_4 &= 12828.3, \\ \nonumber
    a_5 &= -39783.5, \\ \nonumber
    a_6 &= 50.2063 . 
\end{align}
Similarly, following the same procedure but applied to the 1D equal-mass, equal-spin subspace, we find a rational function fit of the form
\begin{align}
    f \left( \hat{S} \right) &= b_0 + \frac{b_1 \hat{S} + b_2 \hat{S}^2 + b_3 \hat{S}^3 + b_4 \hat{S}^4 + b_5 \hat{S}^4}{1 + b_6 \hat{S}^6} ,
\end{align}
\newline
with numerical coefficients
\begin{align}
\label{eq:b_coeff}
    b_0 &= -104.477, \\ \nonumber
    b_1 &= -19.0379, \\ \nonumber
    b_2 &= 15.3476, \\ \nonumber
    b_3 &= -0.419939, \\ \nonumber
    b_4 &= -0.884176, \\ \nonumber
    b_5 &= -0.631487, \\ \nonumber
    b_6 &= -0.729629 .
\end{align}
\newline
This fit is constrained by imposing the limit $\hat{S} \rightarrow 0$, ensuring that there are no discontinuities arise with respect to the non-spinning fit.

\begin{centering}
\begin{figure}
\includegraphics[width=\columnwidth]{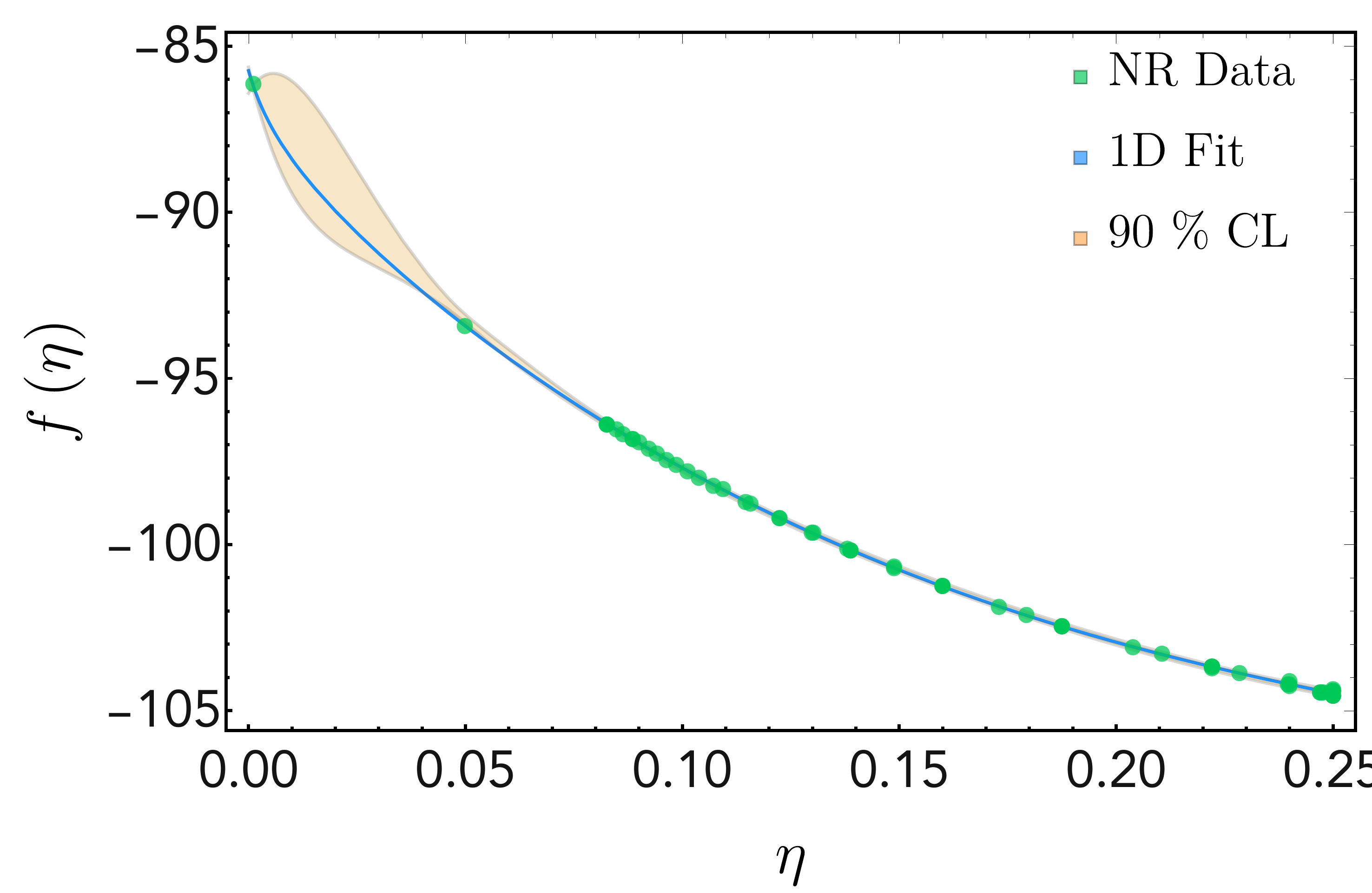}
\includegraphics[width=\columnwidth]{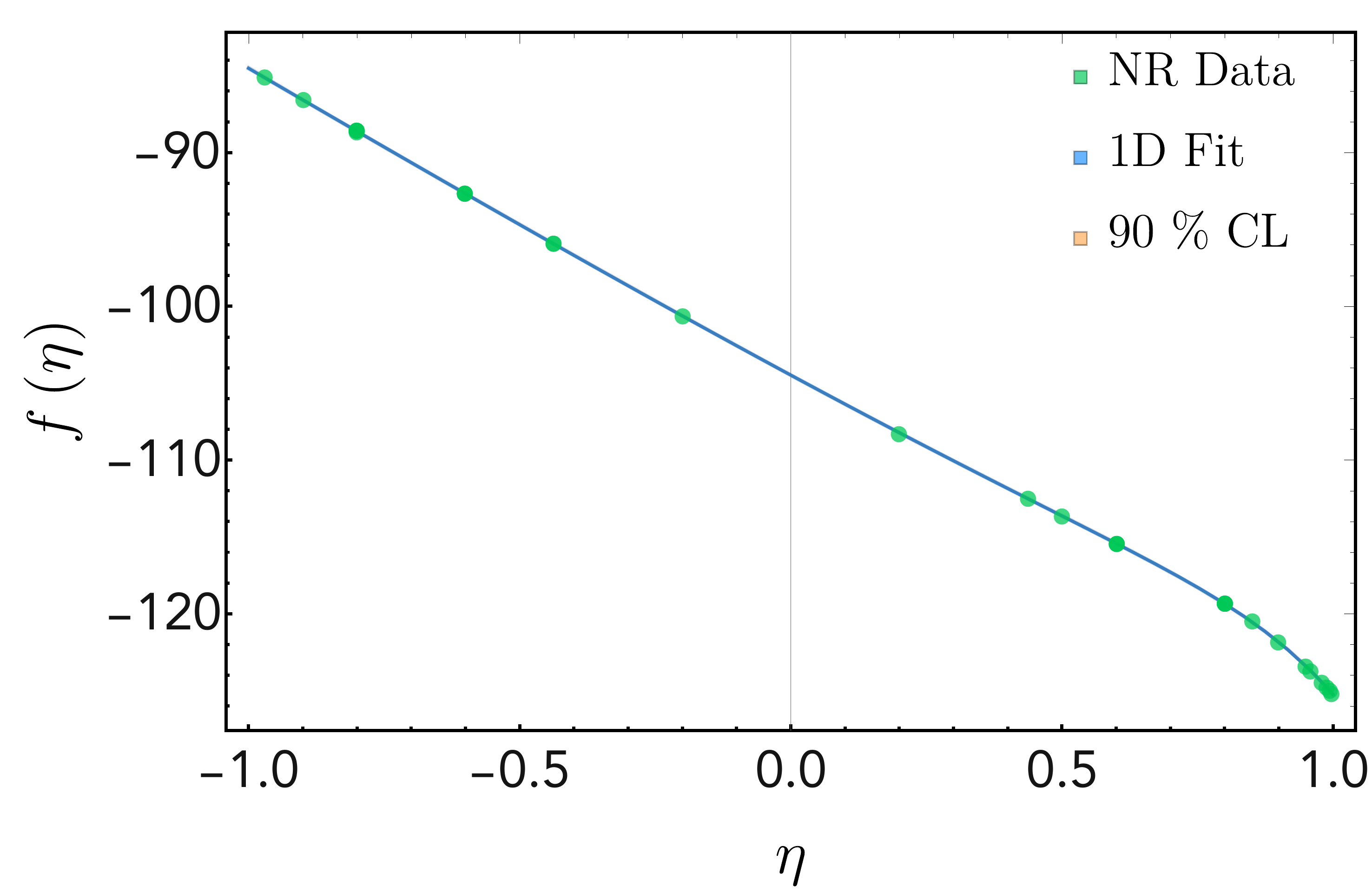}
  \caption{ \label{fig:RD_v4_1D_Fits} %
  The 1D fits to the non-spinning and equal-mass, equal-spin subspaces. The blue curves show the 1D fits to the data and the orange shaded region denotes the 90\% CL for the fit. The green points denote the NR and test-particle datasets. The equal-mass, equal-spin parameter space is extermely well understood. The non-spinning parameter space is densely covered up to $q \sim 10$ with only the non-spinning BAM simulation at $q = 18$. Test-particle data can be used to pin the boundary at $q = 1000$ but there is a clear degree of uncertainty in the intermediate region from $q > 20$. 
}
\end{figure}
\end{centering}

\subsubsection{2D Fits}
\begin{centering}
\begin{figure}
\includegraphics[width=\columnwidth]{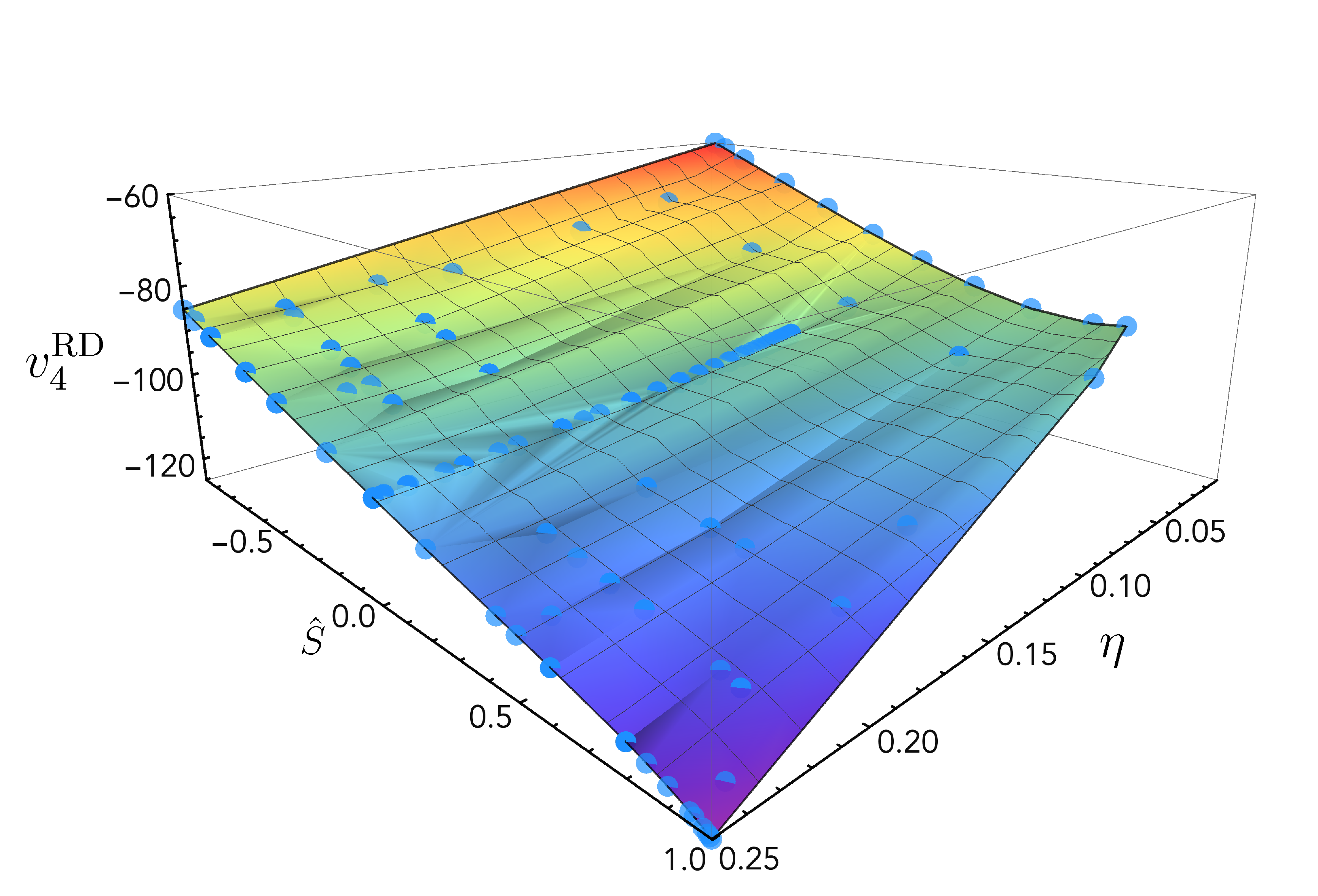}
  \caption{ \label{fig:RD_v4_2D_Fits} %
  Fit $f( \eta, \hat{S} )$ to the two-dimensional subspace $\lbrace \eta , \hat{S} \rbrace$. The blue points denote NR and test-particle data.
}
\end{figure}
\end{centering}

The two-dimensional fits to the $(\eta,\hat{S})$ subspace are constructed by combining both of the 1D subspace fits derived above. As discussed in Sec.~\ref{sec:2d_fits}, we generalize the $\hat{S}$-dependent fits by inserting a polynomial of order $J$ in $\eta$ for each coefficient in the 1D fit \cite{Jimenez-Forteza:2016oae,Keitel:2016krm}. Here we opt to use a fourth order in $\eta$ expansion $(J = 4)$ and kill the least constrained coefficients (i.e. p-values near unity) as determined by the non-linear model fit. In addition, we fix all coefficients on the denominator to avoid singularities. The constrained 2D fit $f(\eta,\hat{S})$ against the input data is shown in Fig.~\ref{fig:RD_v4_2D_Fits}.

\begin{centering}
\begin{figure}
\includegraphics[width=\columnwidth]{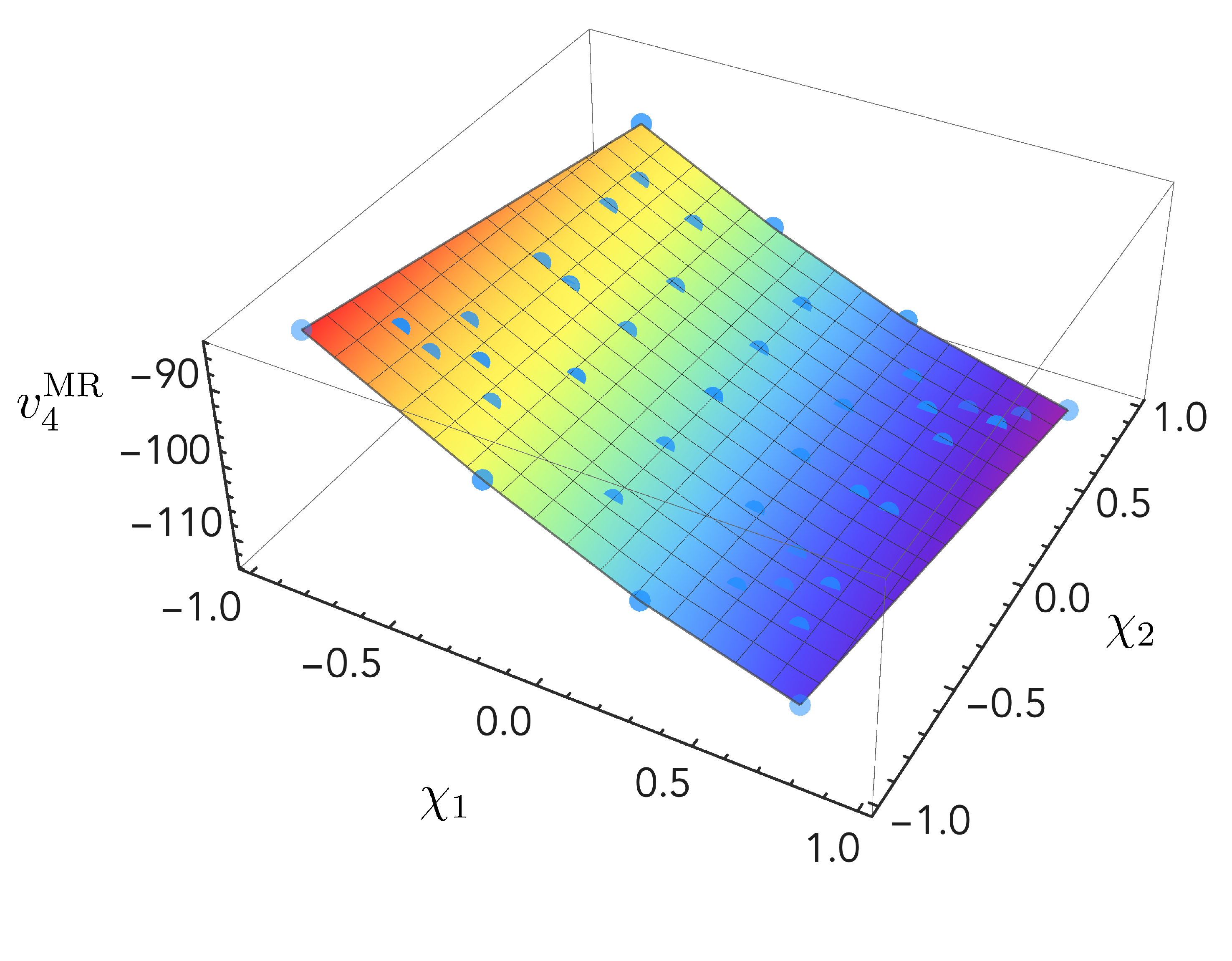}
\includegraphics[width=\columnwidth]{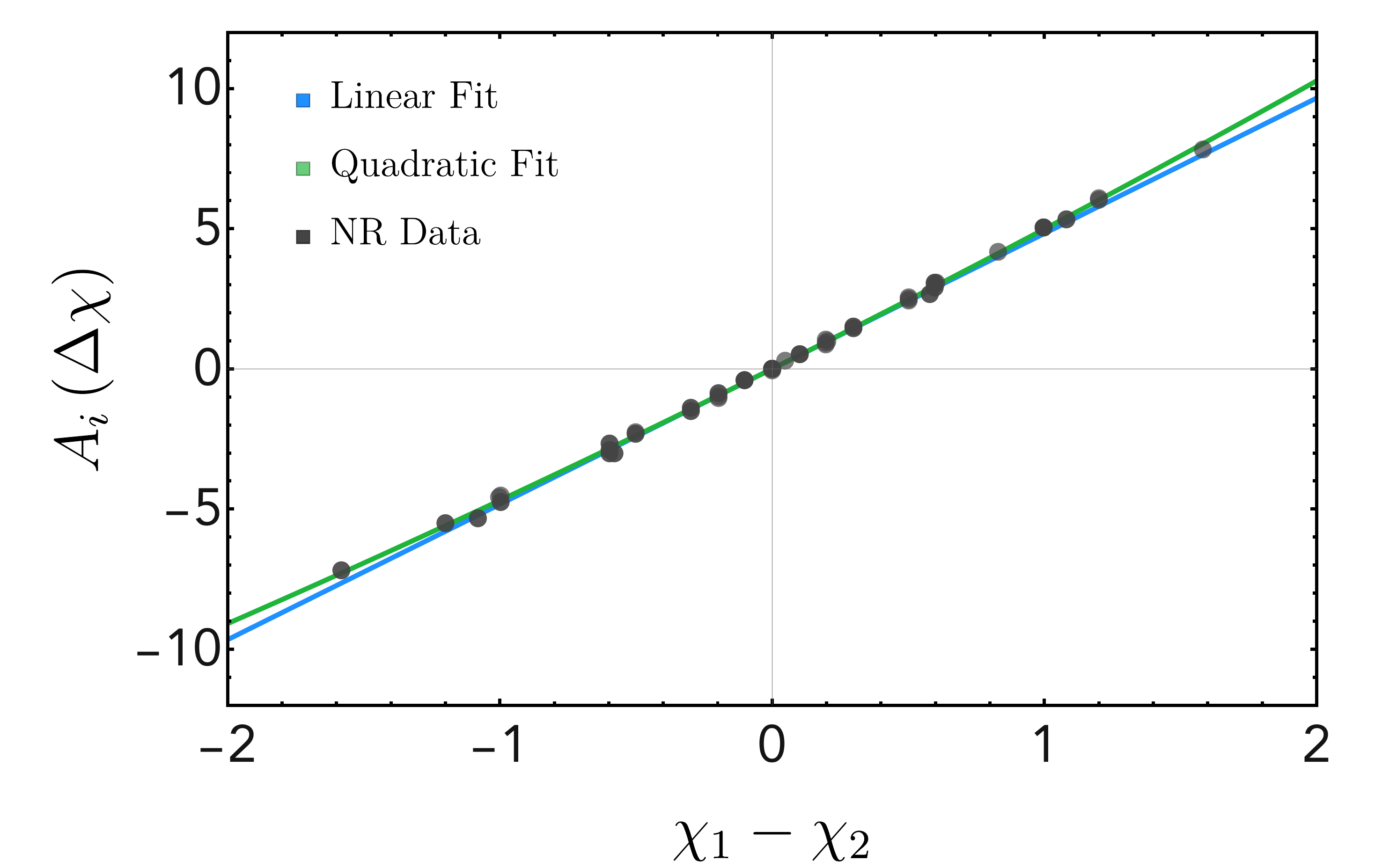}
  \caption{ \label{fig:RD_v4_2D_Fits_q3_data} %
  Example of the spin-difference behaviour at a mass ratio $q = 3$. At such mass ratios the surface is very close to flat and the linear-in-spin difference term dominates. 
}
\end{figure}
\end{centering}

\subsubsection{Full 3D Fits}
\begin{centering}
\begin{figure}
\includegraphics[width=\columnwidth]{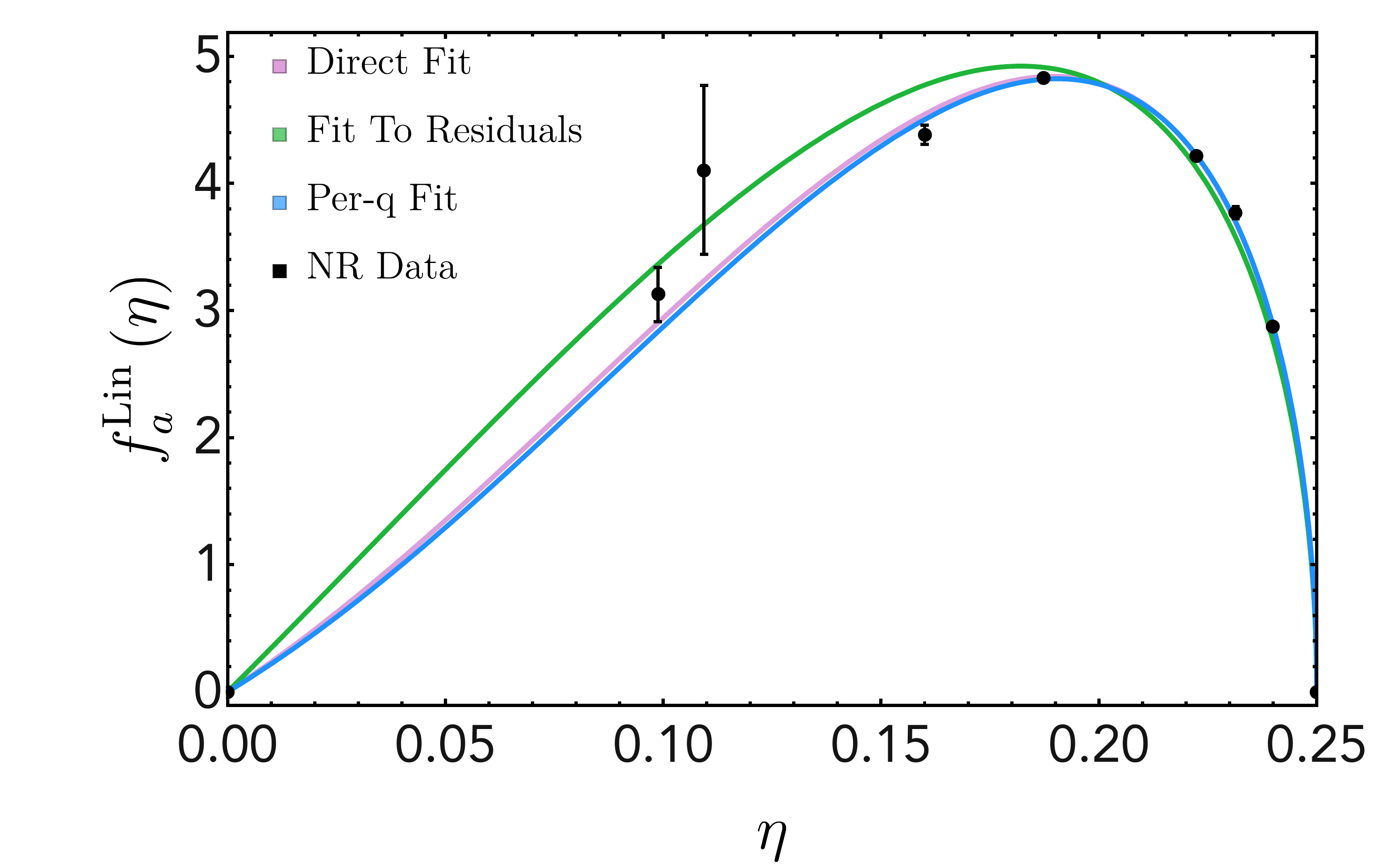}
\includegraphics[width=\columnwidth]{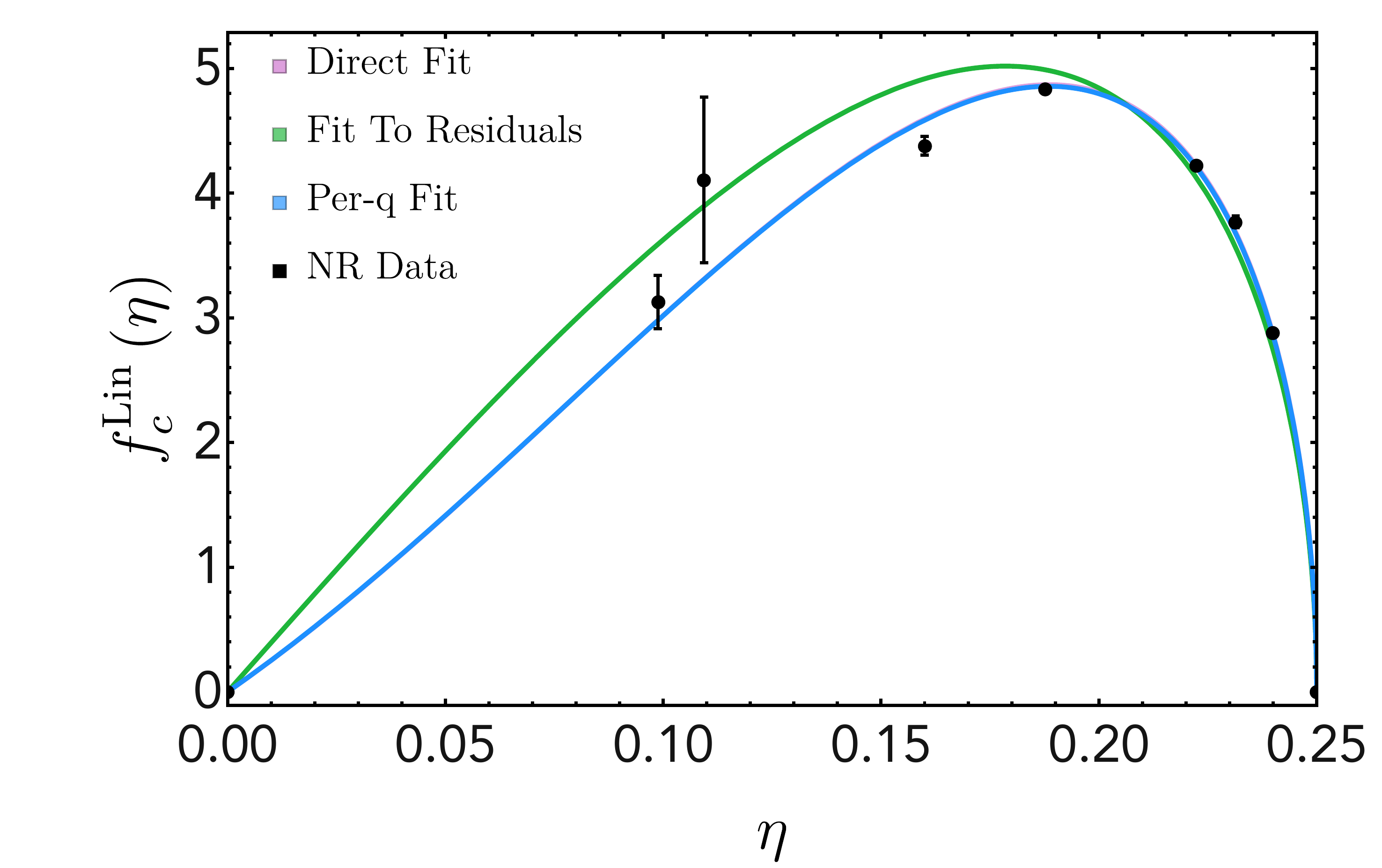}
\includegraphics[width=\columnwidth]{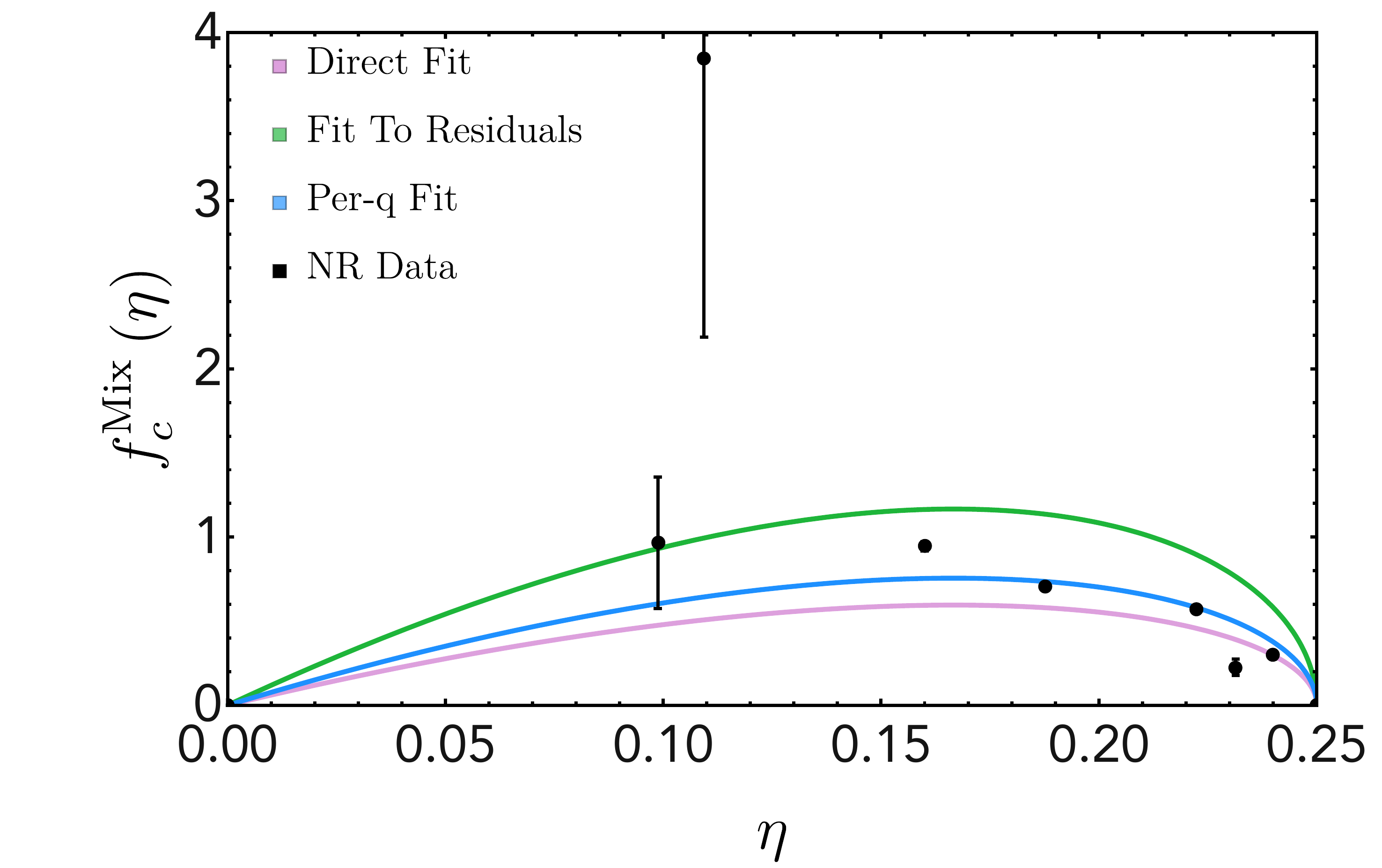}
\includegraphics[width=\columnwidth]{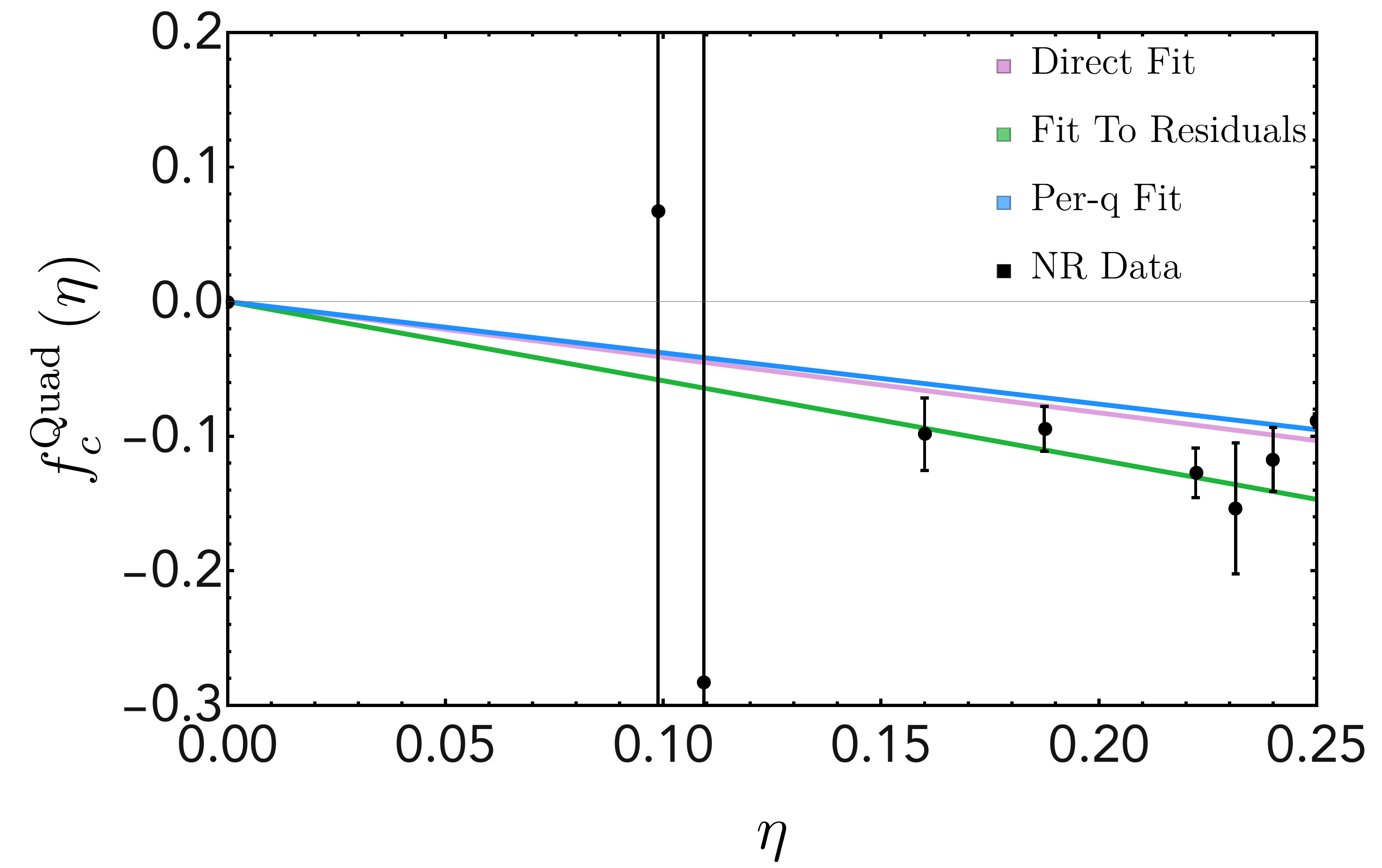}
  \caption{ \label{fig:unequalspinfit} %
  Fits to the unequal spin data. The first plot shows the linear-in-spin difference fit. The second plot shows the linear-in-spin difference contribution to the full unequal spin ansatz in Eq.\ref{eq:full_spin_diff_ansatz}, $f^{\rm{Lin}}_c(\eta) \Delta \chi$ fit. The third plot the mixed-spin fit $f^{\rm{Mix}}_c(\eta) \, \hat{S} \, \Delta \chi $ and the final plot shows the quadratic in spin-difference fit $f^{\rm{Quad}}_c(\eta) \, \left( \Delta \chi \right)^2$. Whilst the linear-in-spin difference is relatively well captured, the second order unequal spin effects are less resolved with the coefficients for the fits becoming poorly constrained.
}
\end{figure}
\end{centering}

The final step in the procedure is to fit the unequal spin effects, parameterized by $\Delta \chi = \chi_1 - \chi_2$, to the residual data, as discussed in Sec.~\ref{sec:3d_fits}. In Fig.~\ref{fig:RD_v4_2D_Fits_q3_data}, we show the unequal spin subspace for all $q = 3$ NR data. As anticipated, the data is dominated by a linear-in-spin-difference terms with only weak evidence for higher order corrections. Fits are performed at all mass ratios for which we have sufficient data to constrain the ansatz. Note that the ansatz per-mass-ratio is used to inform the full 3D ansatz and as a consistency check. The 3D fit implemented in the model is constructed by fitting the constrained 2D ansatz plus the unequal spin terms to the full NR dataset. Here we demonstrate two approaches to constraining the unequal-spin fit. In the first approach, we restrict our analysis to the linear-in-spin-difference term $\Delta \chi$
\begin{align}
    \Delta v^{\rm{MR}}_4 (\eta, \hat{S}, \Delta \chi ) &= A_1 (\eta) \Delta \chi .
\end{align}
\n 
In the second approach we use the full quadratic in unequal spin-correction ansataz
\begin{align}
    \Delta v^{\rm{MR}}_4 (\eta, \hat{S}, \Delta \chi ) &= f^{\rm{Lin}} (\eta) \Delta \chi + f^{\rm{Quad}} (\eta) \Delta \chi^2 + f^{\rm{Mix}} (\eta) \hat{S} \, \Delta \chi.
\end{align}
\n 
Based on the symmetry arguments outlined in \cite{Jimenez-Forteza:2016oae,Keitel:2016krm} and Sec.~\ref{sec:3d_fits}, we adopt an ansatz for the linear in spin-difference term of the form 
\begin{align}
    f^{\rm{Lin}} (\eta) &= d_{10} \eta \left(1 + d_{11} \right) \sqrt{1 - 4 \eta} .
\end{align}
\n 
For the quadratic in spin-difference and mixed spin-difference ans{\"a}tze, we adopt simpler fits of the form
\begin{align}
    f^{\rm{Quad}} (\eta) &= d_{20} \eta ,\\
    f^{\rm{Mix}} (\eta) &= d_{30} \eta \sqrt{1 - 4 \eta} .
\end{align}
\n 
Whilst more complicated ans{\"a}tze could be pursued, we typically find that the systematic errors in the NR data and strong correlations lead to poor constraints on the coefficients. As can be seen in the top two plots of Fig.~\ref{fig:unequalspinfit}, the shape and numerical value of the linear term is robust when adding different combinations of the unequal spin terms. Applying the fits to all data, we find
\begin{align}
    d_{10} &= 22.3632, \\
    d_{11} &= 6.9794 ,
\end{align}
\n 
for the linear-only ansatz and 
\begin{align}
    d_{10} &= 24.1579, \\
    d_{11} &= 6.1330, \\
    d_{20} &= -0.4132, \\
    d_{30} &= 6.1896 
\end{align}
\n 
for the full ansatz. Though the data shows some evidence for quadratic-in-spin-difference corrections, third plot of Fig.~\ref{fig:unequalspinfit}, and mixed spin-difference terms, second plot of Fig.~\ref{fig:unequalspinfit}, systematic errors prevent a robust fit to the data. As such, for \phX we opt to use the linear-only ansatz in the final 3D fit.

The full 3D fit to the data is 
\begin{widetext}
\begin{align}
    \Delta v^{\rm{MR}}_4 (\eta, \hat{S}, \Delta \chi ) &= \frac{a_0 + a_1 \eta + a_2 \eta^2 + a_3 \eta^3 + a_4 \eta_4 + a_5 \eta_6}{1 + a_6 \eta}
    + \frac{1}{1 + b_6 \hat{S}} \Bigg[ \hat{S} \left( c_0 + c_1 \eta + c_2 \eta^2 + c_3 \eta^3 + c_4 \eta^4 \right) 
    \\ \nonumber &\quad + 
    \hat{S}^2 \left( d_0 + d_1 \eta + d_2 \eta^2 + d_3 \eta^3 + d_4 \eta^4 \right) +
    \hat{S}^3 \left( e_0 + e_1 \eta + e_2 \eta^2 + e_3 \eta^3 + e_4 \eta^4 \right) \\ \nonumber
    &+
    \hat{S}^4 \left( f_0 + f_1 \eta + f_2 \eta^2 + f_3 \eta^3 + f_4 \eta^4 \right) +
    \hat{S}^5 \left( g_0 + g_1 \eta + g_2 \eta^2 + g_3 \eta^3 + g_4 \eta^4 \right) \Bigg] + h_0 \left( 1 + h_1 \eta \right) \sqrt{1 - 4 \eta} \Delta \chi
     .
\end{align}
\end{widetext}
The coefficients $a_i$ are defined in Eqs.~\ref{eq:a_coeff} and $b_6$ in Eq.~\ref{eq:b_coeff}. The additional coefficients are 
\begin{align*}
    c_0 &= -24.32, & c_1 &= 50.49 \\ 
    c_2 &= -68.32, & c_3 &= 0.0 \\ 
    c_4 &= 784.98 & \phantom{a} &\phantom{a} \\ 
    d_0 &= 26.62, & d_1 &= -19.39 \\ 
    d_2 &= 13.27, & d_3 &= 1092.51 \\ 
    d_3 &= 2512.13, & \phantom{a} & \phantom{a} \\
    e_0 &= 2.80, & e_1 &= 11.23 \\ 
    e_2 &= -308.99, & e_3 &= 74.22 \\ 
    e_4 &= 3103.82, & \phantom{a} & \phantom{a}\\ 
    f_0 &= -1.68, & f_1 &= -22.78 \\ 
    f_2 &= 76.14, & f_3 &= 0.0 \\ 
    f_4 &= 443.83, & \phantom{a} & \phantom{a} \\ 
    g_0 &= -1.21, & g_1 &= -71.28 \\ 
    g_2 &= 525.24, & g_3 &= 0.0 \\ 
    g_4 &= 3694.97, & \phantom{a} & \phantom{a} \\
    h_0 &= 22.36, & h_1 &= 6.98 .
\end{align*}
\n 
The full 3D fits for all coefficients required for \phX are given in the supplementary material attached to this paper. 

\section{Model Validation}\label{sec:validation}
\subsection{Mismatches Against Hybrid Dataset}
The agreement between two waveforms $h_1$ and $h_2$ can be quantified by the overlap, the noise-weighted inner product
\begin{align}
\label{eq:mismatch}
\langle h_1 , h_2 \rangle &= 4 \ {\rm{Re}} \int^{f_{\rm{max}}}_{f_{\rm{min}}} \, \frac{\tilde{h}_1 (f) \; \tilde{h}^{\ast}_2 (f)}{S_n (f)} \, df .
\end{align}
\n
The \textit{match} is defined as the normalised ($\hat{h} = h / \sqrt{\langle h,h \rangle}$) inner product maximised over time and phase shifts
\begin{align}
M (h_1 , h_2) = \max\limits_{t_0 , \phi_0} \, \langle \hat{h}_1 , \hat{h}_2 \rangle .
\end{align}
\n
The \textit{mismatch} is then defined as
\begin{align}
\mathcal{M} (h_1 , h_2) = 1 - M(h_1 , h_2) .
\end{align} 
\n
In all matches presented here, we use the zero-detuned high-power (zdethp) PSD \cite{TheLIGOScientific:2014jea,dcc:2974}. We use a low frequency cut-off of $20 \rm{Hz}$ and an upper cut-off frequency of $8192 \rm{Hz}$. 

\begin{centering}
\begin{figure}
  \includegraphics[width=\columnwidth]{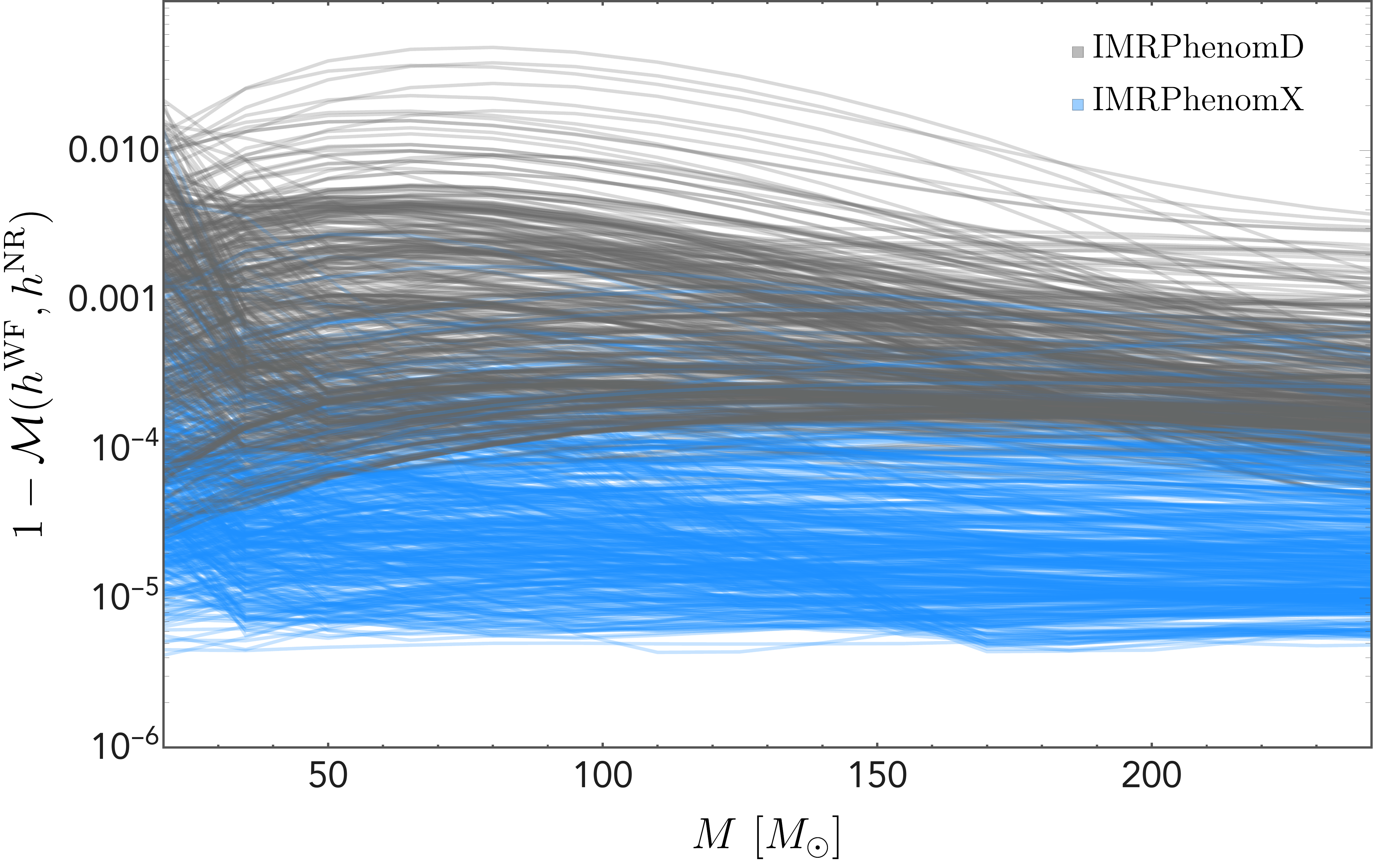}
  \caption{ \label{fig:mismatches} %
  Mismatches for \phX (blue) and \phD (grey) against all SXS NR hybrids. We use the Advanced LIGO design sensitivity PSD and a lower frequency cutoff of $20$Hz. We see a dramatic decrease in the mismatch by $1$ to $2$ orders of magnitude across the parameter space.
}
\end{figure}
\end{centering}

\begin{centering}
\begin{figure}
  \includegraphics[width=\columnwidth]{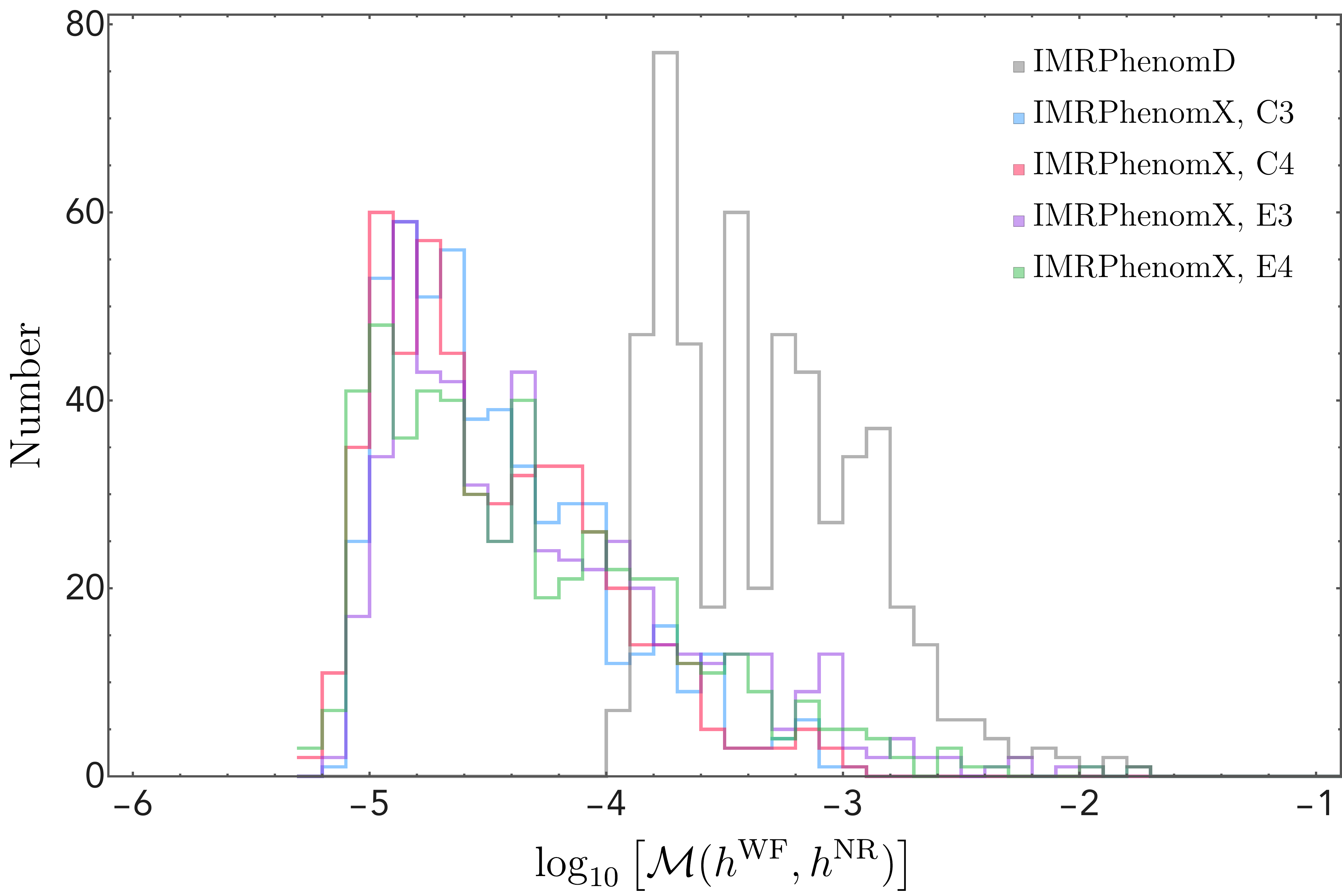}
  \caption{ \label{fig:mismatcheshist} %
  Mass averaged mismatches for \phX and \phD against all SXS NR hybrids. We use the Advanced LIGO design sensitivity PSD and a lower frequency cutoff of $20$Hz. We showcase four variants of \phX corresponding to different inspiral models. The C denotes the canonical TaylorF2 baseline at 3.5PN and the E denotes the extension to $4$ and $4.5$PN discussed in the Appendix. The number, 3 or 4, denotes the number of pseudo-PN terms used in the model.
}
\end{figure}
\end{centering}

\begin{centering}
\begin{figure}
  \includegraphics[width=\columnwidth]{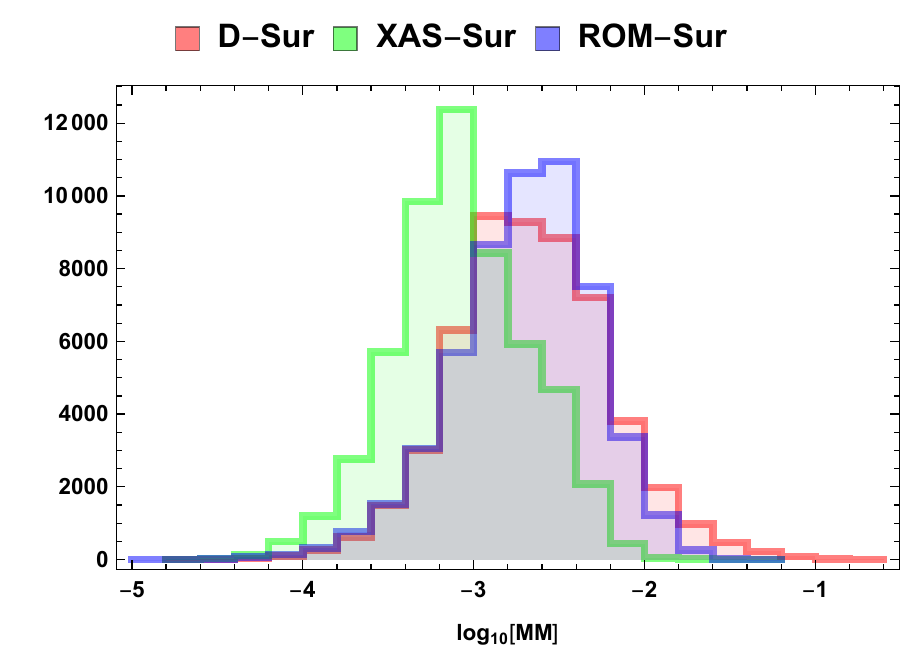}
  \caption{ \label{fig:mismatches_NRHybSur3dq8} %
  Mismatches ($\rm{MM} = \mathcal{M})$ for \phX (green), \phD (red) and \seobnr (blue) against NRHybSur3dq8, the NR hybrid surrogate valid up to a mass ratio $q = 8$ and spins $\chi_i = \pm 0.8$. We compute the matches at random points in the parameter space, including points that fall between the calibration datasets used to construct \phX. Here we clearly see that \phX offers a significant improvement in performance in comparison to \phD or \seobnr. 
}
\end{figure}
\end{centering}

In order to assess the accuracy of our model, we compute the mismatch against all SXS hybrids produced for \phX. As shown in Fig.~\ref{fig:mismatches}, \phX shows 1 to 2 orders of magnitude improvement over \phD across the entire parameter space. Figure~\ref{fig:mismatcheshist} shows mass-averaged mismatches for the performance of the four calibrated inspiral models discussed in Sec.~\ref{sec:phase_insp}. The inclusion of additional pseudo-PN coefficients demonstrates mild performance improvements, though not at a significantly appreciable level. In Fig.~\ref{fig:mismatches_NRHybSur3dq8} we show mismatches for \phX, \phD and \seobnrrom against NRHybSur3dq8 \cite{Varma:2018mmi} for mass ratios below 9.09 and dimensionless spin magnitudes up to 0.8. 

\subsection{Time Domain Conversion}
\begin{centering}
\begin{figure*}
\includegraphics[width=0.8\textwidth]{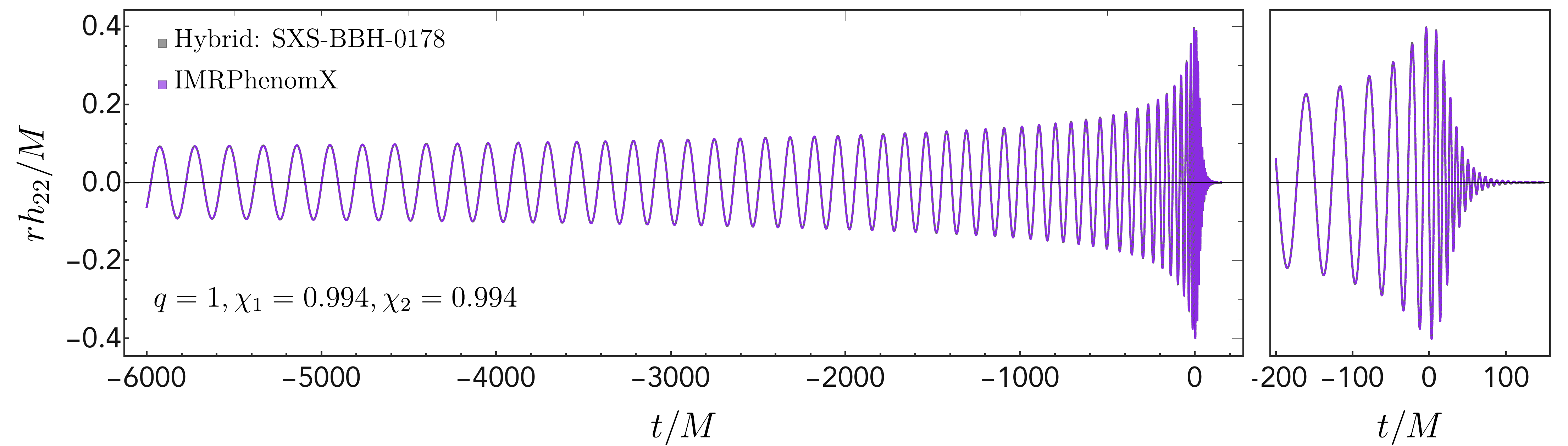}
\includegraphics[width=0.8\textwidth]{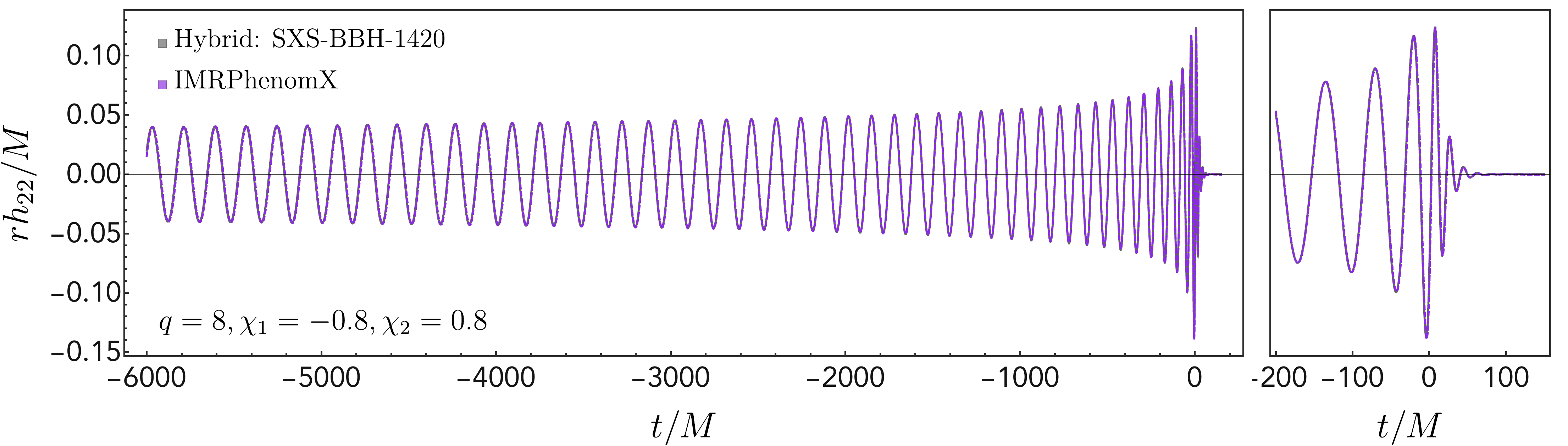}
\includegraphics[width=0.8\textwidth]{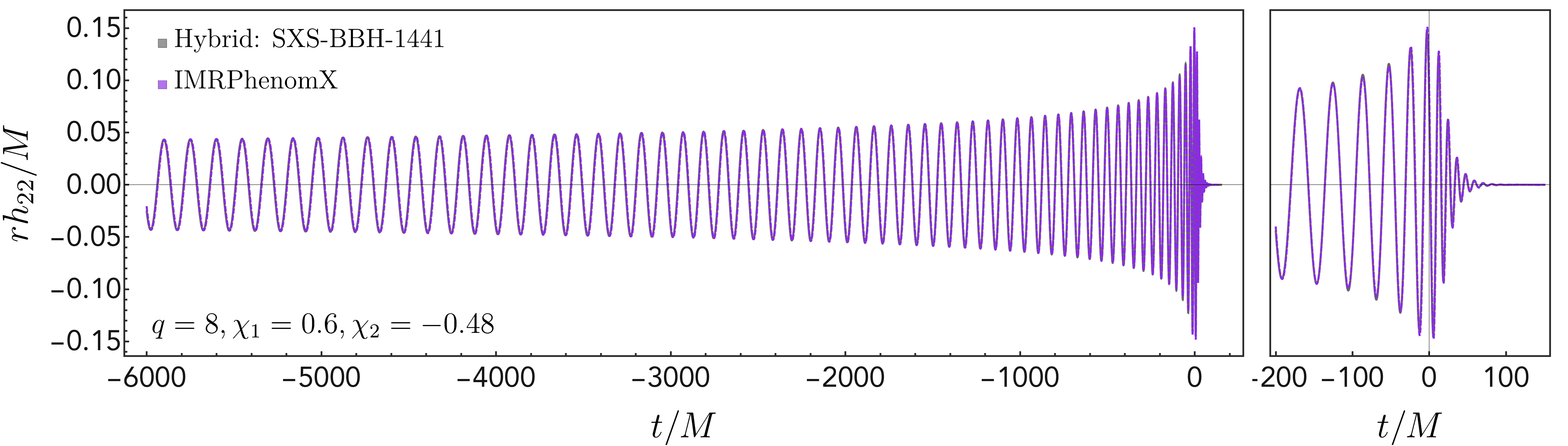}
\includegraphics[width=0.8\textwidth]{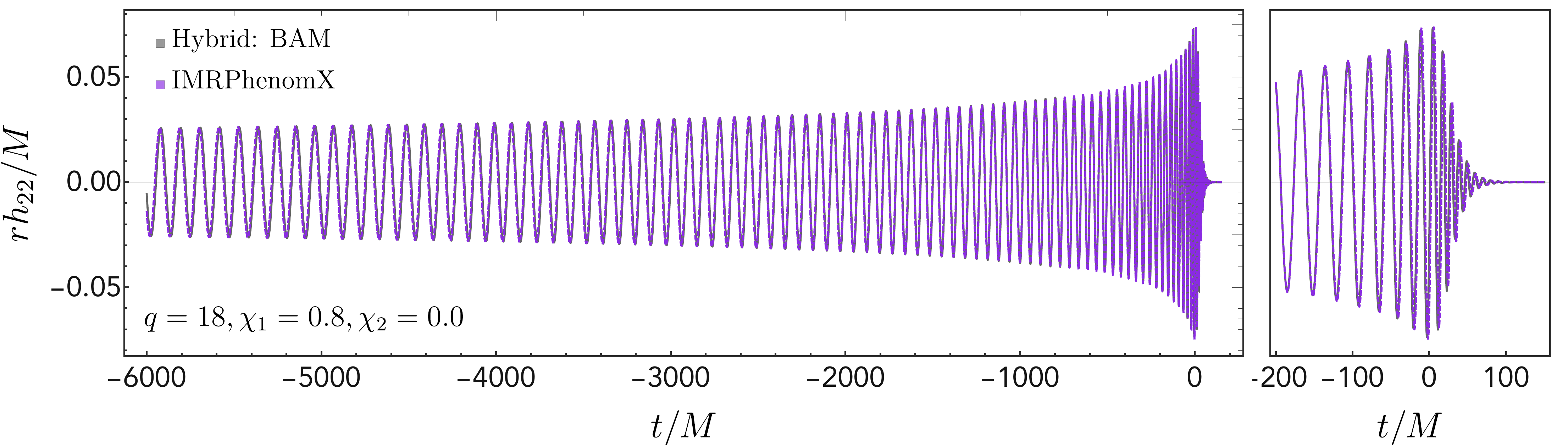}
\includegraphics[width=0.8\textwidth]{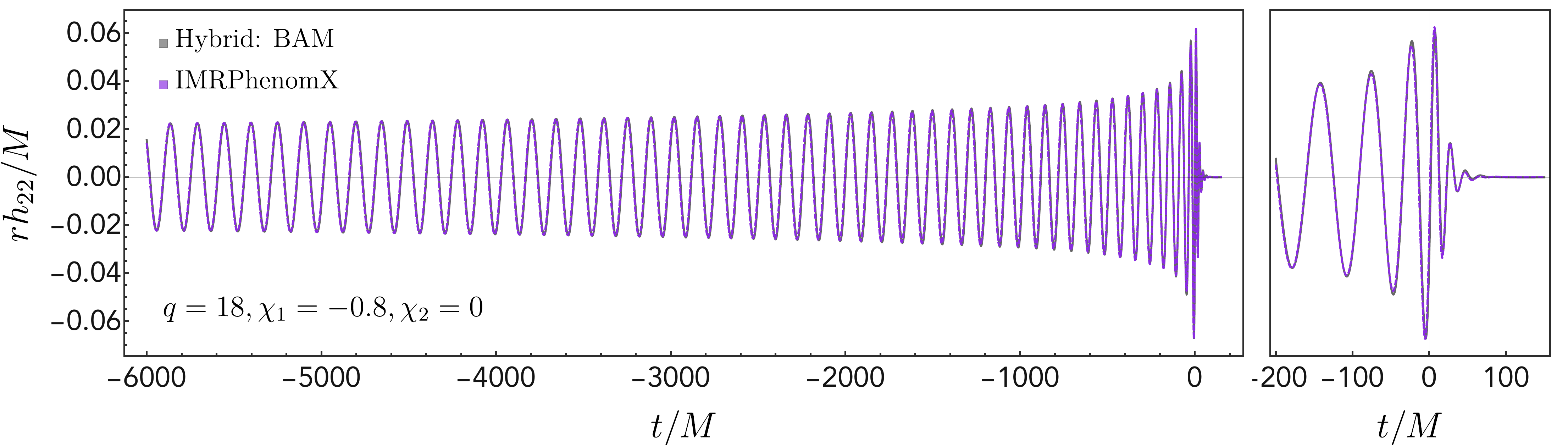}
  \caption{ \label{fig:TD_Comparisons} %
  Time-domain \phX waveforms (violet) and SEOBNRv4-NR hybrids (grey) for configurations at the edge of the calibration domain.
}
\end{figure*}
\end{centering}

Although \phX is expressed in terms of closed-form frequency domain expressions, the input calibration data and output from NR are time-domain function. It is therefore useful and illustrative to check the behaviour of the model when transformed from the frequency-domain back to the time-domain via an inverse Fourier transformation. In particular, the model should be a smooth function in both the frequency- and time-domain. Such comparisons are often useful as an additional consistency check on the physical accuracy of the model. In Fig.~\ref{fig:TD_Comparisons} we plot the time-domain reconstruction of \phX against selected SXS or BAM waveforms at the boundary of the calibration region for NR. We find excellent agreement between \phX and input NR data, even when considering near extremal spin configurations (first panel) as well as at large mass ratios and relatively large spins (last two panels). Note that we have optimized over a time and phase shift when aligning the waveforms. Such comparisons provide further evidence, in addition to the mismatches, that our end-to-end pipeline for hybridization, calibration and model reconstruction are faithfully reproducing the input data. 

As in \cite{Khan:2015jqa}, the frequency domain data is tapered and an optimal sampling rate chosen through the stationary phase approximation. 

\subsection{Parameter Estimation}
\subsubsection{GW150914}
As an example of the application of \phX to gravitational-wave data, we re-analyze GW150914 and demonstrate broad agreement between \phX, \phD and SEOBNRv4. We use coherent Bayesian inference methods to determine the posterior distribution $p(\vec{\theta} | \vec{d})$ for the parameters that characterize the binary. We use the nested sampling algorithm implemented in \texttt{LALInference} \cite{Veitch:2014wba} and the public data from the Gravitational Wave Open Science Center (GWOSC) \cite{GWOSC,gwtc1calib,gwtc1psd}. Following \cite{LIGOScientific:2018mvr}, we marginalize over the frequency dependent spline calibration envelopes that characterize the uncertainty in the detector amplitude and strain. We analyze four seconds of strain data, with a lower cutoff frequency of 20Hz. Our choice of priors is as detailed in Section I of Appendix C in \cite{LIGOScientific:2018mvr}. 

\begin{centering}
\begin{figure}[tbhp]
  \includegraphics[width=0.495\textwidth]{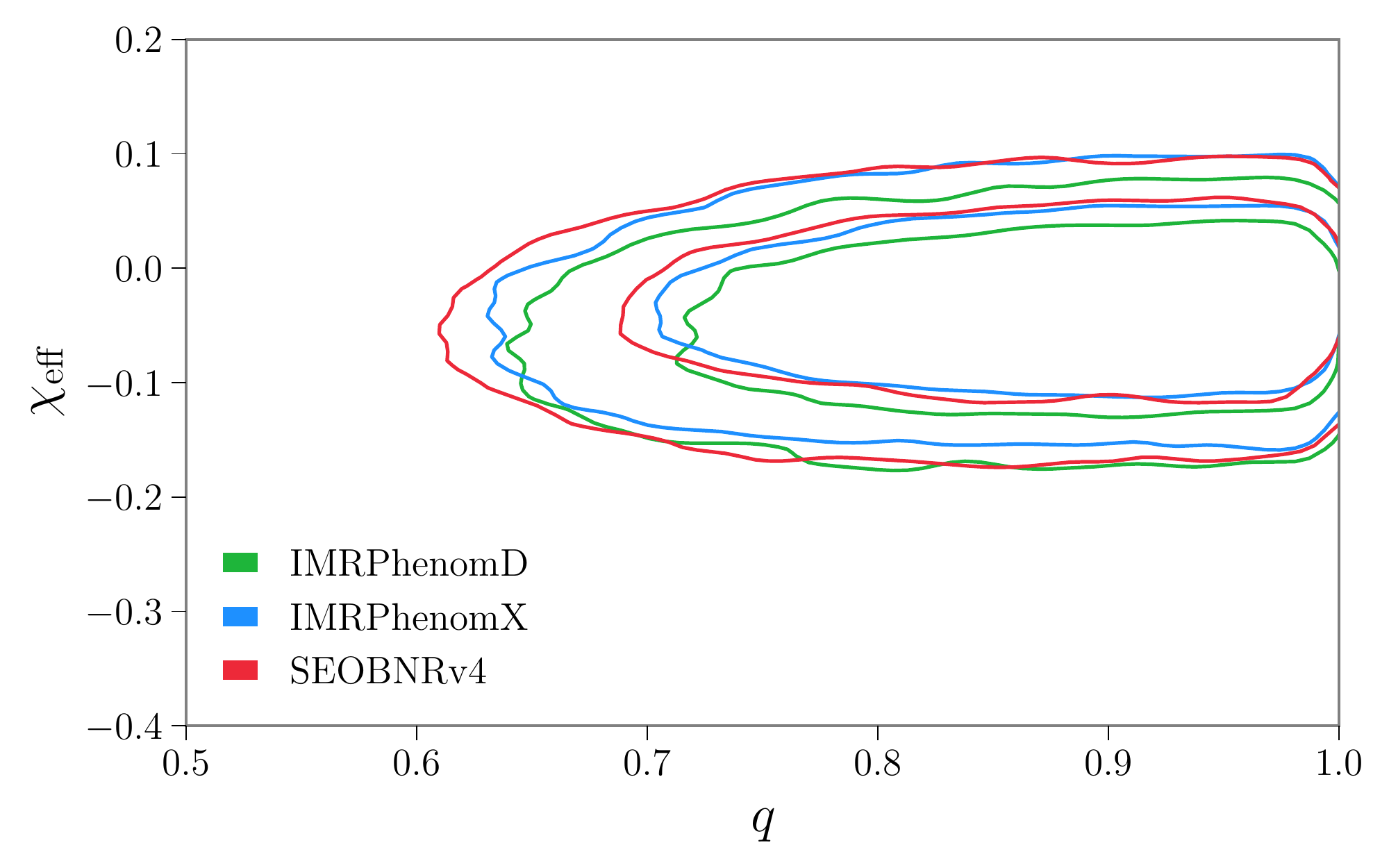}
  \includegraphics[width=0.495\textwidth]{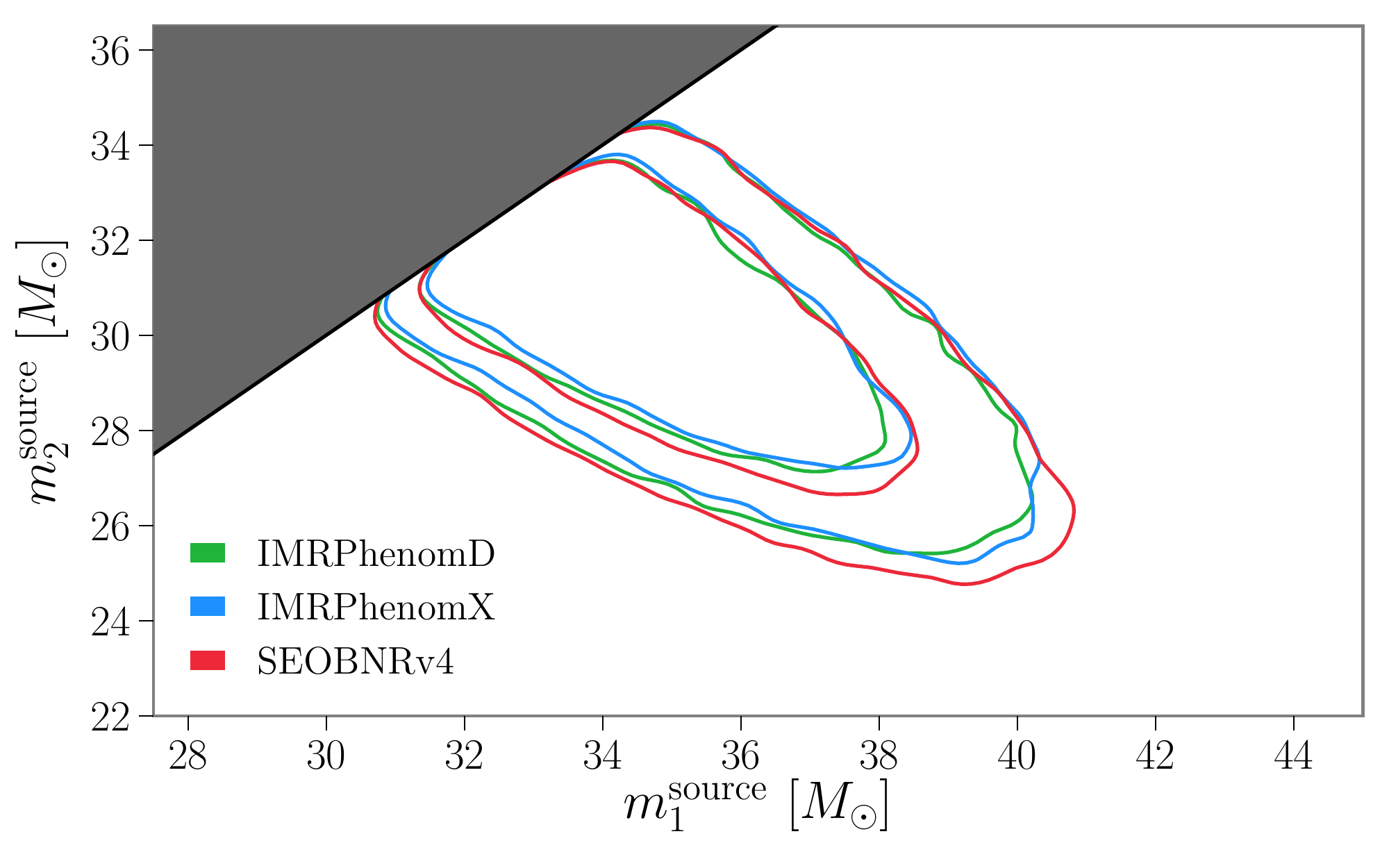}
  \caption{ \label{fig:GW150914_PE} %
  The top panel shows the $q-\chi_{\rm{eff}}$ 2D posteriors recovered by \phX, \phD and \texttt{SEOBNRv4} when analysing GW150914. All models show excellent agreement. The bottom panel shows the recovered component masses in the source frame using the same waveform models. Note that the black line denotes the equal mass limit and we enforce $m_1 > m_2$. 
}
\end{figure}
\end{centering}

Figure \ref{fig:GW150914_PE} shows the posterior densities for the $\mathcal{M}-q$ and $q-\chi_{\rm{eff}}$ subspaces. The consistency between the three waveform models is in agreement with previous studies, demonstrating that systematic errors were below the statistical errors for this event \cite{TheLIGOScientific:2016wfe,Abbott:2016wiq,Abbott:2016izl}.  

\subsubsection{NRHybSur3dq8}
In the second example, we inject a NRHybSur3dq8 waveform into a HLV detector network assuming zero-noise and using the Advanced LIGO and Advanced VIRGO design sensitivity PSDs \cite{adligopsd,TheLIGOScientific:2014jea,TheVirgo:2014hva}. The injected waveform was taken to have a mass-ratio of $q = 3$, chirp mass of $\mathcal{M}_c = 20 M_{\odot}$ and spins of $\chi_1 = 0.6$ and $\chi_2 = -0.3$. The luminosity distance was $d_L = 1 \rm{Gpc}$ and the sky location, polarization and coalesence phase were arbitrarily chosen. Priors are again taken to be as detailed in Section I of Appendix C in \cite{LIGOScientific:2018mvr}. In Fig.~\ref{fig:NRHybSur3dq8_PE}, we highlight the reduced bias provided by \phX over \phD, demonstrating how the advances implemented in \phX will help tighten and improve our constraints on the source properties of astrophysical black holes. A detailed study of waveform systematics and parameter biases is beyond the scope of this paper and will be presented in a forthcoming paper.

\begin{centering}
\begin{figure}[tbhp]
  \includegraphics[width=0.495\textwidth]{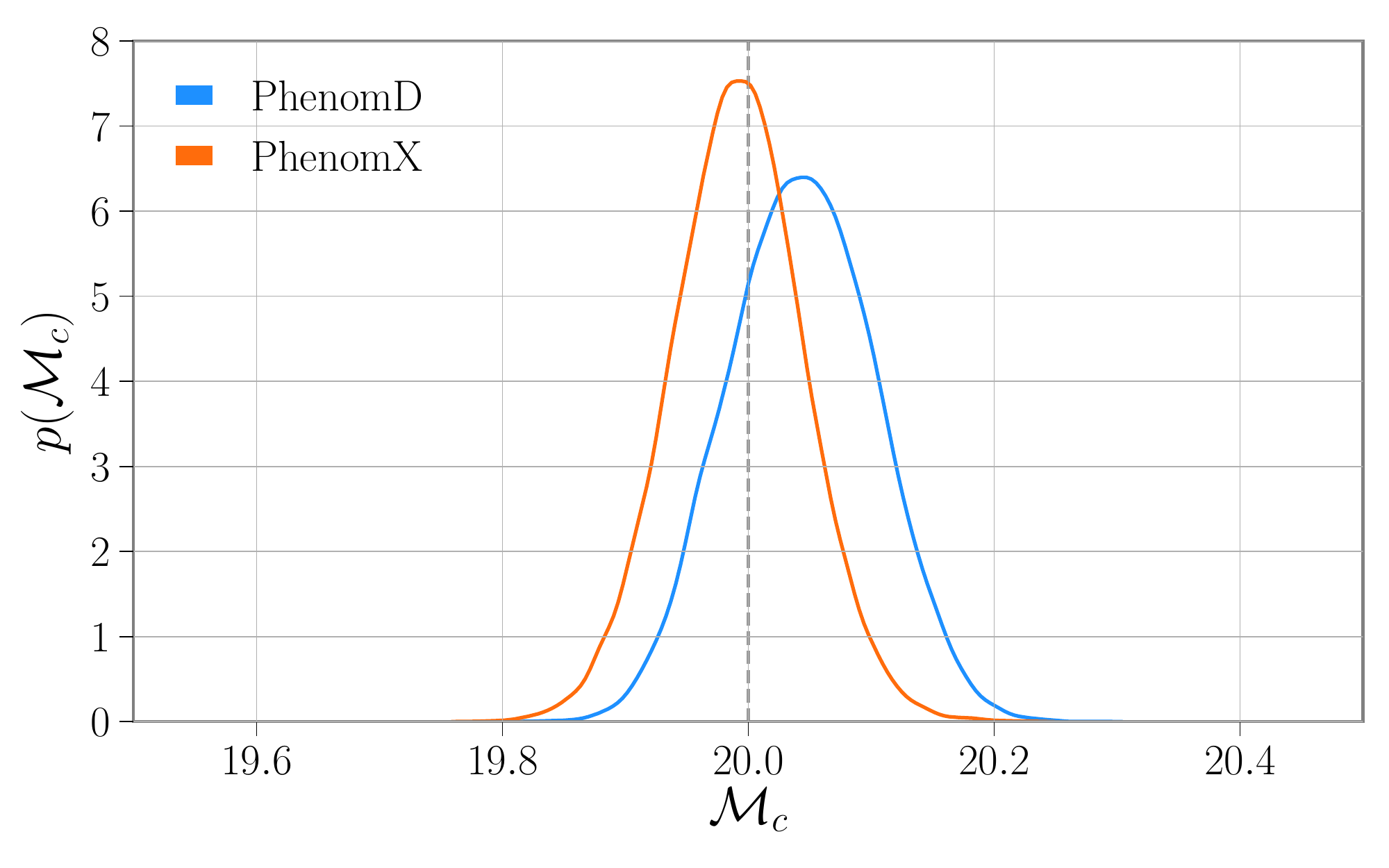}
  \includegraphics[width=0.495\textwidth]{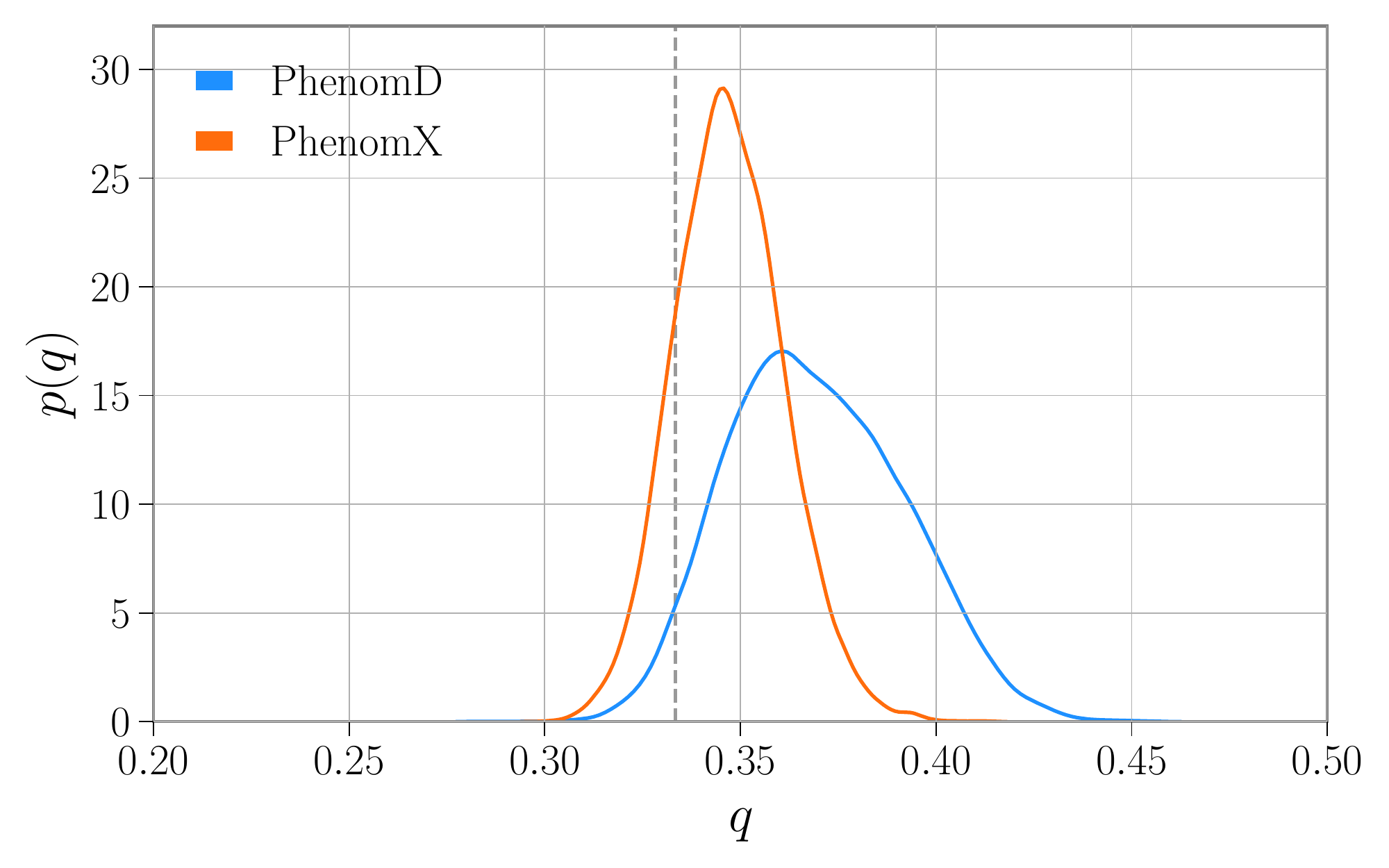}
  \includegraphics[width=0.495\textwidth]{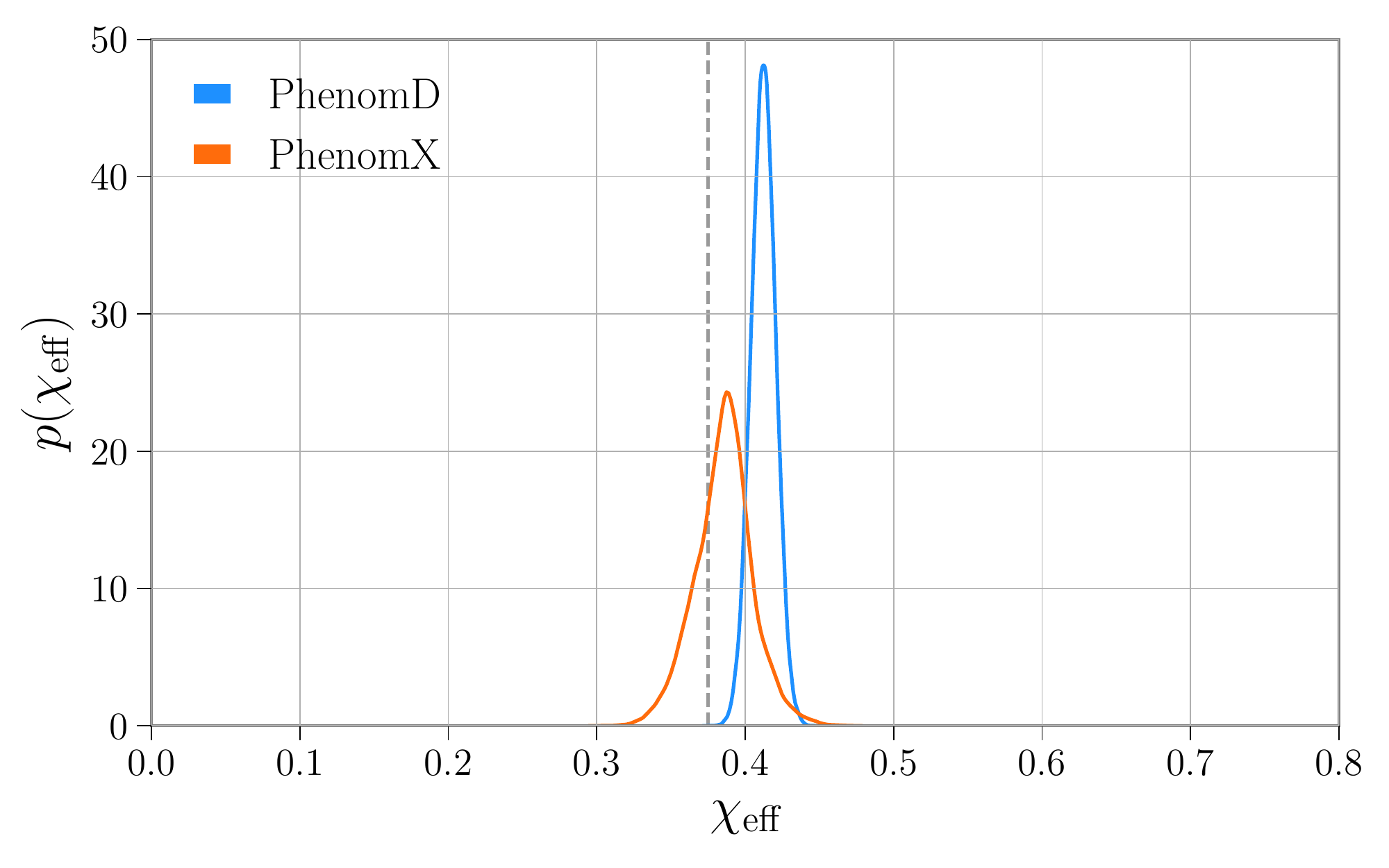}
  \caption{ \label{fig:NRHybSur3dq8_PE} %
  We show the 1D posterior distributions for the chirp mass $\mathcal{M}_c$, mass ratio $q$ and effective spin $\chi_{\rm{eff}}$ as recovered by \phX and \phD against an injected NRHybSur3dq8 waveform. The dashed line denotes the injected values. \phX demonstrates excellent recovery of the injected parameters with significantly smaller biases and tighter posteriors than those exhibited by \phD. 
}
\end{figure}
\end{centering}

\section{Conclusions}
\label{sec:summary}

In this paper we have presented a new model for the dominant $(2,\pm 2)$ spherical harmonics of the gravitational-wave signal from non-precessing, non-eccentric coalescing black holes. In gravitational wave data analysis, this model will most typically be used as part of a multi-mode waveform, where subdominant modes are included using the \phXHM model, which we present in a companion paper \cite{PhenXHM}. Furthermore, a technique to accelerate waveform evaluation is available, which drastically reduces computational cost in particular for low masses (or low start frequencies), as discussed in a second companion paper \cite{our_mb}. 
Details of how to use the LALsuite \cite{lalsuite} implementation of the model are given in Appendix C of \cite{PhenXHM}.

In the following we summarize the main improvements of \phX over \phD.
Several of the changes affect all of the three frequency regions:
\begin{itemize}
    \item The number of NR waveforms the model is calibrated to has increased from 19 to 652. While
    \phD was only calibrated to waveforms up to mass ratio 18. \phX includes waveforms up to mass ratio 1000, which were computed in the in an appropriate test-particle limit as in \cite{Keitel:2016krm}, see Sec.~\ref{sec:input}.
    \item \phD modelled 2-dimensional parameter spaces of symmetric mass ratio and effective spin (choosing different effective spins in different frequency regimes as appropriate). \phX models the complete 3-dimensional parameter space without effective spin approximations.
    \item Ad-hoc parameter space fits have been replaced by a systematic procedure designed to avoid both underfitting and overfitting \cite{Jimenez-Forteza:2016oae,Keitel:2016krm}, which proceeds by hierarchically treating sub-spaces like the non-spinning, or equal-spin systems.
    \item A dramatic improvement in the accuracy of \phX over \phD. Notably, we find 1 to 2 orders of magnitude improvement in the mismatches against the SEOBNRv4-NR hybrids across the parameter space.
    \item \phX incorporates physically motivated transition regions, with the key frequencies obeying a natural hierarchy.
    \item The improvements mentioned above also facilitated a more systematic study regarding the optimal parameterization of a given waveform model, in particular concerning the use of coefficients of basis functions versus the use of collocation points, see Sec.~\ref{sec:collocationPoints}.
    \item Finally, the implementation in the LAL software library for gravitational wave data analysis \cite{lalsuite} has been modularized, to allow independent updates for the inspiral, intermediate and ringdown regions of teh phase or amplitude models.
\end{itemize}

Our description of the inspiral region has improved due to a lower cutoff frequency of 74\% of the \phD value for the inspiral calibration, which corresponds to approximately doubling the length of the waveform in the time domain. In addition, the transition frequency from inspiral to the intermediate region is now carefully chosen as a function of parameter space, instead of set to a fixed frequency, as discussed in Sec.~\ref{sec:regions}, and different post-Newtonian orders of the underlying TaylorF2 approximant have been compared.
Modelling of the intermediate frequency region also greatly benefits from our careful choice of transition frequencies. In addition we have added further degrees of freedom for more accurate fits.

Finally, several changes affect mostly the ringdown, or more generally the highest frequencies:
\begin{itemize}
    \item Hybrid waveforms have been built from the Newman-Penrose scalar $\psi_4$ (see e.g.~\cite{hybrids}) instead of from the strain, which results in a significantly cleaner ringdown waveform.
    \item The time resolution for hybrid waveforms has been increased from $M \Delta t = 1$ to $M \Delta t = 0.5$, which benefits high spin cases with high ringdown frequencies.
    \item The fits for final spin and final mass, which are required for computing the complex ringdown frequency, have been changed from using the 2-dimensional effective spin fits of \cite{Khan:2015jqa} to modelling the full 3-dimensional parameter space dependency, which eliminates the necessity to model the discrepancy between the value from ringdown frequency according to the fits with an extra parameter.
\end{itemize}

An important challenge for the future is to improve the modelling of extreme mass ratios, and to study the transition between comparable and extreme mass ratios. An important element here will be to extend the catalogue of accurate and sufficiently long numerical relativity waveforms beyond mass ratio 18.

\section*{Acknowledgements} 

We thank the internal reviewers of the LIGO and Virgo collaboration
for their careful checking of our \texttt{LALSuite} code implementation
and their valuable feedback.
We thank Patricia Schmidt for useful discussions. 
We thank Alessandro Nagar, Sebastiano Bernuzzi and Enno Harms for giving us access to $\it{Teukode}$ \cite{Harms:2014dqa,Harms:2015ixa,Harms:2016ctx}, which was used to generate our extreme-mass-ratio waveforms. 

This work was supported by European Union FEDER funds, the Ministry of Science, 
Innovation and Universities and the Spanish Agencia Estatal de Investigación grants FPA2016-76821-P,        
RED2018-102661-T,    
RED2018-102573-E,    
FPA2017-90687-REDC,  
Vicepresid`encia i Conselleria d’Innovació, Recerca i Turisme, Conselleria d’Educació, i Universitats del Govern de les Illes Balears i Fons Social Europeu, 
Generalitat Valenciana (PROMETEO/2019/071),  
EU COST Actions CA18108, CA17137, CA16214, and CA16104, and
the Spanish Ministry of Education, Culture and Sport grants FPU15/03344 and FPU15/01319.
MC acknowledges funding from the European Union's Horizon 2020 research and innovation programme, under the Marie Skłodowska-Curie grant agreement No. 751492.
The authors thankfully acknowledge the computer resources at MareNostrum and the technical support provided by Barcelona Supercomputing Center (BSC) through Grants No. AECT-2019-2-0010, AECT-2019-1-0022, AECT-2019-2-0017, AECT-2019-1-0014, AECT-2018-3-0017, AECT-2018-2-0022, AECT-2018-1-0009, AECT-2017-3-0021, AECT-2017-3-0013, AECT-2017-2-0017, AECT-2017-1-0017, AECT-2016-3-0014, AECT2016-2-0009,  from the Red Española de Supercomputación (RES) and PRACE (Grant No. 2015133131). BAM and ET simulations were carried out on the BSC MareNostrum computer under PRACE and RES (Red Española de Supercomputación) allocations and on the FONER computer at the University of the Balearic Islands. 
Benchmarks calculations were carried out on the cluster CIT provided by LIGO Laboratory and supported by National Science Foundation Grants PHY-0757058 and PHY-0823459.

\appendix
\section{TaylorF2}\label{appendix:TaylorF2}

 Here we incorporate non-spinning corrections to $3.5$PN order, spin-orbit corrections to $3.5$PN, spin-orbit tail corrections to $4$PN, quadratic-in-spin corrections to $3$PN and the cubic-in-spin $3.5$PN corrections.

\subsection{Amplitude}
The inspiral amplitude is based on the re-expanded PN amplitude TaylorF2
\begin{align}
A_{\rm{PN}} (f ; \Xi) = A_0 \displaystyle\sum^{6}_{i=0} \, \mathcal{A}_i \, \left( \pi f \right)^{i/3} ,
\end{align}
\newline
where $\Xi = \lbrace \eta, \chi_1, \chi_2 \rbrace$. The expansion coefficients are given by
\begin{align}
\mathcal{A}_0 = 1 ,
\end{align}

\begin{align}
\mathcal{A}_1 = 0 ,
\end{align}

\begin{align}
\mathcal{A}_2 = -\frac{323}{224} + \frac{451 \eta }{168} .
\end{align}

\begin{align}
\mathcal{A}_3 ={\chi_1} \left(\frac{27 \delta }{16}-\frac{11 \eta }{12}+\frac{27}{16}\right)+{\chi_2} \left(-\frac{27 \delta }{16}-\frac{11 \eta }{12}+\frac{27}{16}\right)
\end{align}

\begin{align}
\mathcal{A}_4 &= {\chi_1}^2 \left(-\frac{81 \delta }{64}+\frac{81 \eta }{32}-\frac{81}{64}\right) \\ 
&\quad \; + {\chi_2}^2 \left(\frac{81 \delta }{64}+\frac{81 \eta }{32}-\frac{81}{64}\right) \nonumber \\ 
&\quad \; + \left( \frac{105271}{24192} \eta^2 - \frac{1975055}{338688}\eta - \frac{27312085}{8128512} \right) \nonumber \\
&\quad \; - \frac{47}{16} \, \eta \, {\chi_1} \,{\chi_2} \nonumber
\end{align}

\begin{align}
\mathcal{A}_5 &= {\chi_1}^3 \left(\delta  \left(\frac{3}{16}-\frac{3 \eta }{16}\right)-\frac{9 \eta }{16}+\frac{3}{16}\right) + \\
&\quad \, {\chi_1} \left(\delta  \left(\frac{287213}{32256}-\frac{2083 \eta }{8064}\right)-\frac{2227 \eta ^2}{2016}-\frac{15569 \eta }{1344}+\frac{287213}{32256}\right) + \nonumber \\
&\quad \, {\chi_2}^3 \left(\delta  \left(\frac{3 \eta }{16}-\frac{3}{16}\right)-\frac{9 \eta }{16}+\frac{3}{16}\right) \nonumber + \\
&\quad \, {\chi_2} \left(\delta  \left(\frac{2083 \eta }{8064}-\frac{287213}{32256}\right)-\frac{2227 \eta ^2}{2016}-\frac{15569 \eta }{1344}+\frac{287213}{32256}\right) \nonumber \\
&\quad \, -\frac{85 \pi }{64}+\frac{85 \pi  \eta }{16} \nonumber
\end{align}

\begin{widetext}
\begin{align}
\mathcal{A}_6 &=  {\chi_1} \left(-\frac{17 \pi  \delta }{12}+\left(-\frac{133249 \eta ^2}{8064}-\frac{319321 \eta }{32256}\right) {\chi_2}+\frac{5 \pi  \eta }{3}-\frac{17 \pi }{12}\right) \\
&\quad \, \nonumber
+ {\chi_1}^2 \left(\delta  \left(-\frac{141359 \eta }{32256}-\frac{49039}{14336}\right)+\frac{163199 \eta ^2}{16128}+\frac{158633 \eta }{64512}-\frac{49039}{14336}\right) \\
&\quad \, \nonumber
+ {\chi_2}^2 \left(\delta  \left(\frac{141359 \eta }{32256}+\frac{49039}{14336}\right)+\frac{163199 \eta ^2}{16128}+\frac{158633 \eta }{64512}-\frac{49039}{14336}\right)
+{\chi_2} \left(\frac{17 \pi  \delta }{12}+\frac{5 \pi  \eta }{3}-\frac{17 \pi }{12}\right) \\
&\quad \, \nonumber
-\frac{177520268561}{8583708672}
+ \left( \frac{545384828789}{5007163392} - \frac{205 \pi ^2 }{48} \right)\eta
-\frac{3248849057 \eta ^2}{178827264}
+\frac{34473079 \eta ^3}{6386688}
\end{align}
\end{widetext}

\subsection{Phase}
The underlying frequency-domain phasing model in IMRPhenomX is based on the TaylorF2 post-Newtonian approximant constructed via the application of the stationary phase approximation (SPA). For quasi-circular, non-precessing binaries, the input ingredients are the center-of-mass energy $E$ and the energy flux $F$. The canonical TaylorF2 approximant used in IMRPhenomX implements recent tail-induced spin-orbit terms at 4PN, cubic-in-spin corrections at 3.5PN and quadratic-in-spin corrections at 3PN. Schematically, the energy can be written as
\begin{align}
E &= -\frac{\eta}{2} x \left[ E_{\rm{NS}} + x^{3/2} E_{\rm{SO}} + x^2 E_{\rm{SS}} +  x^{7/2} E_{\rm{SSS}} \right]
\end{align}
\n
where $E_{\rm{NS}}$, $E_{\rm{SO}}$, $E_{\rm{SS}}$ and $E_{\rm{SSS}}$ the non-spinning, spin-orbit, quadratic-in-spin and cubic-in-spin corrections to the energy. Although the non-spinning contributions are currently known to 4PN, the baseline model presented here restricts the non-spinning contributions to 3PN. The spin-orbit terms begin at 1.5PN order are currently known to 3.5PN \cite{Marsat:2014xea,Bohe:2015ana}. The quadratic-in-spin corrections are known at next-to-leading order, coresponding to 3PN \cite{Bohe:2015ana}. The cubic-in-spin terms are currently known to leading order and enter the energy and flux at 3.5PN \cite{Marsat:2014xea}.

Similarly, the flux can be written as
\begin{align}
\mathcal{F} &= \frac{32}{5} \eta x^{5} \left[ \mathcal{F}_{\rm{NS}} + x^{3/2} \mathcal{F}_{\rm{SO}} + x^{2} \mathcal{F}_{\rm{SS}} +  x^{7/2} \mathcal{F}_{\rm{SSS}} \right]
\end{align}
\n
where  $\mathcal{F}_{\rm{NS}}$, $\mathcal{F}_{\rm{SO}}$, $\mathcal{F}_{\rm{SS}}$ and $\mathcal{F}_{\rm{SSS}}$ denote the non-spinning, spin-orbit, quadratic-in-spin and cubic-in-spin corrections to post-Newtonian energy flux.

The frequency-domain phase from the TaylorF2 terms is given by
\begin{align}
\varphi_{\rm{TF}2} \, (f ; \Xi) &= 2 \pi f t_c - \varphi_c - \frac{\pi}{4} \\
\nonumber &\qquad + \frac{3}{128 \eta} \left( \pi f M  \right)^{-5/3} \, \displaystyle\sum^7_{i=0} \, \varphi_i \left( \Xi \right) \, \left( \pi f M \right)^{i/3} .
\end{align}
\n
In \phD, the TaylorF2 baseline was based on non-spinning corrections to 3.5PN, linear spin-orbit corrections to 3.5PN and quadratic spin corrections to 2PN. In addition, upon re-expanding the PN energy and flux in deriving the TaylorF2 phase, all quadratic and higher spin corrections beyond 2PN were implicitly dropped. The coefficients used in \phX incorporate relative 1PN quadratic-in-spin corrections, the leading-order cubic-in-spin corrections and a tail-induced SO term entering at 4PN, $\varphi_8$. The coefficients detailed below define the \textit{canonical} TaylorF2 model discussed in Sec.~\ref{sec:phase_insp}
\begin{widetext}
\begin{align}
\label{eq:phase_coeff_0}
\varphi_0 &= 1 \\
\varphi_1 &= 0 \\
\varphi_2 &=\frac{55 \eta }{9}+\frac{3715}{756} \\
\varphi_3 &= \frac{113 \delta  \chi_a}{3}+\left(\frac{113}{3}-\frac{76 \eta }{3}\right) \chi_s-16 \pi \\
\varphi_4 &= -\frac{405}{4}\delta \,  \chi_a \, \chi_s + \left(200 \eta -\frac{405}{8}\right) \chi_a^2+\left(\frac{5 \eta
   }{2}-\frac{405}{8}\right) \chi_s^2+\frac{15293365}{508032} + \frac{27145 }{504}\eta + \frac{3085}{72}\eta^2 \\
\varphi_5 &= \chi_a \left(-\frac{140 \delta  \eta }{9}-\frac{732985 \delta }{2268}+\left(-\frac{140 \delta  \eta }{9}-\frac{732985 \delta }{2268}\right) \log (\pi 
   f)\right) \\
   \nonumber &\qquad +\chi_s \left(\left(\frac{340 \eta ^2}{9}+\frac{24260 \eta }{81}-\frac{732985}{2268}\right) \log (\pi  f)+\frac{340 \eta ^2}{9}+\frac{24260 \eta
   }{81}-\frac{732985}{2268}\right) \\
    \nonumber &\qquad +\left(\frac{38645 \pi }{756}-\frac{65 \pi  \eta }{9}\right) \log (\pi  f)-\frac{65 \pi }{9}\eta + \frac{38645 \pi }{756} \\
\varphi_6 &= \chi_s \left(\chi_a \left(\frac{75515 \delta }{144}-\frac{8225 \delta  \eta }{18}\right)-520 \pi  \eta +\frac{2270 \pi }{3}\right)+\frac{2270 \pi 
   \delta  \chi_a}{3}-\frac{6848}{63} \log (\pi  f)-\frac{127825 \eta ^3}{1296} \\
   \nonumber &\qquad +\left(-480 \eta ^2-\frac{263245 \eta }{252}+\frac{75515}{288}\right)
   \chi_a^2+\left(\frac{1255 \eta ^2}{9}-\frac{232415 \eta }{504}+\frac{75515}{288}\right) \chi_s^2 \\
   \nonumber &\qquad +\frac{76055 \eta ^2}{1728}+\frac{2255 \pi ^2
   \eta }{12}-\frac{15737765635 \eta }{3048192}-\frac{640 \pi ^2}{3}-\frac{6848 \gamma_E }{21}+\frac{11583231236531}{4694215680}-\frac{13696 \log (2)}{21} \\
\varphi_7 &= \chi_a \left(-\frac{1985 \delta  \eta ^2}{48}+\frac{26804935 \delta  \eta }{6048}-\frac{25150083775 \delta }{3048192}\right) \\
\nonumber &\qquad +\chi_s \left(-1140 \pi 
   \delta  \chi_a+\frac{5345 \eta ^3}{36}+\left(80 \eta ^2-7270 \eta +\frac{14585}{8}\right) \chi_a^2-\frac{1042165 \eta ^2}{3024}+\frac{10566655595
   \eta }{762048}-\frac{25150083775}{3048192}\right) \\
   \nonumber &\qquad +\chi_a^3 \left(\frac{14585 \delta }{24}-2380 \delta  \eta \right)+\chi_s^2 \left(\text{$\chi
   $a} \left(\frac{14585 \delta }{8}-\frac{215 \delta  \eta }{2}\right)+40 \pi  \eta -570 \pi \right)+\left(\frac{100 \eta ^2}{3}-\frac{475 \eta
   }{6}+\frac{14585}{24}\right) \chi_s^3  \\
   \nonumber &\qquad -\frac{74045 \pi  \eta ^2}{756}+(2240 \pi  \eta -570 \pi ) \chi_a^2+\frac{378515 \pi  \eta
   }{1512}+\frac{77096675 \pi }{254016} \\
\varphi_8 &= \pi \Bigg[ \chi_a \left(-\frac{99185}{252}   \delta  \eta +\frac{233915   \delta }{168}+\left(\frac{99185   \delta  \eta }{252}-\frac{233915   \delta
   }{168}\right) \log (\pi  f)\right) \\
   \nonumber &\qquad +\chi_s \left(\left(-\frac{19655   \eta ^2}{189}+\frac{3970375   \eta }{2268}-\frac{233915  }{168}\right) \log
   (\pi  f)+\frac{19655   \eta ^2}{189}-\frac{3970375   \eta }{2268}+\frac{233915  }{168}\right) \Bigg] .
   \label{eq:phase_coeff_8}
\end{align}
\end{widetext}

\subsection{Extending Results to $4.5$PN}
An implicit and powerful feature of the current generation of phenomenological waveform models is the implicit modularity. By separating the waveform into three key regimes we are free to recalibrate or improve aspects of the waveform model in reaction to the latest developments in the literature. A worked example of this would be the extension of the results to include the latest 4PN and 4.5PN results in the literature. For the non-spinning sector, the equations of motion for compact binaries has been derived to $4$PN \cite{Damour:2014jta,Jaranowski:2015lha,Damour:2017ced,Bernard:2015njp,Marchand:2017pir} leading to an additional non-spinning term of the form 
\begin{widetext}
\begin{align}
E^{\rm{4PN}} &= x^4 \Bigg[ \frac{77}{31104}\eta ^4 +\frac{301}{1728}\eta ^3 +\left(\frac{3157 \pi
   ^2}{576}-\frac{498449}{3456}\right) \eta ^2 +\eta  \left(\frac{448}{15} \log (16 x)+\frac{9037 \pi
   ^2}{1536}+\frac{896}{15}\gamma_E -\frac{123671}{5760}\right)-\frac{3969}{128} \Bigg] .
\end{align}
\end{widetext}
As well as the $4$PN derivation above, higher non-linear tail effects associated to quartic non-linear interactions have recently been derived from first principles in the MPM formalism \cite{Marchand:2016vox} as well as an independent derivation from the PN re-expansion of the factorized and resummed EOB fluxes \cite{Messina:2018ghh}. Such interactions lead to a $4.5$PN contribution to the flux
\begin{widetext}
\begin{align}
\mathcal{F}^{\rm{4.5PN}}_{\rm{Tail}} &= \pi \, x^{9/2} \left[ -\frac{3719141 \eta ^3}{38016}-\frac{133112905 \eta
   ^2}{290304}+\left(\frac{2062241}{22176}+\frac{41 \pi ^2}{12}\right) \eta -\frac{3424}{105} \log (16
   x)-\frac{6848 \gamma }{105}+\frac{265978667519}{745113600}\right] .
\end{align}
\end{widetext}
Another interesting contribution derived from the PN re-expansion of the EOB fluxes is the identification of a leading-order tail-induced spin-spin term in the flux \cite{Messina:2018ghh}\footnote{Using an appropriate change in spin variables from the notation of \cite{Messina:2018ghh}.}

\begin{align}
\mathcal{F}^{\rm{3.5PN}}_{\rm{LO-SS,Tail}} &= \pi \, x^{7/2} \Bigg[ \left(8  \delta^2 + \frac{1}{8}\right) \chi_a^2 
\\ &\qquad \qquad \qquad \nonumber +\left(\frac{\delta^2}{8} + 8 \right) \chi_s^2 
   + \frac{65}{4}  \delta \,  \chi_a  \,\chi_s \Bigg] ,
\end{align}
which coincides with the known test-particle limit \cite{Tagoshi:1996gh}. Adding these terms to the PN flux and energy, we find the following higher order contributions to the PN phasing
\begin{widetext}
\begin{align}
\label{eq:phase_coeff_7}
\varphi_7^{N} &= -\pi x^{7/2} \left[ -325 \, \delta \,  \chi_a \, \chi_s+\left(640 \eta
   -\frac{325}{2}\right) \, \chi_a^2 + \left(10 \eta -\frac{325}{2}\right) \,
   \chi_s^2\right] \\
\varphi_9^{N} &= \pi x^{9/2} \Bigg[ \frac{10323755}{199584}\eta^3 + \frac{45293335}{127008}\eta^2 + \left(-\frac{1492917260735}{134120448}+\frac{2255 \pi ^2}{6}\right) \eta \label{eq:phase_coeff_9} \\ \nonumber &\qquad \qquad -\frac{6848 \log (x)}{21}-\frac{640 \pi ^2}{3}-\frac{13696
   }{21}\gamma_E + \frac{105344279473163}{18776862720}-\frac{27392 \log (2)}{21} \Bigg] ,
\end{align}
\end{widetext}
in agreement with \cite{Messina:2018ghh} and \cite{Nagar:2018plt}. A somewhat more vexing task is how to incorporate, in a fully-consistent way, the incomplete knowledge at $4$PN. One possible approach, as taken in \cite{Messina:2018ghh}, is to construct an approximant that depends on as of yet unknown analytical coefficients $c_N$, allowing the incomplete $4$ and $4.5$PN terms to be included in a fully-consistent way, complete with $\eta$ dependence. Here, however, we choose to drop the unknown analytical information and instead absorb these into the pseudo-PN calibration. In practice, we do not find any significant difference between the canonical TaylorF2 approximant used and the higher order PN expressions given here after the pseudo-PN calibration is taken into account. The \textit{extended} TaylorF2 approximant discussed in Sec.~\ref{sec:phase_insp} uses the coefficients detailed in Eqs.~\ref{eq:phase_coeff_0} to \ref{eq:phase_coeff_8} plus the additional terms in Eqs.~\ref{eq:phase_coeff_7} and \ref{eq:phase_coeff_9}.

\section{Stationary Phase Approximation}
\label{appendix:SPA}
Here we overview the stationary phase approximation (SPA) applied to a time domain signal \cite{Finn:1992xs,Cutler:1994ys,Droz:1999qx}
\begin{align*}
h_{\ell m} (t) = A_{\ell m} (t) \; e^{- i \; m \; \varphi (t) } .
\end{align*}
\n
The orbital phase $\varphi$ is related to the orbital frequency by $\omega = \dot{\varphi}$. The SPA approximation is formally valid if the following criteria are met \cite{Cutler:1994ys,Finn:1992xs,Marsat:2018oam}
\begin{equation*}
\left| \frac{\dot{A} / A}{\omega} \right| \ll 1  , \quad \left| \frac{\dot{\omega}}{\omega^2} \right| \ll 1  , \quad \left| \frac{ (\dot{A} / A )^2 }{\dot{\omega}} \right| \ll 1.
\end{equation*}
\n
The SPA approximation works as the Fourier transform of a signal is highly oscillatory and unless there are strong cancellations between the orbital phase $\varphi(t)$ and the $2 \pi f t$ term, the Fourier transform will have support that is roughly centered on the point of stationary phase. This enables us to define a time as a function of the frequency 
\begin{align*}
m \; \omega \, (t_f ) &= 2 \pi f ,
\end{align*}
\n
where $t_f$ is strictly only valid in the SPA regime.  Assuming a monotonically increasing orbital phase, such that $\omega > 0$ and $\dot{\omega} > 0$, then we can expand the signal about the SPA time 
\begin{align*}
\tilde{h}_{\rm SPA} (f) \simeq A_{\ell m} (t_f) \; e^{2 \pi i f t_f - m i \varphi \, (t_f)} \; \int e^{- i \, (t - t_f)^2 \, m \, \dot{\omega} \, (t_f) \, / 2} dt
\end{align*} 
\n
Noting that $| (\dot{A}/A)^2 / \dot{\omega} | \ll 1$, we can treat the amplitude as being approximately constant. Performing the Gaussian integration, we find
\begin{align*}
\tilde{h}{\ell m} (f) &\simeq A_{\ell m} (f) \; e^{-i \Psi_{\ell m} (f)} ,\\
A_{\ell m} (f) &\simeq A_{\ell m} (t_f) \; \sqrt{ \frac{2 \pi}{m \dot{\omega} (t_f)} } , \\
\Psi_{\ell m} (f) &\simeq m \; \varphi (t_f) - 2 \pi f t_f + \frac{\pi}{4} ,
\end{align*}
\n 
where we have made use of the standard integral $\int^{\infty}_{-\infty} dx \; e^{- i x^2} = \sqrt{\pi} e^{-i \pi/4}$.
This can now be expressed in terms of the TaylorF2 phase $\varphi_{\ell m}^{\rm TF2}$ and a phase shift $\varphi_{0,\ell m}$
\begin{align*}
\Psi_{\ell m} (f) &\simeq - 2 \pi f t_0  + \frac{\pi}{4} + \varphi_{0,\ell m} + \varphi_{\ell m}^{\rm TF2} (f)
\end{align*}
\n
where 
\begin{align*}
\varphi_{0,\ell m} = \frac{m}{2} \; \varphi_{0,22} + \varphi_{\ell m}^{\rm Amp}  ,
\end{align*}
\n
and
\begin{align}
\varphi_{\ell m}^{\rm TF2} (f) &= \frac{m}{2} \varphi_{22}^{\rm TF2} \left( \frac{2 f}{m} \right) .
\end{align}
\n
The term $\varphi^{\rm Amp}_{\ell m}$ corresponds to phase corrections arising from the complex PN amplitudes and $\varphi_{0,\ell,m}$ a gauge freedom associated to phase shifts. Collecting this all together, we can write the SPA of the time domain mode as 
\begin{align}
\tilde{h}_{\ell m} (f) &= A_{\ell m} (t_f) \; \sqrt{ \frac{2 \pi }{m \dot{\omega} } } \; e^{i \left[ 2 \pi f t_f - m \varphi (t_f) - \pi / 4 \right] } . 
\end{align}
\begin{align*}
\end{align*}

%


%





\bibliography{phenomx,phenom_refs,eob_refs,nr_refs,postnewtonian,surrogates,gravwaves}





\end{document}